\documentclass[12pt]{article}
\pdfoutput=1
\usepackage{amssymb,amsmath}
\usepackage{graphicx}
\usepackage{epsfig}
\usepackage{epstopdf}
\usepackage{latexsym}
\usepackage{psfrag}
\usepackage{booktabs}
\usepackage{bbm}
\usepackage[toc]{appendix}
\usepackage{color}
\usepackage{datetime}
\usepackage[
      colorlinks=true,
      linkcolor=darkblue,  
      urlcolor=blue,    
      filecolor=blue,     
      citecolor=red,
      linktocpage=true,
      pdfstartview=FitV,
      bookmarksopen=true    
      ]{hyperref}

\definecolor{darkblue}{rgb}{0.2, 0, 0.8}

\numberwithin{equation}{section}

\newcommand{\tth}{\text{th}}
\newcommand{\inv}{^{-1}}
\newcommand{\ang}[1]{\langle #1 \rangle}

\newtheorem{theorem}{Theorem}

\newcommand{\two}[1]{\langle #1 \rangle}
\newcommand{\four}[1]{\big\langle #1 \big\rangle}
\newcommand{\five}[1]{\big[ #1 \big]}

\newcommand{\lat}{\tilde\lambda}
\newcommand{\mut}{\tilde\mu}

\newcommand{\tfour}[1]{\langle #1 \rangle}

\renewcommand{\a}{\alpha}
\renewcommand{\^}[1]{\hat{#1}}

\newcommand{\A}{\mathcal{A}}

\newcommand{\beq}{\begin{equation}}
\newcommand{\eeq}{\end{equation}}

\newcommand{\compactsubsection}[1]{\vspace{2mm}\noindent {\bf #1}\\[1mm]}

\newcommand{\CP}{\mathbb{CP}}

\newcommand{\C}{\mathbb{C}}

\newcommand{\cn}{{\cal N}}

\newcommand{\reef}[1]{(\ref{#1})}

\newcommand{\be}{\begin{equation}}
\newcommand{\ee}{\end{equation}}

\newcommand{\lda}{\langle\!\langle}
\newcommand{\rda}{\rangle\!\rangle}
\newcommand{\dtwo}[1]{\lda #1 \rda}

\def\be{\begin{equation}}
\def\ee{\end{equation}}
\def\eq{\begin{equation}}
\def\eqe{\end{equation}}
\def\eqa{\begin{eqnarray}}
\def\eqae{\end{eqnarray}}
\def\bea{\begin{eqnarray}}
\def\eea{\end{eqnarray}}
\def\ba{\begin{array}}
\def\ea{\end{array}}
\def\bd{\begin{displaymath}}
\def\ed{\end{displaymath}}

\def\a{\alpha}

\def\l{\lambda}
\def\m{\mu}

\def\s{\sigma}                                   

\def\pa{\partial}                              
\def\>{\rangle} 
\def\<{\langle} 
\def\Dsl{D \hskip-.6em \raise1pt\hbox{$ / $ } }
\def\to{\rightarrow}
\def\pa{\partial}

\newcommand{\eps}{\epsilon}
\newcommand{\lra}{\longleftrightarrow}

\begin{document}  


\begin{titlepage}

 \begin{flushright}
{\tt MCTP-14-36} \\
\end{flushright}

\vspace*{1cm}

\begin{center}

{\Large \bf  Grassmannians for scattering amplitudes \\[4mm]
in 4d $\mathcal{N}=4$ SYM and 3d ABJM}

\vspace*{1.2cm}

\hspace{-0.14in}{\bf Henriette Elvang$^{1}$, Yu-tin Huang$^{2}$, 
 Cynthia Keeler$^{1}$, Thomas Lam$^{3}$, 
\\
Timothy M.~Olson$^{1}$, Samuel B.~Roland$^{1}$, David E~Speyer$^{3}$}

\vspace{5mm}
$^{1}$Michigan Center for Theoretical Physics \&
Randall Laboratory of Physics \\
Department of Physics,\\
University of Michigan, 
450 Church St, 
Ann Arbor, MI 48109, USA\\[4mm]

$^{2}$Department of Physics and Astronomy, \\
National Taiwan University,
Taipei 10617, Taiwan, ROC\\[4mm]

$^{3}$Department of Mathematics,\\
University of Michigan, 
530 Church St, 
Ann Arbor, MI 48109, USA

\bigskip
\texttt{elvang, keelerc, tfylam, timolson, rolandsa, speyer@umich.edu, yutinyt@gmail.com} \\
\end{center}

\begin{abstract}  
Scattering amplitudes in 4d $\mathcal{N}=4$ super Yang-Mills theory (SYM)  can be described by Grassmannian contour integrals whose form depends on whether the external data is encoded in momentum space, twistor space, or momentum twistor space. After a pedagogical review, we present a new, streamlined proof of the equivalence of the three integral formulations.  
A similar strategy allows us to derive a new Grassmannian integral for 3d $\mathcal{N}=6$ ABJM theory amplitudes in momentum twistor space: it is a contour integral in an orthogonal Grassmannian with the novel property that the internal metric depends on the external data. The result can be viewed as a central step towards developing an amplituhedron formulation for ABJM amplitudes.  
Various properties of Grassmannian integrals are examined, 
including boundary properties, pole structure, and a homological interpretation of the global residue theorems for  $\mathcal{N}=4$ SYM.

\end{abstract}

\end{titlepage}

\setcounter{tocdepth}{2}
{\small
\setlength\parskip{-0.5mm}
\tableofcontents
}

\newpage

\section{Introduction}
\label{s:Introduction}
Recent years have brought remarkable progress in our understanding of the mathematical structure of scattering amplitudes, especially in the planar limit of $\cn=4$ super Yang-Mills theory (SYM).  Among the new approaches are 
the  Grassmannian formulations \cite{ArkaniHamed:2009dn,Mason:2009qx,ArkaniHamed:2009vw}, on-shell diagrams \cite{ArkaniHamed:2012nw}, and the geometrization of amplitudes in the ``amplituhedron" \cite{Arkani-Hamed:2013jha,Arkani-Hamed:2013kca,ArkaniHamed:2010gg}. 
 Many of the ideas from planar $\mathcal{N}=4$ SYM carry over to the  superconformal 3d $\cn=6$ Chern-Simons matter theory constructed by Aharony, Bergman, Jafferis and Maldacena (ABJM)~\cite{WestCSM}; see also \cite{EastCSM}. In this paper we study the Grassmannian descriptions of amplitudes in both 4d $\cn=4$ SYM  and in 3d ABJM theory. 

The Grassmannian G$(k,n)$ is the set of all $k$-planes in $n$-dimensional space; in the context of scattering amplitudes, $n$ counts the number of external particles while $k$ refers to a classification of amplitudes. The Grassmannian description of amplitudes depends on how the external data ---  particle momenta and type --- are encoded. 
For  $n$-particle N$^k$MHV amplitudes in 4d planar $\mathcal{N}=4$ SYM  
there are three formulations \cite{ArkaniHamed:2009dn,Mason:2009qx}:
\begin{itemize}
\item The  {\bf \em momentum space} formulation uses the spinor helicity representation for the external momenta, e.g.~$p_{a\dot{b}} = \lambda_a \tilde{\lambda}_{\dot{b}}$. The relevant Grassmannian  is G$(k+2,n)$. 
\item The {\bf \em twistor space} formulation encodes the data via a half-Fourier transform of the external momenta, i.e.~a Fourier transform of $\lambda$, but not $\tilde{\lambda}$. This representation makes the superconformal symmetry $SU(2,2|4)$ manifest. As in momentum space, the Grassmannian is G$(k+2,n)$. 
\item  The {\bf \em  momentum twistor} formulation is applicable in the planar limit and  makes the {\it dual} superconformal symmetry $SU(2,2|4)$ manifest. The relevant Grassmannian is  G$(k,n)$.
\end{itemize}
In Section \ref{s:backgr}, we provide a pedagogical review of these three representations of the 4d external data and present the corresponding Grassmannian integrals explicitly.

The three Grassmannian descriptions are directly related. The relation between the 4d SYM twistor space and momentum space integrals was utilized already in the early literature \cite{ArkaniHamed:2009dn} on the subject. The momentum twistor Grassmannian was introduced shortly after in \cite{Mason:2009qx}. Its relation to the two other formulations was given in \cite{ArkaniHamed:2009vw} using a set of intricate integral manipulations. In particular, the argument of  \cite{ArkaniHamed:2009vw} uses a gauge fixing that  
breaks little group scaling and therefore results in very complicated Jacobians that are difficult to write explicitly. In Section \ref{s:TandMT}, we present a new version of the proof, valid for all $n$ and $k$, that manifestly preserves the little group scaling at each step of the calculation and yields all Jacobians as simple explicit expressions. 

In Section \ref{s:res} we address how the Grassmannian integrals are evaluated as contour integrals and demonstrate this with the explicit computation of the residues in the $n$-point NMHV sector using the momentum twistor Grassmannian integral. The results for the individual residues are 
known to be 
the dual superconformal invariant building blocks of the tree amplitudes, i.e.~the ``$R$-invariants" \cite{Drummond:2008vq} or ``5-brackets" \cite{Mason:2009qx}. These invariants obey a set of linear relations, which in the Grassmannian formulation simply follow from global residue theorems. 
We  provide a homological interpretation of the residue theorems for the NMHV residues. We then review the structure of physical versus spurious (unphysical) poles of the residues and how this gives a specification of contours for which the  Grassmannian integral exactly produces the tree-level amplitudes. 
The results are then rephrased in the context of on-shell diagrams and the positroid stratification of the Grassmannian, and we show how the `boundary operation' in the Grassmannian integral is related to both the residue theorems and the pole structure. 

In Section \ref{sec:3DGrass}, we turn to the Grassmannian descriptions of amplitudes in 3d ABJM theory. We briefly review the momentum space spinor helicity formalism in 3d and the associated Grassmannian integral for ABJM  amplitudes, which was introduced previously in \cite{LeeOG,HW,HWX,KimLee}. It encodes the $n{=}(2k{+}4)$-point N$^k$MHV 
amplitudes of ABJM theory as contour integrals 
in the {\it orthogonal Grassmannian} OG$(k+2,2k+4)$. The space OG$(k,n)$ is equipped with the metric $g_{ij}=\delta_{ij}$, $i,j=1,2,\dots,n$, and consists of $k$-dimensional null planes in $\mathbb{C}^{n}$; in other words, the 
$k \times n$ matrices $B \in \text{OG}(k,n)$ satisfy $B g B^T = 0$.

A momentum twistor version of the ABJM Grassmannian integral has not previously been constructed. We achieve this goal in Section \ref{sec:3DGrass}, which is an important first step towards developing an amplituhedron for the ABJM theory.  To this end, we first introduce another  formulation of the 3d spinor helicity formalism that facilitates the definition of 3d momentum twistors. 
 They are simply the 4d momentum twistors $Z_i^A$ ($A=1,2,3,4$) subject to the bi-local $SO(2,3) \sim Sp(4)$-invariant constraint $Z_i^A Z_{i+1}^B \Omega_{AB}=0$ for all $i=1,2,\dots,n$.

Next, our streamlined proof of the relation between the 4d Grassmannian integral representations in Section \ref{s:TandMT} allows us to derive the desired momentum twistor version of the Grassmannian integral for ABJM theory. Just like the momentum space Grassmannian integral, the new integral has an orthogonality constraint, but a novel feature is that the $k$-dimensional planes are now null with respect to a metric defined by $Sp(4)$-invariant inner products of the momentum twistors, $Z_i^A Z_j^B \Omega_{AB}$. 

Orthogonal Grassmannians defined by non-trivial metrics for Grassmannians have been encountered previously in the mathematics literature in the context of electrical networks and related combinatorics \cite{HS,Lam}.  However, the dependence of the metric on external data appears to be a new property that would be exciting to explore further.  In particular, it plays a crucial role for the boundary properties needed for the physical poles of the 6-point ABJM tree-amplitude, as we discuss in some detail. 

We end in Section \ref{s:out} with a brief outlook to Grassmannians beyond the NMHV level and open questions. A few technical results are relegated to  appendices.

\section{$\cn=4$ SYM and the Grassmannian}
\label{s:backgr}
This section is intended as a  short review of  the Grassmannian formulation of amplitudes in planar $\cn=4$ SYM.  Sections \ref{s:extdata1} and  \ref{s:extdata2} introduce the basic definitions and concepts needed for (super)amplitudes in $\cn=4$ and present the three different forms of the external data: momentum space, twistor space, and momentum twistor space. See \reef{input} for an overview. Section \ref{s:grassint} presents the Grassmannian integrals. Experts can skip ahead to  Section \ref{s:TandMT}.

\subsection{External data: momentum space and superamplitudes}
\label{s:extdata1}
The external data for an amplitude encodes the information about the initial and final state particles in the scattering process; practically we take all states to be outgoing. We are here considering only amplitudes in $\cn=4$ SYM, so each of the $n$ external particles --- labeled $i=1,2,\ldots,n$ --- is specified by a null momentum $p_i$, i.e.~$p_i^2 = 0$, along with a specification of the particle type (e.g.~gluon, gluino, or scalar). The scattering amplitude $A_n$  takes this external data as input and returns a complex number on the support of a delta function that enforces momentum conservation $\delta^4(p_1 + \ldots + p_n)$. 

In 4d, a null vector $p_i^\mu$ is conveniently written as a $2\times 2$ matrix $p_i^{\dot{a}a}$ with vanishing determinant. Because it has rank 1, the matrix can be expressed as a product of two 2-component vectors:\footnote{The spinor helicity conventions used in this paper are chosen to conform with much of the literature on Grassmannians. They differ from those used in the recent review \cite{Elvang:2013cua}.} 
$p_i^{\dot{a}a} = \lambda^a \tilde{\lambda}^{\dot{a}}$.  Thus, for an $n$-particle amplitude with $n$ external massless particles, the on-shell momenta $p_i$ with $p_i^2=0$ are specified as $\big(\lambda_i,\tilde\lambda_i \big)$. For the purpose of exploring the mathematical properties of amplitudes, it is useful to work with complex-valued momenta. In that case, $\lambda_i$ and $\tilde\lambda_i$ are independent. (Alternatively, we can keep $p_i$ real and work with a metric with signature $(-,-,+,+)$.)

The scattering amplitudes are built from Lorentz-invariant contractions of the spinors, such as the angle bracket
\be
  \<ij\> := \eps_{ab}\lambda_i^a\lambda_j^b\,,
  \label{defAngle}
\ee  
constructed with the help of the antisymmetric Levi-Civita symbol, here  $\eps^{12} =-\eps^{21} = 1 = -\eps_{12}  = \eps_{21}$, of the $SL(2)$ subgroup of the 4d Lorentz group $SO(3,1)$. Lorentz indices are often suppressed in our presentation. 

The physical spectrum of $\cn=4$ SYM consists of 16 massless particles: the gluon $g^{\pm}$ with helicity states $h=\pm1$, four gluinos $\Lambda^A$ and $\Lambda_A$ with $h=\pm\tfrac{1}{2}$, and six scalars $S^{AB}$ with $h=0$;  $A,B=1,2,3,4$. The helicity $h$ states transform in rank $r=2{-}2h$ fully antisymmetric representations of the global $SU(4)$ $R$-symmetry of $\cn=4$ SYM.  
It is very convenient to encode the states using anticommuting Grassmann variables $\tilde{\eta}_{iA}$, with fundamental  $SU(4)$ index $A=1,2,3,4$ and particle label $i=1,2,\dots,n$; Grassmann monomials are in one-to-one correspondence with the states, e.g.~$\tilde{\eta}_{31} \tilde{\eta}_{33} \tilde{\eta}_{34}$ means that particle 3 is a negative helicity gluino 
$\Lambda^{134} \sim \Lambda_2$. 

The $n$-point component amplitudes $A_n$ combine into {\bf \em superamplitudes} $\mathcal{A}_n^\text{N$^k$MHV}$ (or more generally $\mathcal{A}_n$), which are  polynomials of degree $4(k+2)$ in the Grassmann variables.   $R$-symmetry requires $\mathcal{A}_n$  to be an $SU(4)$ singlet, hence the Grassmann degree of each term must be a multiple of 4. The label N$^k$MHV stands for (Next-to)$^k$ Maximally Helicity Violating --- this sector of amplitudes consists of all gluon amplitudes with $k+2$ negative helicity gluons and $n-k-2$ positive helicity gluons, as well as all amplitudes related to those via supersymmetry. The sector with $k=0$ is simply called MHV. The coefficient of a given Grassmann monomial in $\mathcal{A}_n$ is a component amplitude whose external states are those dictated by the Grassmann variables; for example, the coefficient of the monomial $(\tilde{\eta}_{11} \tilde{\eta}_{12} \tilde{\eta}_{13} \tilde{\eta}_{14})( \tilde{\eta}_{23} \tilde{\eta}_{24})(\tilde{\eta}_{41} \tilde{\eta}_{42})$ is the component amplitude $A_n\big[g^-(p_1) S^{34}(p_2) g^+(p_3) S^{12}(p_4)  g^+(p_6)  \ldots g^+(p_n)\big]$.

The $SU(N)$ gauge group of $\cn=4$ SYM dresses the amplitudes with a color-structure that factorizes from the kinematic information. Amplitudes in the {\em planar} theory have a single trace of $SU(N)$ generators,\footnote{For further details, see Section 2.5 of the review \cite{Elvang:2013cua}.} and as a result the planar $n$-particle superamplitudes are invariant 
under {\em cyclic permutations} of the external labels, i.e.~under $i \to i+1$ mod $n$. 

The MHV sector is the simplest. The tree-level MHV superamplitude is given by the supersymmetrization of the Parke-Taylor gluon amplitude \cite{Parke:1986gb,Nair:1988bq}:
\be
  \mathcal{A}_n^\text{MHV} 
  =
  \frac{\delta^4\big( \sum_{i=1}^n\lambda_i \tilde{\lambda}_i \big)
  \, \delta^{(8)}\big(  \sum_{i=1}^n \lambda_i \tilde{\eta}_i \big)}
  {\<12\> \<23\> \cdots \<n1\>}\,.
  \label{AMHV}
\ee
The four bosonic delta functions in \reef{AMHV} encode momentum conservation via $ p_i=\lambda_i \tilde{\lambda}_i$, while the Grassmann delta function,\footnote{\label{footie:delta}The bosonic delta functions are defined as standard in distribution theory \cite{MR0435831}, i.e.~they have the property that $\int_{\mathbb{R}^m} d^m x\,\delta^m(x-x_0) \,f(x) = f(x_0)$ for any suitable test function $f$. Similarly, we use
$\int d^{m}x\, d^{n}y \, \delta^{n}\big( g(x,y) \big) f(x,y)
=  \int d^m x\,  \sum_{g(x,y)=0} f(x,y)/\det(dg/dy)$. 
The Grassmann delta functions are defined by the same property,  
$\int d\eta \, \delta^{(1)}(\eta-\eta_0) \, f(\eta) = f(\eta_0)$, but using the Berezin integral $\int d\eta \, \eta = 1$ and $\int d\eta \,  1 = 0$. Thus, the Grassmann delta function is simply $\delta^{(1)}(\eta) = \eta$ and 
$\delta^{(2)}\big(\sum_i\lambda_i \eta_i\big) = \tfrac{1}{2}\sum_{i,j} \<i j\> \eta_{i} \eta_{j}$. The superscript on the Grassmann delta-function indicates its polynomial Grassmann degree.
}
defined as
\be
  \delta^{(8)}\Big(  \sum_{i=1}^n \lambda_i \tilde{\eta}_i \Big)
  := 
  \frac{1}{2^4} \prod_{A=1}^4 \sum_{i,j=1}^{n}
    \<ij\>\, \tilde{\eta}_{iA} \tilde{\eta}_{jA} \,,
\ee
ensures conservation of $\cn=4$ supermomentum,  
$q_{iA} := \lambda_i \tilde{\eta}_{iA}$. The superamplitude \reef{AMHV} clearly has cyclic symmetry.

To summarize, for $n$-particle superamplitudes in $\cn=4$ SYM, the external data is specified in terms of the set $\big(\lambda_i,\tilde\lambda_i \,|\, \tilde{\eta}_{iA}\big)$ for $i=1,2,\dots,n$. We call this 
the {\bf\em momentum space} representation of the external data (or sometimes `on-shell superspace').

The momentum space external data has a redundancy known as {\bf \em little group scaling}:
\begin{align}
\l_i\to t_i\l_i\,,~~~~~
\tilde\l_i \to t_i^{-1}\tilde\l_i\,,~~~~~
\tilde{\eta}_i \to t_i^{-1}\tilde{\eta}_i\,,
 \label{litgrp}
\end{align}
for each $i=1,2,\dots,n$. A component amplitude scales homogeneously under little group scaling with weight $t_i^{-2h_i}$, where $h_i$ is the helicity of the  $i^\tth$ particle. The scaling of the Grassmann variables ensures uniform weight for all external states in a superamplitude:
\be
  \label{litgrpAmp}
  \mathcal{A}_n \to t_i^{-2}  \mathcal{A}_n   \,.
\ee
for each $i=1,2,\dots,n$. Little group scaling plays a key role in several explorations of scattering amplitudes, including the work we present in this paper. 

 \subsection{External data: twistor space and momentum twistor space}
\label{s:extdata2}
In addition to the momentum space representation, we will be using two other formulations for the 4d external data, namely twistor space and momentum twistor space. We describe each in turn. 

{\bf \em Twistor space} is obtained from momentum space via a Fourier transform of 
$\lambda_j$, formally via 
\be
  \int d^2 {\lambda}_j \, \exp(-i \tilde{\mu}^{{a}}_j{\lambda}_{j{{a}}})\, \bullet \,,
  \label{FTdef}
\ee 
for  each $j=1,2,\ldots,n$. The bullet indicates the expression that is Fourier transformed. (We are ignoring factors of $2\pi$ in all Fourier transforms here and henceforth as these only amount to overall normalizations.) The external data is then encoded in the 4-component twistor $W_i = (\tilde{\mu}_i,\tilde{\lambda}_i)$ and its companion, the supertwistor 
$\mathcal{W}_i= (\tilde{\mu}_i,\tilde{\lambda}_i\, |\, \tilde{\eta}_i)$. Under little group scaling \reef{litgrp}, we have $\tilde{\mu}_i \to t_i^{-1} \tilde{\mu}_i$, so the (super)twistor scales uniformly, e.g.~$\mathcal{W}_i \to t_i^{-1} \mathcal{W}_i$. 
The measure in the integral \reef{FTdef} scales as $t_i^{2}$, so this exactly compensates the little group scaling of the superamplitude \reef{litgrpAmp}. Thus, after the half-Fourier transformation for all $j=1,2,\ldots,n$, the superamplitude is invariant under little group scaling. In other words, the superamplitude in twistor space is defined projectively, 
and the twistors $W_i$ and supertwistors $\mathcal{W}_i$ are  homogeneous coordinates of projective space, $\mathbb{CP}^3$ and  $\mathbb{CP}^{3|4}$, respectively.

The third description of the external data uses the 4-component {\bf \em momentum twistors} $Z_i = (\lambda_i, \mu_i)$ \cite{Hodges:2009hk} and their momentum supertwistor extensions $\mathcal{Z}_i = (\lambda_i, \mu_i \,|\,\eta_i)$.   The 2-component spinors $\mu_i$ 
are defined via incidence relations\footnote{For a more comprehensive review of dual space and momentum twistors, see Section 5.4 of \cite{Elvang:2013cua}.}
\be
  \mu_i := \lambda_i y_i = \lambda_i y_{i+1}\,,
  \label{incidence4d}
\ee 
where the dual space coordinates $y_i$  are defined in terms of the momenta as
\be
  p_i = y_i - y_{i+1}\,.
  \label{pandy}
\ee
The second relation in \reef{incidence4d} follows from the Weyl equation, $p_i \lambda_i = 0$. The definition \reef{pandy} makes momentum conservation automatic via the identification $y_{n+1}=y_1$. The on-shell condition $p_i^2=0 $ requires `adjacent' points $y_i$ and $y_{i+1}$ to be null separated. Dual conformal symmetry acts on the dual space variables $y_i$ in the familiar way, e.g.~under dual inversion we have
$y_i \to y_i/y_i^2$. 

The geometric interpretation of the incidence relations \reef{incidence4d} is that a point $Z_i = (\lambda_i, \mu_i)$ in momentum twistor space corresponds to a null line defined by the points $y_i$ and $y_{i+1}$ in dual space. Similarly, the line defined by $Z_{i-1}$ and $Z_i$  in momentum twistor space maps to a point in dual space via
\be
  y_i = \frac{\lambda_i \mu_{i-1}-\lambda_{i-1} \mu_{i} }{\<i-1,i\>} \,.
  \label{y-lambda-mu}
\ee
This follows from \reef{incidence4d}. 

In our applications, we need to be able to map directly from momentum space variables $(\lambda_i, \tilde\lambda_i \,|\, \tilde{\eta}_i)$ to momentum twistor variables $\mathcal{Z}_i = (\lambda_i, \mu_i \,|\,\eta_i)$. This is done via the relations
\be
  \begin{split}
  \tilde{\lambda}_{i} 
  &=~
  \frac{\two{i+1,i}\mu_{i-1} 
     + \two{i,i-1}\mu_{i+1}
     + \two{i-1,i+1}\mu_{i} 
      } 
      {\two{i-1,i}\two{i,i+1}}\,,
   \\[2mm]
  \tilde{\eta}_{iA} 
  &=~
  \frac{\two{i+1,i}\eta_{i-1,A} 
      + \two{i,i-1}\eta_{i+1,A}
      + \two{i-1,i+1}\eta_{iA} } 
      {\two{i-1,i}\two{i,i+1}}\,.
  \end{split}
  \label{inc}
\ee
It follows from \reef{inc} that both $\mu_{i}$ and $\eta_i$  scale linearly with $t_i$  under little group transformations, so the momentum (super)twistors scale uniformly, e.g.~$\mathcal{Z}_i \to t_i \mathcal{Z}_i$. Therefore, the $\mathcal{Z}_i$ naturally live in projective space, $\mathbb{CP}^3$ and  $\mathbb{CP}^{3|4}$. With the external data given in momentum twistor space, the superamplitude still scales uniformly as  in \reef{litgrpAmp}. However, as we shall see, one can split off the  MHV superamplitude \reef{AMHV} as an overall factor; it takes care of the scaling properties 
and leaves behind an object that is invariant under little group scaling and therefore projectively well-defined.

The relations between the three different forms of the external data can be summarized compactly as follows:
\be
   \hspace{-2mm}
     \boxed{
   \begin{array}{ccccccc}
   \text{twistor space}  &\multicolumn{3}{c}{\hspace{6mm}\text{momentum space}} & \text{momentum twistor space}\\[3mm]
   \mathcal{W}_i = (\tilde{\mu}_i,\tilde{\lambda}_i \, | \, \tilde{\eta}_i) 
   &\!\!\longleftrightarrow\!\!& 
  (\lambda_i,\tilde{\lambda}_i \, | \, \tilde{\eta}_i) ~~
   &\!\!\longleftrightarrow\!\!& 
   \mathcal{Z}_i = (\lambda_i, \mu_i \,  | \, \eta_i)\\[-2mm]
   &
   \!\!\text{\tiny Fourier transform}\!\!
   &&
   \!\!\!\!\!\!\!\!\!\text{\tiny incidence relations}\!\!\!\!\!\!\!\!\!
   \\[-3mm]
   &
   \text{\tiny eq \reef{FTdef}}
   &&
   \text{\tiny eq \reef{inc}}
   \end{array}}
   \label{input}
\ee
Superamplitudes in $\cn=4$ SYM enjoy superconformal symmetry $SU(2,2|4)$; the action of this symmetry is linearized in  (super)twistor variables. In the planar limit, the superamplitudes (at tree-level or more generally the loop-{\em integrands}) also have dual superconformal symmetry $SU(2,2|4)$ whose action is linearized in the momentum twistor description. The generators of the `ordinary' and dual superconformal symmetries can be arranged to generate an infinite-dimensional algebra called the $SU(2,2|4)$ Yangian. 
Further details of the representation of the amplitudes and their symmetries can be found in \cite{Elvang:2013cua}.

\subsection{Grassmannian integrals}
\label{s:grassint}

The complex Grassmannian G$(k,n)$ is the space of $k$-planes in $\mathbb{C}^n$. A $k$-plane can be described as a collection of $k$ $n$-component vectors. Since any $GL(k)$ rotation of the vectors yield the same $k$-plane, the Grassmannian G$(k,n)$ can be given equivalently in terms of $k \times n$ matrices modulo $GL(k)$. The dimension of G$(k,n)$ is therefore $kn -k^2 = k(n-k)$.  Here, we will list and briefly describe the three Grassmannian integrals relevant for amplitudes in $\mathcal{N}=4$ SYM; the actual connection to the amplitudes is made  in Section \ref{s:res}.

\subsubsection{Grassmannian with twistor space data $\mathcal{W}$}
In terms of twistor variables, $\mathcal{W}$, the relevant Grassmannian integral was first presented in \cite{ArkaniHamed:2009dn}. For the  N$^k$MHV sector of $n$-point superamplitudes, the associated Grassmannian is G$(k+2,n)$ and in this space we study the integral 
\begin{equation}
  \tilde{\mathcal{L}}_{n;k}(\mathcal{W})
  =
  \int \frac{d^{\tilde{k}\times n} B}{GL(\tilde{k})}\,
  \frac{\delta^{4\tilde{k}|4\tilde{k}}\big(B\cdot\mathcal{W}\big)}{m_1 m_2 \dots m_n}\,.
  \label{GrassW}
\end{equation}
Here $\tilde{k} = k+2$
and 
the $m_i$'s are the $\tilde{k} \times \tilde{k}$ consecutive  minors of the matrix $B$, i.e.~$m_1 = (1\,2\ldots \tilde{k})_B$, $m_2 = (2\, 3 \ldots \tilde{k}+1)_B$, $\dots$,
$m_n = (n\,1\ldots \tilde{k}-1)_B$. The integral \reef{GrassW} should be understood as a contour integral; this will be discussed in Section \ref{s:contours} and more concretely in Section \ref{s:NMHVeval}.\footnote{The integral \reef{GrassW} 
exhibits two conventions typical of this field. First, we write an integral over a parameter space with $r$ complex parameters to mean that the integral will be taken over a real $r$-dimensional contour to be specified later. Second, let $X$ be a parameter space on which some connected group $G$ acts, let $d X$ be a $G$-invariant volume form on $X$ and choose a left invariant volume form $\mu_G$ on $G$. We write $d X/G$ for the volume form on $X/G$ so that, if we locally identify a patch on $X$ with a product of a patch on $X/G$ and a patch on $G$, then $dX = (dX/G) \times \mu_G$. We do not actually specify the measure $\mu_G$, since it only adds a global constant factor.}  
The external data  enters the integral \reef{GrassW} only via the argument of the delta-functions  $B \cdot \mathcal{W} = \sum_{i=1}^n B_{\alpha i}\mathcal{W}_i$ with $\alpha = 1,2, \dots, \tilde{k}$. Specifically, we have
\be
   \delta^{4\tilde{k}|4\tilde{k}}\big(B\cdot\mathcal{W}\big)
   =
   \prod_{\alpha=1}^{\tilde{k}}  
   \delta^{4}\Big( \sum_i B_{\alpha i} W_i\Big) \,
   \delta^{(4)}\Big( \sum_i B_{\alpha i} \tilde{\eta}_i\Big) \,
\ee
with  the sum over $i=1,2,\ldots, n$. Note two simple properties of \reef{GrassW}:
\begin{itemize}
 \item Little group scaling, $\mathcal{W}_i \to t_i^{-1} \mathcal{W}_i$, can be absorbed  via a scaling of the $i^\tth$ column of $B$: $B_{\alpha i} \to t_i B_{\alpha i}$ for all $\alpha =1,\ldots,\tilde{k}$. The $i^\tth$ column is included in exactly $\tilde{k}$ minors, so the scaling of the product of minors is $t_i^{\tilde{k}}$ and this precisely cancels the scaling of the measure $d^{\tilde{k}\times n}B$. Thus, $\tilde{\mathcal{L}}_{n;\tilde{k}}$ is invariant under little group scaling; it is projectively defined, just as are  the superamplitudes in twistor space. 
 \item
 $\tilde{\mathcal{L}}_{n;\tilde{k}}$ produces objects of Grassmann degree $4 \tilde{k}=4(k+2)$ which is the same as for superamplitudes in the N$^{k}$MHV sector.
\end{itemize}

\subsubsection{Grassmannian with momentum space data $(\lambda,\tilde{\lambda}\,|\,\tilde{\eta})$}
\label{s:grassmMom}
In momentum space, the Grassmannian for $n$-point N$^k$MHV amplitudes is also G$(k+2,n)$. The integral can be written
\be
  {\mathcal{L}}_{n;k}\big(\lambda,\tilde{\lambda},\tilde{\eta}\big) 
  =
  \int \frac{d^{\tilde{k}\times n}B}{{GL}(\tilde{k})}\,
  \frac{
\delta^{2\tilde{k}}\big(B_{\alpha i}\, \tilde{\lambda}_i\big)\,
\delta^{2(n-\tilde{k})}\big({B}^\perp_{\alpha i} \lambda_i\big)\,
\delta^{(4\tilde{k})}\big(B_{\alpha i}\,\tilde{\eta}_{iA}\big)
}{m_1 m_2 \cdots m_n}
\,,\\
\label{GrassM}
\ee 
where $\tilde{k}=k+2$ and $B^\perp$ is the $(n-\tilde{k})\times n$ matrix parameterizing the $(n-\tilde{k})$-plane orthogonal to the $\tilde{k}$-plane defined  by $B$; i.e.~$B(B^\perp)^T = 0$.\footnote{$B^\perp$ is defined only up to a $GL(n-\tilde{k})$ redundancy, but after fixing the $GL(\tilde{k})$ of $B$ we can choose a canonical $B^\perp$ to avoid ambiguities.}

The momentum space Grassmannian integral \reef{GrassM} has $2n$ bosonic delta-functions whereas the twistor space version \reef{GrassW} has $4(k+2)$; the difference arises from the Fourier transformations that relate \reef{GrassW} and \reef{GrassM}, as we review in detail in Section \ref{s:TandMT}.
The first $2\tilde{k}$ delta functions in \reef{GrassM} require that the 2-plane defined by the $n$ $\tilde{\lambda}_i$'s must lie in the orthogonal complement to the $\tilde{k}$-plane defined by $B$. The remaining  $2(n-\tilde{k})$ delta functions require the $\lambda$ 2-plane to be in the orthogonal complement of $B^\perp$; i.e.~the $\lambda$-plane must be contained in $B$. Hence, the bosonic delta functions require  the 2-planes defined by $\lambda$  and $\tilde{\lambda}$ to be orthogonal: $\sum_i \lambda_i \tilde{\lambda}_i = 0$.  This is just momentum conservation. We conclude that 4 of the $2n$ bosonic delta-functions in \reef{GrassM} simply enforce a condition on the external data, thus leaving constraints only on $2n-4$ of the integration variables $B$. 


\subsubsection{Grassmannian with momentum twistor space data $\mathcal{Z}$}
The G$(k,n)$  Grassmannian integral with external data given in momentum twistor space was introduced in \cite{Mason:2009qx}. For the N$^k$MHV sector with $n$ external particles it is 
\begin{equation}
  \mathcal{L}_{n;k}(\mathcal{Z})
  = 
  \mathcal{A}_n^{\text{MHV}}
  \int \frac{d^{k\times n} C}{GL(k)}\,\frac{\delta^{4k|4k}\big(C\cdot\mathcal{Z}\big)}{M_1 M_2 \cdots M_n}\,.
  \label{GrassZ}
\end{equation}
The $k \times k$ minors of the matrix $C$ are $M_1 = (1\,2\ldots k)_C$, etc., and the overall factor is the MHV superamplitude \reef{AMHV}.

As above we note that
\begin{itemize}
 \item The integral on the RHS of \reef{GrassZ} is invariant under little group scaling $\mathcal{Z}_i \to t_i \mathcal{Z}_i$ after a compensating scaling by $t_i^{-1}$ of the $i^\tth$ column of $C$. However, the MHV factor scales as 
 $t_i^{-2}$. Thus $\mathcal{L}_{n;k}(\mathcal{Z}) \to t_i^{-2}\mathcal{L}_{n;k}(\mathcal{Z})$; this is precisely the scaling \reef{litgrpAmp} needed for superamplitudes in momentum twistor space. 
 \item
 ${\mathcal{L}}_{n;{k}}$ produces objects of Grassmann degree $4 k + 8$, with the ``$+8$" arising from the MHV factor. This is the correct count for superamplitudes in the N$^{k}$MHV sector.
\end{itemize}

\subsubsection{Contours}
\label{s:contours}

Beyond the comments about little group scaling and Grassmann degrees, we have not yet established the connection between the Grassmannian integrals and superamplitudes. The first step is to define what is actually meant by the integrals. The idea is the same in all three cases, so we focus on the momentum twistor integral  \reef{GrassZ}. 

Fixing the $GL(k)$ invariance  of ${\mathcal{L}}_{n;{k}}$ in  \reef{GrassZ} leaves an integral over $k(n-k)$ variables. Of these, the bosonic delta functions localize $4k$. Thus, we are left with $k(n-k-4)$ variables to be integrated. The prescription is to interpret the integrals as $k(n-k-4)$-dimensional contour integrals. We can consider contours that select  $k(n-k-4)$ simultaneous zeros of the minors. For each such contour $\gamma$, the integral \reef{GrassZ} computes a $k(n-k-4)$-dimensional residue
${\mathcal{L}}_{n;{k}}^{(\gamma)}$. The sum of certain sets of such residues turns out to be exactly the N$^k$MHV tree superamplitude in momentum twistor space: denoting the corresponding contour $\Gamma_\text{tree}$, we therefore have ${\mathcal{L}}_{n;{k}}^{(\Gamma_\text{tree})}(\mathcal{Z}) = \mathcal{A}_{n,\text{tree}}^\text{N$^k$MHV}(\mathcal{Z})$. We demonstrate the explicit calculation of the individual NMHV residues in Section \ref{s:res} and discuss the associated global residue theorems in Section \ref{s:resrels}. 
The NMHV `tree-contour' $\Gamma_\text{tree}$ is described in Section \ref{s:resapp}.

The Grassmannian integrals \reef{GrassW}, \reef{GrassM}, and \reef{GrassZ} are directly related. This was argued in \cite{ArkaniHamed:2009vw} and we now provide a streamlined proof.

\section{Relating the three Grassmannian formulations}
\label{s:TandMT}
The twistor space and momentum space Grassmannian integrals \reef{GrassW} and \reef{GrassM} are easily related via the half-Fourier transform \reef{FTdef}; for completeness we review this below. The derivation of the momentum twistor Grassmannian integral \reef{GrassZ} from either of the other two integrals requires more effort since one needs to reduce the Grassmannian G$(k+2,n)$ to G$(k,n)$. As noted in the Introduction, this was first done in  \cite{ArkaniHamed:2009vw}. We present here a streamlined and more explicit version of the proof; this will be useful for deriving the equivalent momentum twistor Grassmannian integral for ABJM theory in Section \ref{sec:3DGrass}.

\subsection{From twistor space to  momentum space}
The Grassmannian integral in twistor space $\widetilde{\mathcal{L}}_{n;k}\big(\mathcal{W}\big)$ is converted to  momentum space via the inverse of the Fourier transform \reef{FTdef} that relates momentum space and twistor space. 
Thus, the momentum space Grassmannian integral is given as 
\be
  {\mathcal{L}}_{n;k}\big(\lambda,\tilde{\lambda},\tilde{\eta}\big) 
  =
  \bigg( \prod_{i=1}^n \int d^2 \tilde\mu_i\,
  \,e^{i {\lambda}_i.\tilde{\mu}_i}\bigg)
  \widetilde{\mathcal{L}}_{n;k}\big(\mathcal{W}\big)\, . 
  \label{FTofLnk}
\ee
Since $\widetilde{\mathcal{L}}_{n;k}\big(\mathcal{W}\big)$ is invariant under little group scaling, the  expression
 ${\mathcal{L}}_{n;k}$ scales as $t_i^{-2}$ thanks to the scaling of the measure of the Fourier transform.

The only $\mut$-dependent part of $\widetilde{\mathcal{L}}_{n;k}\big(\mathcal{W}\big)$ is $\delta^{2\tilde{k}}\big(B\cdot\mut\big)$, as can be seen from \reef{GrassW}. 
This $\delta$-function enforces that $B$ must be orthogonal to the 2-plane defined by $\mut_a$ (viewed as two $n$-component vectors). It is convenient to introduce the $B^\perp$ as the $(n-\tilde{k})\times n$ matrix parameterizing the $(n-\tilde{k})$-plane orthogonal to the $\tilde{k}$-plane defined  by $B$; i.e.~it satisfies $B(B^\perp)^T = 0$. The constraints of $\delta^{2\tilde{k}}\big(B\cdot\mut\big)$ can then be reformulated as $\mut \subset B^\perp$. In other words, $\mut_a$ is some linear combination of the rows of $B^\perp$:
\beq
\delta^{2k}\big(B_{\alpha i}\mut_i\big) = \int d^{2(n-\tilde{k})}\sigma_{\bar\alpha}\, \delta^{2n}\big(\mut_i - \sigma_{\bar\alpha} B^\perp_{\bar\alpha i}\big)\,,
\label{sigmaInt}
\eeq
where $\bar\alpha = 1,\ldots,n-\tilde{k}$. We can now easily perform the inverse-Fourier transform back to momentum space. The delta functions \reef{sigmaInt} localize the Fourier integral \reef{FTofLnk} to give
$e^{i \sigma \cdot B^\perp\cdot \lambda}$, so that integration of the $\sigma$'s then yields $2(n-\tilde{k})$ new delta functions $\delta^{2(n-\tilde{k})} \big( B^\perp \cdot \lambda\big)$. The result is the momentum space Grassmannian integral \reef{GrassM}.

\subsection{Derivation of the momentum twistor Grassmannian}

Having derived the momentum space integral \reef{GrassM} from the twistor space one \reef{GrassW}, we now continue to momentum twistor space. The  key step is the reduction of the integral from G$(k+2,n)$ to G$(k,n)$.

The bosonic delta functions $\delta^{2(n-\tilde{k})} \big( B^\perp \cdot \lambda\big)$ in \reef{GrassM} require  that the $\lambda$ 2-plane lies in the orthogonal complement of $B^\perp$, so 
\be
  \prod_{\beta=1}^{\tilde{k}} \int d^2\rho_\beta \,\delta^{2n} \big( \lambda_j - \rho_\alpha B_{\alpha j}\big) \,,
  \label{2ndeltas}
\ee
where $\rho^a_{\alpha}$ is a $2 \times \tilde{k}$ array of dummy integration variables.

As an aside, let us note that we could easily have found  \reef{2ndeltas} directly from the inverse Fourier integral of the twistor space integral \reef{FTofLnk} by writing $\delta^{2\tilde{k}}\big(B\cdot \tilde{\mu}\big)$ as $\int d^2\rho_\alpha\,e^{-i\rho_\alpha B_{\alpha j} \tilde{\mu}_j}$ and then carrying out the $2n$ Fourier integrals in \reef{FTofLnk} to find \reef{2ndeltas}.

The $GL(\tilde{k})=GL(k+2)$ redundancy of the $B$'s is transferred to the $\rho$'s. So we can go ahead and fix part of  $GL(k+2)$  by choosing   
\be
   \rho = \left( 
   \begin{array}{ccccccc}
       0 &\cdots &0 & 1 & 0\\
       0 &\cdots &0 & 0 &1 
   \end{array}
   \right)\,.
\ee  
The $2n$ delta functions \reef{2ndeltas} then fix the last two rows of $B$ to be the $\lambda$'s:
\beq
B=\left( \begin{array}{cccc}
B_{11} & B_{12} & \cdots  & B_{1n} \\
\vdots & \vdots & \ddots  & \vdots \\
B_{k1} & B_{k2} & \cdots & B_{kn} \\
\lambda_{1}^1 & \lambda_{2}^1 & \cdots  & \lambda_{n}^1 \\
\lambda_{1}^2 & \lambda_{2}^2 & \cdots  & \lambda_{n}^2
\end{array}
\right)\,.
\label{Cfixed}
\eeq
Thus, after evaluating the $\rho$ integrals, we find
\begin{equation}
\label{momgrass}
 {\mathcal{L}}_{n;k}\big(\lambda,\tilde{\lambda},\tilde{\eta}\big) 
  =
\delta^4\big(\l_i \lat_i\big)  \delta^{(8)}\big(\l_i \tilde{\eta}_i\big) \times 
\int
\frac {d^{k\times n} B_{\hat{\alpha} i}} {GL(k)\ltimes T_k} \,
\frac {\delta^{2k}\big(B_{\hat\alpha i}\lat_i\big) \,
\delta^{(4k)}\big(B_{\hat\alpha i}\tilde{\eta}_i\big)} {m_1 m_2 \cdots m_n}\,,
\end{equation}
with $\hat{\alpha}=1,2,\ldots,k$. 
The gauge choice \reef{Cfixed} preserves little group scaling. Note that all the delta functions in \reef{momgrass} are little group invariant using 
$B_{\hat\alpha i} \to t_i B_{\hat\alpha i}$. Again, the $n$ minors scale as $t_i^{\tilde{k}}=t_i^{k+2}$, but now the measure only contributes $t_i^{k}$. So overall, the expression \reef{momgrass} for ${\mathcal{L}}_{n;k}$ scales as 
$t_i^{-2}$, as anticipated.

In \reef{momgrass}, $T_k$ indicates the translational redundancy in the $B_{\hat\alpha i}$-variables. The translational symmetry acts as
\be
  B_{\hat\alpha i} \to B_{\hat\alpha i} + r_{1\hat\alpha} \lambda_i^1 
  + r_{2\hat\alpha} \lambda_i^2 
~~~~~
  \text{for all $i$ simultaneously,}
    \label{translation}
\ee
where $r_{1\hat\alpha}$ and  $r_{2\hat\alpha}$ are any numbers. This is a mixing of the last two rows in the $B$-matrix \reef{Cfixed} with the other rows, and this leaves the 
minors  $m_i$ unchanged. It is also clear that on the support of the two delta functions $\delta^4\big(\l_i \lat_i\big)  \delta^{(8)}\big(\l_i \tilde{\eta}_i\big) $ (that encode momentum and supermomentum conservation), the delta-functions in the integral \reef{momgrass}
are invariant under such a shift. 

So far, what we have done  parallels the work \cite{ArkaniHamed:2009vw}. At this stage, the authors of \cite{ArkaniHamed:2009vw} fix the translation invariance $T_k$ via $2k$ delta functions $\delta(B_{\hat\alpha i } \lambda_i)$. This  breaks the little group scaling and the associated Jacobian is therefore   unpleasant. We proceed here in a way that preserves little group scaling at every step and gives very simple Jacobians that can be presented explicitly.

We change variables in the external data to go from momentum space to momentum twistor space. The momentum supertwistors  
$\mathcal{Z}_i = (\lambda_i, \mu_i \,|\, \eta_i )$ are related to the momentum space variables via the relations \reef{inc}. Using these relations, we directly find for each $\hat\alpha=1,2,\dots,k$:
\be
  \sum_{i=1}^n B_{\hat\alpha i}\tilde{\lambda}_i = 
   - \sum_{i=1}^n C_{\hat\alpha i}\mu_i\,,
   ~~~~~~~~~
  \sum_{i=1}^n B_{\hat\alpha i}\tilde{\eta}_i = 
   - \sum_{i=1}^n C_{\hat\alpha i}\eta_i\,,
\ee
where the reorganization on the RHS directly gives
\be
  \label{DfromC}
  C_{\hat\alpha i}
  =
  \frac{\two{i,i+1}B_{\hat\alpha, i-1} + \two{i-1,i}B_{\hat\alpha, i+1}+ \two{i+1,i-1}B_{\hat\alpha i} } {\two{i-1,i}\two{i,i+1}}\,.
\ee
A sign was absorbed which flipped the angle brackets relative to \eqref{inc}.
The expression \reef{DfromC} implies that $C_{\hat\alpha i} \to t_i^{-1} C_{\hat\alpha i}$ under little group scaling.

We can rewrite the $(k+2) \times (k+2)$ minors $m_i$ of the $B$-matrix in terms of the $k\times k$ minors of the $C$-matrix as \cite{ArkaniHamed:2009vw}
\be
  m_1 =  (B_1 \dots B_{k+2})  = -\<12\> \cdots \<k+1,k+2\> (C_2\dots C_{k+1} )\,~~~\text{etc.}
\label{minors}
\ee  
Defining the $k \times k$ minors of the $k\times n$ $C$-matrix to be 
$M_1 :=  (C_1\dots C_{k})$ etc, we thus have
\be 
   m_1 m_2 \cdots m_n 
   = (-1)^n \big( \two{12}\two{23} \cdots \two{n1} \big)^{k+1}
   M_1 M_2 \cdots M_n  \,.
   \label{minorleague}
\ee
(We drop the signs $(-1)^n$ just as we drop $2\pi$'s in the Fourier transforms.)
Thus, we now have
\begin{equation}
\label{momgrass2}
 {\mathcal{L}}_{n;k}\big(\lambda,\tilde{\lambda},\tilde{\eta}\big) 
  =
\frac{\delta^{4}\big(\l_i \lat_i\big) \,\delta^{(8)}\big(\l_i \tilde{\eta}_i\big) }
{\big( \two{12}\two{23} \cdots \two{1n} \big)^{k+1}}  
\int
\frac{d^{k\times n} B_{\hat\alpha i}} {GL(k)\ltimes T_k} \,
\frac{\delta^{2k}\big(C_{\hat\alpha i}\m_i\big) \,
\delta^{(4k)}\big(C_{\hat\alpha i}\eta_i\big)} {M_1 M_2 \cdots M_n}.
\end{equation}
It is here understood that the $C$'s are functions of the $B$'s as given by \reef{DfromC}.

Note that the Schouten identity guarantees the following two important properties:
\begin{itemize}
\item The $C$'s are invariant under the translations \reef{translation}.
\item The expression \reef{DfromC} implies that  $C_{\hat\alpha i}\lambda_i = 0$.
\end{itemize}
We would now like to do two things: fix the translational redundancy and rewrite the integral in terms of $C$'s instead of $B$'s. 

Because of the  translational invariance, the $B$'s are not independent variables: for example we can use translations to set $2k$ of them to zero (see below). So after fixing translational invariance, we will have 
$kn - 2k = k(n-2)$ variables to integrate over.

\compactsubsection{Step 1: Fixing translation invariance.} Let us use the translation invariance $T_k$ to fix the first two columns in $B_{\hat\alpha i}$ to be zero, i.e.~for all $\hat{\alpha}=1,\ldots,k$ we set  $B_{\hat\alpha 1}=B_{\hat\alpha 2}=0$. This gives
\be
 \frac {d^{k\times n} B_{\hat\alpha i}} {T_k} 
 = \<12\>^{k} \,d^{k\times (n-2)} B_{\hat\alpha i}\,,
 \label{translationfix}
\ee   
where the included prefactor preserves the scaling properties of the measure. (This can be derived more carefully as a Jacobian of the gauge fixing.)

\compactsubsection{Step 2: Changing variables from $B$ to $C$}
We know how $C_{\hat{\alpha}i}$ is related to $B_{\hat{\alpha}i}$ from equation \eqref{DfromC}. We can use that relation to solve for $k(n-2)$ of the components of $C$ in terms of the $k(n-2)$ unfixed components of $B$. Given our choice to set $B_{\hat{\alpha}1}=B_{\hat{\alpha}2}=0$, we have the following system of $k(n-2)$ equations:
\begin{align}
C_{\hat{\alpha}{i}} = \left\{ 
\begin{array}{cll}
  \displaystyle
  \frac{\two{i,i+1}B_{\hat\alpha, i-1} + \two{i+1,i-1}B_{\hat\alpha i} 
      + \two{i-1,i}B_{\hat\alpha, i+1}} {\two{i-1,i}\two{i,i+1}} & \text{for}& 3<i<n 
\\[5mm]
\displaystyle
  \frac{\two{42}B_{\hat\alpha 3} 
      + \two{23}B_{\hat\alpha 4}} {\two{23}\two{34}} & \text{for}& i=3
\\[5mm]
\displaystyle
  \frac{\two{n1}B_{\hat\alpha, n-1} + \two{1,n-1}B_{\hat\alpha n} 
      } {\two{n-1,n}\two{n1}} & \text{for} & i=n
\end{array}\right.\,.
\label{CfromBgauge12}
\end{align}
We can write this as $C_{\^\a\^j} = B_{\^\a\^i}Q_{\^i\^j}$, with $\hat{i},\hat{j} = 3,\ldots, n$, for a square symmetric matrix $Q_{\^i\^j}$ with nonzero entries only on and adjacent to the main diagonal. In Appendix \ref{app:detQ}, we show that
\begin{align}
|\det Q\,| = \frac{\two{12}^2}{\two{12}\two{23}\cdots \two{n1}}\,.\
\label{detQresult}
\end{align}
Therefore the measure transforms as
\begin{align}
d^{k(n-2)} B_{\^\a \^i} ~=~ \frac{d^{k(n-2)}C_{\^\a\^i}}{|\det Q\,|^k} 
~=~ \bigg( \frac{\two{12}\two{23}\cdots \two{n1}} {\two{12}^2} \bigg)^k d^{k(n-2)}C_{\^\a\^i}\,.
\label{measureBC}
\end{align}
We take the absolute value of the determinant since the overall sign is irrelevant. Once again, the little group scaling of $dC$ is compensated by the Jacobian factor so that the overall scaling of the right-hand side matches that of $dB$ on the left.

\compactsubsection{Step 3: Restoring the full set of $C$ variables}
With the help of \reef{translationfix} and \reef{measureBC}, the integral \reef{momgrass2} now takes the form
\be
\label{momgrass3}
 {\mathcal{L}}_{n;k}\big(\mathcal{Z}\big) 
  =
\frac{\delta^{4}\big(\l_i \lat_i\big) \,\delta^{(8)}\big(\l_i \tilde{\eta}_i\big) }
{\two{12}\two{23} \cdots \two{1n}}  
\frac{1}{\<12\>^k}
\int
\frac{d^{k\times (n-2)} C_{\hat\alpha i}} {GL(k)} \,
\frac{\delta^{2k}\big(C_{\hat\alpha i}\mu_i\big) \,
\delta^{(4k)}\big(C_{\hat\alpha i}\eta_i\big)} {M_1 M_2 \cdots M_n}
\bigg|_{C_{\hat\alpha 1,2} = C_{\hat\alpha 1,2}^{(0)}}
\,.
\ee
We recognize the first factor as the MHV superamplitude $\mathcal{A}_n^\text{MHV}$ from \reef{AMHV}. 

The restriction of $C_{\hat\alpha 1}$ and $C_{\hat\alpha 2}$ in \reef{momgrass3} follows from the relation \reef{DfromC} between the $B$ and $C$; it was used above to solve for $k(n-2)$ components of $C$ in terms of the $B$'s, but the remaining $2k$ components of $C$ are then fixed as
\be
C_{\^\a 1}= \frac{B_{\^\a n}}{\two{n1}} 
= \sum_{j=3}^n \frac{\two{2j}}{\two{12}} C_{\^\a j}
=:
  C_{\^\a 1}^{(0)}\,, ~~~~~~
C_{\^\a 2}= \frac{B_{\^\a 3}}{\two{23}} 
= -\sum_{j=3}^n \frac{\two{1j}}{\two{12}} C_{\^\a j}
=:
 C_{\^\a 2}^{(0)}\,.
\label{C1C2}
\ee
To verify the second equality in each relation, use \reef{CfromBgauge12} and rejoice in the beauty of the sum telescoping under the Schouten identity. 

Thus, when evaluating the integral \reef{momgrass3}, $C_{\hat\alpha 1}$ and $C_{\hat\alpha 2}$ are functions of the other $k(n-2)$ $C$-components. This is a restriction of the region of integration that we can also impose via $2k$ delta functions $\delta \big( C_{\^\a i} -  C_{\^\a i}^{(0)} \big)$ for $i=1,2$. 
Moreover, it follows from the explicit solution \reef{C1C2} that the constraints are equivalent to  $C_{\^\a i}\lambda_i = 0$. Thus we can rewrite the delta function restriction as
\be
  \delta \big( C_{\^\a 1} -  C_{\^\a 1}^{(0)} \big)\,
  \delta \big( C_{\^\a 2} -  C_{\^\a 2}^{(0)} \big)
  =
  \<12\> \,\delta^2 \big( C_{\^\a i}\lambda_i\big)\,.
\ee
 for each $\hat{\alpha}=1,2,\ldots,k$.
 
We then have
\be
\label{momgrass4}
 {\mathcal{L}}_{n;k}\big(\mathcal{Z}\big) 
  =
\mathcal{A}_n^\text{MHV}
\int
\frac{d^{k\times n} C_{\hat\alpha i}} {GL(k)} \,
\frac{\delta^{2k} \big( C_{\^\a i}\lambda_i\big)\,
\delta^{2k}\big(C_{\hat\alpha i}\m_i\big) \,
\delta^{(4k)}\big(C_{\hat\alpha i}\eta_i\big)} {M_1 M_2 \cdots M_n}
\,.
\ee
Although we chose to fix the first two columns of $B$ to be zero, 
the answer is independent of that choice; the factors of $\two{12}$ cancel out. 
We can now write the result directly in terms of the momentum supertwistors
$\mathcal{Z}_i=( Z_i | \eta_i )= ( \lambda_i, \mu_i | \eta_i )$
as
\be
\label{momgrass5}
 {\mathcal{L}}_{n;k}\big(\mathcal{Z}\big) 
  =
\mathcal{A}_n^\text{MHV}
\int
\frac{d^{k\times n} C_{\hat\alpha i}} {GL(k)} \,
\frac{\delta^{4k} \big( C_{\^\a i} Z_i\big)\,
\delta^{(4k)}\big(C_{\hat\alpha i}\eta_i\big)} {M_1 M_2 \cdots M_n}
=
\mathcal{A}_n^\text{MHV}
\int
\frac{d^{k\times n} C_{\hat\alpha i}} {GL(k)} \,
\frac{\delta^{4k|4k} \big( C_{\^\a i} \mathcal{Z}_i\big)} {M_1 M_2 \cdots M_n}
\,.
\ee
This completes our derivation of the $\mathcal{N}=4$ SYM Grassmannian integral in momentum twistor space from that in momentum space. 
A very similar procedure leads to an analogous result in 3d ABJM theory as we explain below in Section \ref{sec:3DGrass}. In the intervening section, we demonstrate an explicit evaluation of the $\cn=4$ SYM momentum twistor integral \eqref{momgrass5} in the NMHV sector.

\section{NMHV Integrals and Residues}
\label{s:res}

In this  section  we evaluate the NMHV Grassmannian integral in momentum twistor space and discuss some properties of the residues and their relations to on-shell diagrams. While part of this is review, new material includes recasting the residue theorems in terms of the homology and a precise description of how the residue relations and pole structures relate to the boundary operation and the boundaries of cells in the Grassmannian.  

\subsection{Evaluation of the NMHV residues}
\label{s:NMHVeval}
We  focus on the  momentum twistor Grassmannian, so for NMHV we have $k=1$ and \reef{GrassZ} is a contour integral in the Grassmannian $\text{G}(1,n)$. The elements $C\in \text{G}(1,n)$ are  $1\times n$ matrices modulo a $GL(1)$ scaling,
\begin{align}
   C = \big[\begin{array}{cccc} c_1 & c_2 & \ldots & c_n\end{array}\big], 
   \label{topC}
\end{align}
with complex numbers $c_i$. The Grassmannian integral is 
$\mathcal{L}_{n;1}(\mathcal{Z}) = \mathcal{A}_n^\text{MHV} \mathcal{I}_{n;1}(\mathcal{Z})$ with
\begin{align}
   \mathcal{I}^{(\Gamma)}_{n;1}(\mathcal{Z}) 
   := 
   \oint_\Gamma \frac{d^{1\times n}C}{GL(1)\, c_1 c_2 \cdots c_n}
   \,\delta^4\big(c_i Z_i\big)\, \delta^{(4)}\big(c_i {\eta}_i\big)\,.
\label{NMHVlink}
\end{align}
The oriented volume form on $\mathbb{C}^n$ is $d^nC = \bigwedge\limits_{i=1}^n\,dc^i$, and the contour $\Gamma$ will be specified below. 

The bosonic delta function $\delta^4(c_i Z_i)$ fixes four $c_i$'s, 
and the $GL(1)$ redundancy fixes another. This leaves an integral with $n-5$ variables. Now suppose the contour $\Gamma$ encircles a pole where exactly $n-5$ of the $c_i$'s vanish. Such a contour can be characterized 
 by specifying which five $c_i$'s are non-vanishing at the pole. Let us denote these five non-vanishing $c_i$'s by  $c_a$, $c_b$, $c_c$, $c_d$, and $c_e$, and the corresponding contour $\gamma_{abcde}$.

We now evaluate $\mathcal{I}_{n;1}^{\gamma_{abcde}}(\mathcal{Z})$ ``by inspection". Appendix \ref{app:residues} gives a more careful evaluation that also computes the sign of the residue correctly. It follows from \reef{NMHVlink} that the residue where all $c_i$ vanish for $i\ne a,b,c,d,e$ is, up to a sign, simply
\be
  \mathcal{I}_{n;1}^{\gamma_{abcde}}(\mathcal{Z})
  = \frac{\delta^4\big(c_a Z_a + c_b Z_b + c_c Z_c+ c_d Z_d + c_e Z_e\big)\, \delta^{(4)}\big(c_a \eta_a + c_b \eta_b + c_c \eta_c+ c_d \eta_d + c_e \eta_e \big)}{c_a c_b c_c c_d c_e}\,.
\label{Ires}
\ee   
Now, the constraint enforced by the bosonic delta-function is trivially solved by 
\be
  c_a = \<bcde\>\,,~~~~
  c_b = \<cdea\>\,,~~~~
  c_c = \<deab\>\,,~~~~
  c_d = \<eabc\>\,,~~~~
  c_e = \<abcd\>
  \,, 
\ee
using the 5-term Schouten identity (or Cramer's rule) that states that five 4-component vectors are necessarily linearly dependent:
\begin{align}
\four{ijkl}Z_m +\four{jklm}Z_i + \four{klmi}Z_j + \four{lmij}Z_k + \four{mijk}Z_l=0 \,.
\label{5Schouten}
\end{align}
The 4-brackets are the fully antisymmetric $SU(2,2)$-invariants
\be
\label{def4bracket}
  \<ijkl\> :=  -\eps_{\mathsf{ABCD}}  Z_i^\mathsf{A}Z_j^\mathsf{B}Z_k^\mathsf{C}Z_l^\mathsf{D} = \det\big( Z_i Z_j Z_k Z_l\big) \,.
\ee
We conclude that 
\be
 \mathcal{I}_{n;1}^{\gamma_{abcde}} =
  \frac{\delta^{(4)}\Big(\four{bcde}{\eta}_a + \four{cdea}{\eta}_b+\four{deab}{\eta}_c+\four{eabc}{\eta}_d + \four{abcd}{\eta}_e\Big)}{\four{bcde}
\four{cdea}
\four{deab}
\four{eabc}
\four{abcd} }
~=: \five{abcde}\,,
  \label{resresult}
\ee
The expression \reef{resresult} is manifestly antisymmetric in the five labels $a,b,c,d,e$. This follows from the standard evaluation of higher-dimensional contour integrals, as we review in Appendix \ref{app:residues}. In addition, the general results in the appendix tell us that the residue is also fully antisymmetric in the labels $i\ne a,b,c,d,e$. We can incorporate that by labeling the residue  \eqref{Ires}, including the appropriate signs from the appendix, by the $n-5$ values  $i_1,i_2,\ldots,i_{n-5} \ne a,b,c,d,e$ as $\{i_1,i_2,\ldots,i_{n-5}\}$, which is antisymmetric in its indices. Then the final answer, which includes all of the signs from Appendix \ref{app:residues}, is
\begin{align}
\mathcal{I}_{n;1}^{\gamma_{abcde}} 
=
 \frac{1}{(n-5)!}\, \varepsilon^{ a\,b\,c\,d\,e\,i_1\, i_2\,\ldots\, i_{n-5}}\
\{i_1,i_2,\ldots,i_{n-5}\}
= \five{abcde}  
\label{5res} \,.
\end{align}
This completes the calculation of the residues of the Grassmannian integral 
$\mathcal{L}_{n;1}$ in momentum twistor space. The result, 
\be
  \mathcal{L}_{n;1}^{\gamma_{abcde}} 
  = \mathcal{A}_n^\text{MHV} \,\five{abcde}
  \,,
\ee
shows that the individual residues produced by $\mathcal{L}_{n;1}$ are the 5-brackets 
$\five{abcde}$. These are the known building blocks of  
NMHV amplitudes, both at tree and loop-level.

\subsection{NMHV residue theorems}
\label{s:resrels}

Since the residues $\five{abcde}$ of the NMHV Grassmannian integral \reef{NMHVint} are characterized by five labels, $a,b,c,d,e \in \{1,2,3,\ldots,n\}$, as in \reef{5res}, it follows that there are a total of ${n \choose 5}$ NMHV residues. These, however, are not independent.  While it is difficult to derive the residue relations --- or even verify  them --- by direct computations, the constraints among them follow quite straightforwardly from the Grassmannian residue theorems, as first noted in \cite{ArkaniHamed:2009dn}. In this section, we count the number of independent  NMHV residues $\five{abcde}$ and examine the linear relations among them. Since the only input is residue theorems, these relationships are also true off the support of the external momentum and supermomentum delta functions in the overall MHV factor in the momentum twistor Grassmannian integral. 

Let us begin by taking an abstract view of the integral \reef{NMHVint}. 
We are integrating over $C = \big[ c_1 \ldots c_n \big]$ modulo a $GL(1)$ that identifies $C \sim s C$ for any $s \in \C-\{0\}$. Thus $C$ can be viewed as homogeneous coordinates of $\CP^{n-1}$. We are interested in the residues associated with simultaneously vanishing `minors' $c_i$. Each condition $c_i=0$ defines a hyperplane in $\CP^{n-1}$. In other words, we are interested in the $n$ hyperplanes 
$h_i := \{ C \in \CP^{n-1} | c_i = 0 \}$. This is called a {\em hyperplane arrangement} in $\CP^{n-1}$. Specifically, the residue $\five{abcde}$ corresponds to picking up the residue from the $(n-5)$-dimensional toroidal contour $(S^1)^{n-5}$ surrounding the intersection of the $n-5$ hyperplanes $h_i$ with $i \ne a,b,c,d,e$.

The bosonic delta functions in the momentum twistor integral \reef{NMHVint} impose four conditions among the $n$ components of $C$. These homogeneous linear relations respect the $GL(1)$ scaling, so they reduce the space of interest from $\CP^{n-1}$ to $\CP^{n-5}$. Consequently, we are interested in the  arrangement of $n$ hyperplanes in  $\CP^{n-5}$.

Now, suppose we focus on the complement of $h_n$, i.e. $c_n\neq0$. We fix $c_n$ to be some non-vanishing value to eliminate the projective freedom, so $\CP^{n-5}\to \C^{n-5}$. The problem then reduces to the study of $n-1$ hyperplanes $\{h_i\}_{i=1,\ldots,n-1}$ in $\C^{n-5}$. This step is equivalent to fixing the $GL(1)$ redundancy in the Grassmannian integral. The $(n-5)$-dimensional contours of \eqref{NMHVint} must therefore 
live in the hyperplane arrangement complement 
\be
 X= \C^{n-5} - \bigcup_{i<n} h_i\,.
\ee 
  A residue does not change under continuous deformation of the contour, so the result only depends on the homology class of the contour. Thus, the key observation is that the number of possible independent residues is the dimension of the homology class $H_{n-5}(X,\C)$.

The geometry of hyperplane arrangements is well-studied in the mathematics literature and the results include the following theorem \cite{OrlikTerao}:
\begin{theorem}
Let $X = \C^N-\bigcup\limits_{i=1}^r h_i$ be a hyperplane arrangement complement.  Then
\begin{enumerate}
\item
The cohomology $H^*(X,\C)$ is generated by the forms $\frac{d\alpha_i}{\alpha_i}$.
\item
Suppose $\{h_i\}$ is generic.\footnote{When the ambient space is $\C^N$, we say that the hyperplane arrangement is {\it generic} if $h_{i_1} \cap h_{i_2} \cap \cdots \cap h_{i_r}$ has dimension $N-r$, for $r \leq n$. In other words, a hyperplane arrangement is generic if all intersections of hyperplanes have the expected dimension.}  Then $H^k(X,\C)$ is zero for $k \geq N$, and for $k \in \{0,1,\ldots,N\}$, a basis for $H^k(X,\C)$ is given by the forms
$$
\frac{d\alpha_{i_1}}{\alpha_{i_1}} \wedge \cdots \wedge \frac{d\alpha_{i_k}}{\alpha_{i_k}}
$$
ranging over subsets $\{i_1,i_2,\ldots,i_k\} \subset \{1,2,\ldots,r\}$.  In particular,
\be
\dim H^k(X,\C) = {{r}\choose{k}}\,.
\ee
\end{enumerate}
\end{theorem}
The algebra $H^*(X,\C)$ (generic arrangement or otherwise) can be described in a combinatorial fashion and is called the Orlik-Solomon algebra.

For our purpose, the ambient space has dimension $N=n-5$, there are $r=n-1$ hyperplanes, and 
we are interested in the dimension of the homology $H_{n-5}(X,\C)$. It is of course the same dimension as the corresponding cohomology $H^{n-5}(X,\C)$. Hence, by the above theorem, the number of independent residues is
\begin{align}
R=\dim H_{n-5}(X) = {{n-1}\choose{n-5}} = {{n-1}\choose{4}} \,.
\label{indepRes}
\end{align}

The residue relations have a simple  geometrical interpretation (see also \cite{ArkaniHamed:2009dn}). Let $\{i_1,i_2,\ldots,i_{n-5}\}$ denote the residue corresponding to the intersection of $n-5$ hyperplanes ${h_{i_1} \cap h_{i_2} \cap \cdots \cap h_{i_{n-5}}}$. As explained in the Section \ref{s:NMHVeval}, the residue is fully antisymmetric in its labels. For example, when $n=6$, the ``hyperplanes'' are just individual points $\{i\}$ in $\CP^1$; there are six such points.
Since $\CP^1$ is isomorphic to a two-sphere $S^2$, any contour surrounding all six can be contracted to a point. Hence the residue theorem states that the sum of the six residues is zero. Thus there is one relation among six residues, leaving five independent. This clearly agrees with the counting \reef{indepRes} for $n=6$.

Let us now use this to understand the relations under which only $n-1 \choose 4$ of the $n \choose 5$ residues are independent. Consider a choice of $n-6$ hyperplanes, $h_{i_k}$ with $k=1,2,\ldots,n-6$, in $\CP^{n-5}$. Imagine that we take  the $S^1$ contours surrounding each of these $h_i$ very small so that we effectively look at the subspace $\CP^1 = S^2$ of the intersection of those $n-6$ hyperplanes. This subspace is (generically) intersected by the other hyperplanes $h_j$ at $6$ distinct points. Just as for the $n=6$ case, a contour in  $\CP^1$ that surrounds these six points can be contracted a point, and the sum of the six residues must vanish: the resulting residue theorem is 
\begin{align}
\sum_{j=1}^n \{i_1,i_2,\ldots,i_{n-6},j\} = 0 \,.
\label{resCon}
\end{align}
This holds for any choice of $n-6$ labels $i_1,i_2,\ldots,i_{n-6}$; 
 hence we get a web of linear relations among the $n \choose 5$ residues. The statement \reef{indepRes} is that under these relations, only $n-1 \choose 4$ residues are independent.

The counting of independent residues can be verified directly from the relations \reef{resCon}. While there may appear to be $n \choose {n-6}$ constraints in \reef{resCon}, some of them are redundant. 
Without loss of generality, consider only those for which all 
$i_k\neq n$, $k=1,2,\ldots,n-6$. There are ${n-1} \choose {n-6}$ distinct constraints of that sort. In each such sum, the index $n$ will appear exactly once, namely when $j=n$, so we can solve for each residue that includes $n$ in terms of residues which do not:
\begin{align}
\{i_1,i_2,\ldots, i_{n-6},n\} = - \sum\limits_{j=1}^{n-1}\{i_1,i_2,\ldots,i_{n-6},j\}
\label{nthRes}
\end{align}
where $i_1,i_2,\ldots, i_{n-6} \ne n$. 
This determines all of the residues labeled by $n$ in terms of all of the others. Furthermore, since the first $n-6$ indices form a unique set, all ${n-1}\choose{n-6}$ equations in \eqref{nthRes} are independent.

The remaining equations in \eqref{resCon} have  $i_k =n$ for some $k$, but they do not provide any further constraints. To see this, use the antisymmetry and \eqref{nthRes} to eliminate the index $n$:
\begin{align}
\sum\limits_{j=1}^{n}\{i_1,i_2,\ldots,i_{n-7},n,j\}&=-\sum\limits_{j=1}^{n}\sum\limits_{m=1}^{n-1}\{i_1,i_2,\ldots,i_{n-7},m,j\} \nonumber\\
&=-\sum\limits_{m=1}^{n-1} \Big(\sum\limits_{j=1}^{n}\{i_1,i_2,\ldots,i_{n-7},m,j\}\Big)=0\,.
\end{align}
 We conclude that all of the constraints with a fixed index $n$ are redundant with the  ones in \reef{nthRes}. Hence, the number of independent constraints are
${{n-1}\choose{n-6}}$ and therefore the number of independent residues is
\begin{align}
R = {{n}\choose{n-5}} - {{n-1}\choose{n-6}} = {{n-1}\choose{n-5}} = {{n-1}\choose{4}}\,,
\end{align}
in agreement with the dimension of the homology \reef{indepRes}.

\subsection{Applications}
\label{s:resapp}

\subsubsection{Residue theorems as boundary operations}
Let us now consider some applications of the NMHV residue theorems. The case of $n=6$ is very well-known. There is just one constraint from \reef{resCon},
\begin{align}
\{1\}+\{2\}+\{3\}+\{4\}+\{5\}+\{6\}=0\,,
\end{align}
 or via \reef{5res} in terms of the 5-brackets it is the six-term identity
\be
  \five{23456} -  \five{13456} + \five{12456} - \five{12356} + \five{12346} - \five{12345} = 0\,.
\ee
The LHS of this identity 
can be succinctly abbreviated as defining the
{\bf \em boundary operation} $\partial \big[123456\big]$;  the relation to boundaries is explain in Section \ref{s:bdr}. 
More generally, we can write the boundary operation as
\be
 \partial \big[abcdef\big] = 0
 ~~~~\longleftrightarrow~~~~ 
 \sum_{a',b',c',d',e',f'=1}^n \eps^{i_1 i_2 \ldots i_{n-6} a'b'c'd'e'f'} [a'b'c'd'e'] = 0\,,
\label{bdrOp}
\ee
where $\{ i_1, \ldots ,i_{n-6} \}$ are the complement of $\{a,b,c,d,e,f\}$ in the set $\{1,2,\ldots,n\}$. The relation \reef{5res} between the 5-brackets and the residues now makes it clear that the boundary conditions are equivalent to the residue theorems \reef{resCon}:
\be
 \sum_{j=1}^n \{i_1,i_2,\ldots,i_{n-6},j\} = 0  
 ~~~~
 \raisebox{3mm}{$\underleftrightarrow{~~
 \scriptstyle  \{i_1,i_2,\ldots,i_{n-6}\}\; = \; \overline{\{a,b,c,d,e,f\}}~~}$}
 ~~~~
 \partial\big[abcdef\big] = 0\,.
\ee

\subsubsection{Identities among $R$-invariants}
Prior to the introduction of momentum twistors, the momentum space versions of the 5-brackets were denoted as {\bf \em $R$-invariants} \cite{Drummond:2008vq}:
\be
  R_{ijk} := \five{i,j-1,j,k-1,k}\,.
\ee
It was observed  that the $R$-invariants obey the two identities
\be
  R_{i,i+2,j} = R_{i+2,j,i+1}\,
  ~~~~\text{and}~~~~
  \sum_{s=3}^{k-2} \sum_{t=s+2}^k R_{1st}
  = \sum_{s=2}^{k-3} \sum_{t=s+2}^{k-1} R_{kst} \,.
  \label{Ridentities}
\ee
for any $k=1,2, \dots, n$. These identities have been used in various applications, such as proving dual conformal invariance of the 1-loop ratio function in $\cn=4$ SYM \cite{Drummond:2008vq,Brandhuber:2009xz,Elvang:2009ya}. Let us now review how these arise as a consequence of the symmetries and residue theorems of the 5-brackets.

The first identity in \reef{Ridentities} follows straightforwardly from the antisymmetry of the 5-bracket \cite{Mason:2009qx}:
\be
  R_{i,i+2,j}
  = \five{i,i+1,i+2,j-1,j}
  =\five{i+2,j-1,j,i,i+1}
  = R_{i+2,j,i+1} \,.
\ee
For the second identity in \reef{Ridentities}, note that  $R_{1st} = \five{1,s-1,s,t-1,t}$ vanishes for $s=2$, so on the LHS of \reef{Ridentities}  the sum can trivially be extended to include $s=2$. 
Then using the six-term identity resulting from $\partial \five{1,s-1,s,t-1,t,k} =0$, the 5-bracket  $\five{1,s-1,s,t-1,t}$ can be eliminated
 in favor of the five other 5-brackets appearing in the identity.
This includes  
$\five{s-1,s,t-1,t,k} = \five{k,s-1,s,t-1,t}$, which vanishes trivially for $t=k$ and for $s=k-2$. Hence 
this part of the sum gives the desired RHS of \reef{Ridentities}. We are left to show that the sum of the remaining four terms vanishes; they are
\be
  \sum_{s=2}^{k-2} \sum_{t=s+2}^k
  \Big( \five{1,s-1,s,t-1,k} - \five{1,s-1,s,t,k}
   +\five{1,s-1,t-1,t,k}-\five{1,s,t-1,t,k}
  \Big)\,. 
  \label{left4terms}
\ee
 The sum of the first two terms telescopes to 
 $\sum_{s=3}^{k-2}  \five{1,s-1,s,s+1,k}$ while the sum of the last two terms collapses to
 $-\sum_{s=2}^{k-3} \five{1,s,s+1,s+2,k}$. These two sums are  identical and thus the sum \reef{left4terms} vanishes. This completes the derivation of the identities \reef{Ridentities}.

\subsubsection{Locality and the NMHV tree superamplitude}
\label{s:tree}
For $k=n$, the second identity in \reef{Ridentities} can be written
\be
 \sum_{i<j} \five{1,i-1,i,j-1,j} = \sum_{i<j} \five{n,i-1,i,j-1,j}\,.
\ee
Note that in this representation, the first label on the LHS plays no special role and can be replaced with any momentum twistor $Z_*$. Hence the sum of $\five{*,i-1,i,j-1,j}$ over all $i<j$ is independent of $Z_*$.

Let us now study the {\bf \em pole structure} of the 5-brackets. A given 5-bracket has five poles, namely where each of the five 4-brackets in the denominator vanish. Consider two 5-brackets that differ by just one momentum twistor, e.g.~$\five{abcdx}$ and $\five{abcdy}$. They share one common pole, namely $\four{abcd}$. The singularity occurs on the subspace where the four momentum twistors $Z_{a,b,c,d}$ become 
linearly dependent. Since $Z_y \in \CP^3$ can be expressed as a linear combination of any four other (linearly independent) 
momentum twistors, we can write $Z_y = w_x Z_x + w_a Z_a + w_b Z_b + w_c Z_c$. Using this, it is  straightforward to show that the residue at the pole $\four{abcd}=0$ is the same for $\five{abcdx}$ and $\five{abcdy}$. In other words, the residue of the pole  $\four{abcd}$ vanishes in the combination $\five{abcdx}-\five{abcdy}$. 

It is natural to associate a boundary operation with the residues of the poles of the 5-brackets, written as
\be
  \label{bdr5bracket}
  \pa \five{abcde} 
  := 
  \five{bcde} - \five{acde}+\five{abde}-\five{abce}+\five{abcd}
  \,.
\ee
The signs keep track of the relative signs of the residues.

It now follows that the cancellation of $\five{abcd}$ in 
$\pa \big( \five{abcdx}-\five{abcdy} \big)$ is equivalent to the statement that the residue of the pole at $\four{abcd}=0$ vanishes in the difference of the two five-brackets.

Physical poles in color-ordered tree-level  scattering amplitudes are exactly those associated with vanishing Mandelstam invariants $(p_i + p_{i+1} + \ldots)^2$ involving a sum of a subset of adjacent momenta. These are precisely associated with poles in the 5-brackets of the form $\four{i-1,i,j-1,j}$ because of the identity \cite{Hodges:2009hk,Mason:2009qx} 
\be
 \big(p_i + p_{i+1} + \ldots p_{j-1}\big)^2
 = \frac{\four{i-1,i,j-1,j}}{\<i-1,i\>\<j-1,j\>}\,.
\ee
Poles \emph{not} of the form $\four{i-1,i,j-1,j}$ are spurious: they cannot appear 
in the tree-amplitude. 

A straightforward algebraic exercise shows that 
\be
 \pa  \sum_{i<j} \five{*,i-1,i,j-j,j}
 =  \sum_{i<j} \five{i-1,i,j-j,j} \,.
\ee
This means that all spurious poles in the LHS sum telescope to zero, leaving just the  manifestly local poles. 
For $* = q = 1,2,\dots,n$, this sum --- times the MHV superamplitude ---  is exactly the expression one finds \cite{Drummond:2008cr} as the solution to the BCFW recursion relation based on a $[q,q+1\>$-supershift  of the  {\bf \em tree NMHV superamplitude}  
\be  
  \mathcal{A}_n^\text{NMHV} = \mathcal{A}_n^\text{MHV} 
  \sum_{i<j} \five{*,i-1,i,j-j,j} \,,
  \label{NMHVtree}
\ee 
for $Z_* = Z_q$.\footnote{When $Z_*$ is not selected to be one of the $n$ momentum twistors $Z_i$ of the external data, the expression \reef{NMHVtree} is a CSW-like representation of the NMHV superamplitude.} 
From the point of view of the Grassmannian, we see that \reef{NMHVtree} results from a certain choice of contour. Thus, there are choices of contours for the Grassmannian integral such that the result is exactly the NMHV tree amplitude; such a contour what we called the `tree contour'. Note that the insistence of locality, in the sense of having only physical poles, allowed us to identify the tree contours. 

It may seem puzzling that only a small subset of the residues produced by the Grassmannian integral appear in the BCFW-form \reef{NMHVtree} of the NMHV tree superamplitude: residues of the form $\five{q,i-1,i,j-1,j}$ are used, while residues such as $\five{1,2,4,6,8}$ or $\five{1,3,5,7,9}$ do  not seem to play a role, other than through the residue theorems. It would be peculiar if the other residues of the same Grassmann degree were not relevant for NMHV amplitudes; it turns out that they are. It has been conjectured  \cite{ArkaniHamed:2009dn}  that --- in addition to the tree superamplitudes --- the Grassmannian integral also produces all the Leading Singularities of all amplitudes in planar $\cn=4$ SYM at any loop order.
The 1-loop NMHV ratio function 
\cite{Drummond:2008vq,Brandhuber:2009xz,Elvang:2009ya} can be written in terms of the exactly the same types of residues $\five{q,i-1,i,j-1,j}$ as at tree-level, so one has to go to 2-loop order to encounter `non-tree' residues in the Leading Singularities \cite{ArkaniHamed:2009dn}. Also, it has been demonstrated that no new Leading Singularities appear beyond 3-loop order in the NMHV sector \cite{Bullimore:2009cb,Broedel:2010rr}, so the first three loop-orders of the NMHV amplitudes are expected to utilize the full set of residues produced by the Grassmannian integral $\mathcal{L}_{n;1}$.

\subsection{Cells, permutations, and on-shell diagrams}
\label{s:onshelldiag}
So far we have described the evaluation of NMHV amplitudes in the language of contour integrals and residue theorems, but we find that  it is also instructive to take a more abstract view and consider how it fits into the context of on-shell diagrams, permutations, and cells of the Grassmannian. Since the calculations in Section \ref{s:res} were performed in the momentum twistor formulation, we will discuss that case first, and subsequently develop the corresponding story in the momentum space formulation.

Before delving into the details, it will be helpful to quickly review some terminology from \cite{ArkaniHamed:2012nw}. 
Subspaces of the Grassmannian G$(k,n)$ can be classified into {\bf \em cells} by specifying the ranks of cyclically consecutive columns $C_i$, that is ${\rm rank}\big({\rm span}(C_i,C_{i+1},\ldots,C_j)\big)$ for all cyclic intervals $[i,j]$. The {\em dimension} of a cell is the number of parameters it takes to specify a matrix representative modulo the $GL(k)$ redundancy. Cells are uniquely labeled by decorated permutations, which are ``permutations" of the set $\{1,2,\ldots,n\}$ in which $k$ of the elements are shifted beyond 
$n$.\footnote{The decorated permutations will be familiar to practitioners of the juggling arts, where they are also referred to as ``juggling patterns." 
The decoration encodes that balls can only be thrown forward.} Throughout the remainder of this text, we will use `permutation' and `decorated permutation' interchangeably, but we will always mean the latter. Each permutation labels a cell by encoding the linear dependencies of the columns in a representative matrix of the cell. Treating the matrix columns $c_i$ as $k$-vectors, a given permutation
\begin{align}
\sigma = \{\s(1),\s(2),\ldots\} = \{a,b,\ldots\}
\end{align}
encodes  that $c_a$ is the first column with $a>1~($mod $n)$ such that $c_1$ is in the span of 
$\{c_2,c_3,\ldots,c_a\}$. Similarly, $c_2$ is spanned by $\{c_3,\ldots,c_b\}$, and so on. Entries for which $\sigma(i)=i$ imply that the $i^\tth$ column is identically zero. As an example, consider the 5-dimensional cell in G$(2,6)$ with representative matrix
\begin{align}
\left(
\begin{array}{cccccc}
 1 & 0 & c_{11} & c_{12} & 0 & c_{14} \\
 0 & 1 & c_{21} & c_{22} & 0 & 0 \\
\end{array}
\right)\,.
\end{align}
One can easily verify that this cell is labeled by the permutation $\sigma=\{3,4,6,8,5,7\}$. 

The cell with maximal dimension $k(n-k)$ in G$(k,n)$ is known as the {\bf \em top cell}, and it is the unique cell in which at a generic point 
none of the $k\times k$ minors vanish in a representative matrix. Since none of the consecutive minors vanish, each column must be spanned by the next $k$ columns, and therefore the top cell is  labeled by a permutation of the form
\begin{align}
\sigma_\text{top} = \{1+k,2+k,\ldots,n+k\}\,.
\end{align}

\subsubsection{Momentum twistor space}
The $n$-particle NMHV integral in momentum twistor space \eqref{NMHVlink} is an integral over the $(n-1)$-dimensional top cell of G$(1,n)$, which has a representative $1\times n$ matrix $C$, as in \eqref{topC}. Since the minors of $C$ are determinants of $1\times1$ matrices (i.e.~numbers),  
the top cell is represented by matrices with all non-zero entries.

The external data enters through the delta functions $\delta^{4|4}(C\cdot\mathcal{Z})$. The four independent bosonic delta functions fix all degrees of freedom for any 4-dimensional cell of G$(1,n)$.  In order to reach a 4d cell from the top cell, one must set $n-5$ of the minors to zero. This is done in practice by choosing an $(n-5)$-dimensional contour $\gamma_{abcde}$ that encircles a point where only five of the coordinates, i.e. minors, are non-vanishing. For a given choice of $a,b,c,d,e$, the result of evaluating the contour integral 
is  an integral (that will be fully localized by the delta functions) over 
a unique 4d cell 
labeled by the decorated permutation
\begin{align}
& {\color{blue}\,a 
~~~~~~~~~~~~~~b
~~~~~~~ c
~~~~~~~ d
~~~~~~~~ e}
\nonumber\\[-2mm]
\sigma_{abcde} = \{1,2,\ldots,a-1,\, & b,a+1,\ldots,c,\ldots,d,\ldots,e,\ldots,a+n,\ldots,n\}\,,
\end{align}
where all entries are self-identified except those at positions $a,b,c,d,e$ (marked above). The bosonic delta functions fix the remaining degrees of freedom and leave the residue $\{i_1,i_2,\ldots,i_{n-5}\}$, which is related to the five-bracket $\five{abcde}$ via equation \eqref{5res}.
In the momentum twistor Grassmannian integral, all 4d cells meet the support of the delta-function at NMHV level thanks to the 5-term Schouten identity \reef{5Schouten}. (This is specific to NMHV level; it is not the case for higher $k$ amplitudes.)

\subsubsection{Momentum space}
The momentum space version of the Grassmannian integral was given in \reef{GrassM}. For NMHV we have $\tilde{k}=k+2=3$, so the relevant Grassmannian is G$(3,n)$. 
The top cell of G$(3,n)$ is $(3n-9)$-dimensional and is labeled by the decorated permutation
\begin{align}
\widetilde{\sigma}_\text{top} = \{4,5,\ldots,n+3\}\,.
\end{align}
The representative matrices are $3\times n$ and for the top cell  all $3\times3$ minors are non-vanishing. 

The momentum space Grassmannian integral \reef{GrassM} has $2n$ bosonic delta functions, and as we noted in Section \ref{s:grassmMom}, four of them ensure external momentum conservation; similarly, eight of the fermionic delta functions impose supermomentum conservation. The remaining bosonic delta functions fix all degrees of freedom in $(2n-4)$-dimensional cells of G$(3,n)$. Each of these cells is labeled by a unique decorated permutation of the appropriate dimension. Each permutation is also associated with a representative {\bf \em on-shell diagram} (or plabic graph) following the techniques introduced in \cite{ArkaniHamed:2012nw}. For example, the permutation $\widetilde{\sigma}=\{3,5,6,7,9,8,11\}$ is represented by the graph 
\be
\raisebox{-1.5cm}{\includegraphics[height=3cm]{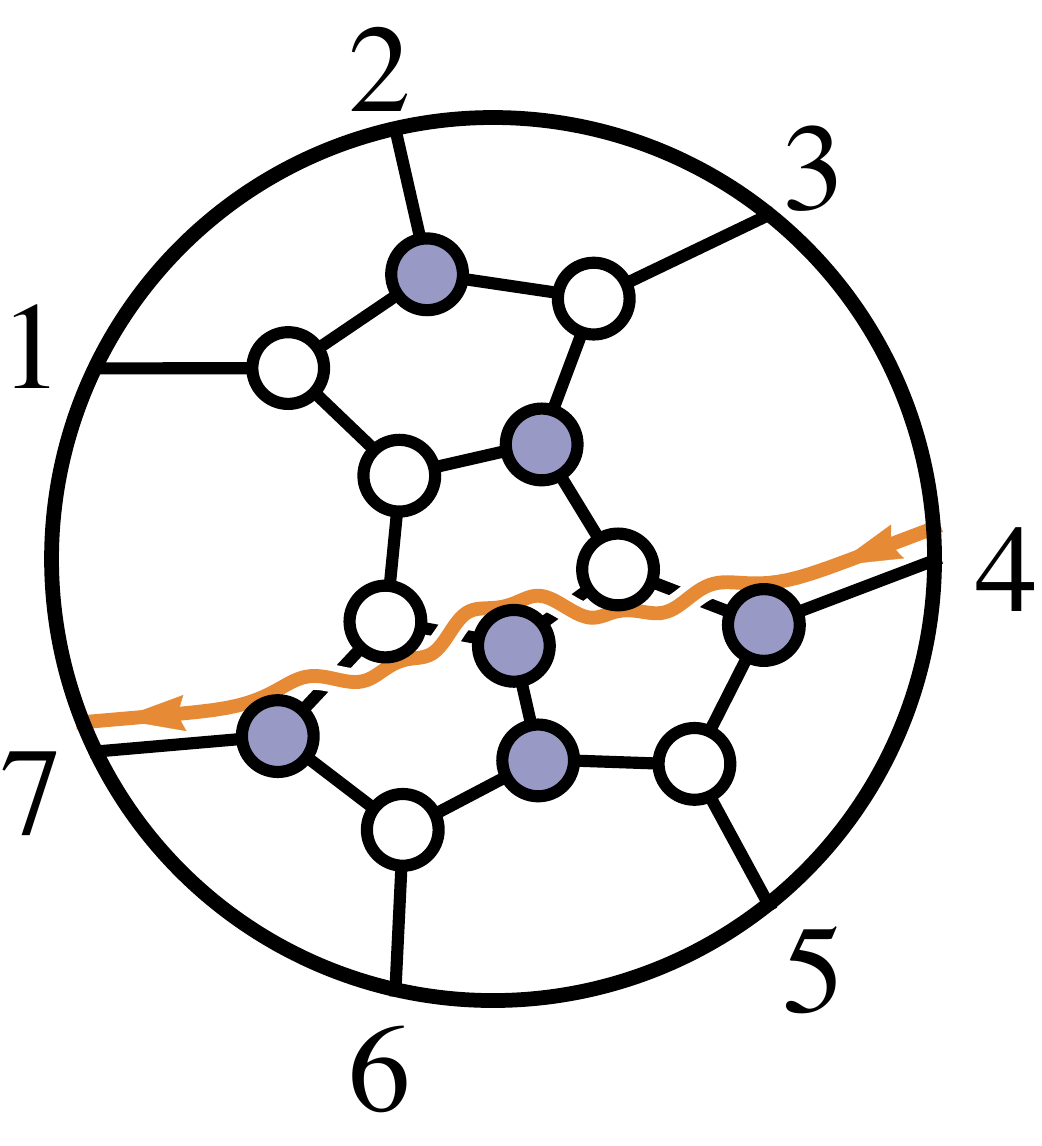}}~\,.
\label{fig:osd}
\ee
The permutation is obtained from the graph by following the `left-right paths' from each external leg, turning left at white vertices and right at black vertices; the figure \eqref{fig:osd} shows one such path, yielding $\widetilde{\sigma}(4)=7$.

The value of each on-shell diagram is computed by associating  each black/white vertex with a 3-point superamplitude, MHV or anti-MHV, respectively. External lines carry the information of the external data while for each internal line an integral must be performed over the corresponding momentum and Grassmann-variables. For details, see \cite{ArkaniHamed:2012nw}. The point that will be important for us in the following is that the vertices enforce special 3-particle kinematics, namely a white vertex with legs has $\lambda_a \propto \lambda_b \propto \lambda_c$ while a black vertex imposes the equivalent condition on the $\tilde{\lambda}$'s. 

In order to go from the $(3n-9)$-dimensional top cell in   G$(3,n)$ to a $(2n-4)$-dimensional cell, we need to eliminate $(3n-9)-(2n-4)=n-5$ degrees of freedom by taking an $(n-5)$-dimensional contour around singularities of the integrand. Of course one could evaluate the integral by first changing variables following the procedure of Section  \ref{s:TandMT} and treating the resulting integral in momentum twistor space, but we would like to treat the integral directly in momentum space. Unfortunately, this turns out to be a difficult problem due to the non-linear nature of the $n$ consecutive minors in the denominator. It is not \emph{a priori} sufficient to simply take $n-5$ of the minors to vanish. While this would land in a cell of the correct dimension, there are many $(2n-4)$-dimensional cells which cannot be reached by this technique. This failure is due to the appearance of so-called `composite singularities', which occur when one or more of the minors factorize on the zero-locus of a subset of the coordinates \cite{ArkaniHamed:2009dn}.  The dlog forms constructed in \cite{ArkaniHamed:2012nw} resolve many of these difficulties, but there are still some challenges associated with such forms as we mention in the Outlook.

\subsubsection{Relating the spaces}
In general, there are more $(2n-4)$-dimensional cells in G$(3,n)$ than $4$-dimensional cells in G$(1,n)$, so it may seem puzzling at first that the twistor and momentum twistor integrals of Section \ref{s:TandMT} are supposed to be equivalent. However, at this point we have not yet imposed the bosonic 
delta functions. NMHV cells are special because all 4d momentum twistor cells meet the support of the delta function. The same is not true of the momentum space cells; all, and only those, cells 
which meet the delta function support also have momentum twistor duals, but there are some momentum space cells of the correct dimension ($2n-4$) which do not intersect the delta function support and therefore have no corresponding cell in the Grassmannian with momentum twistor formulation. 

From the associated on-shell diagrams it is easy to see which NMHV momentum space cells will not be supported for generic momenta: any two external legs that are connected by a path containing vertices of only one color are forced by the delta functions at each vertex to have parallel momenta. This condition is not satisfied for generic external data; hence those residues vanish. For example, consider the 10d cell in G$(3,7)$ labeled by the permutation $\widetilde\sigma = \{4,5,6,7,9,10,8\}$, which is shown below
\be
\centering
\raisebox{-0.5\height}{\includegraphics[height=3cm]{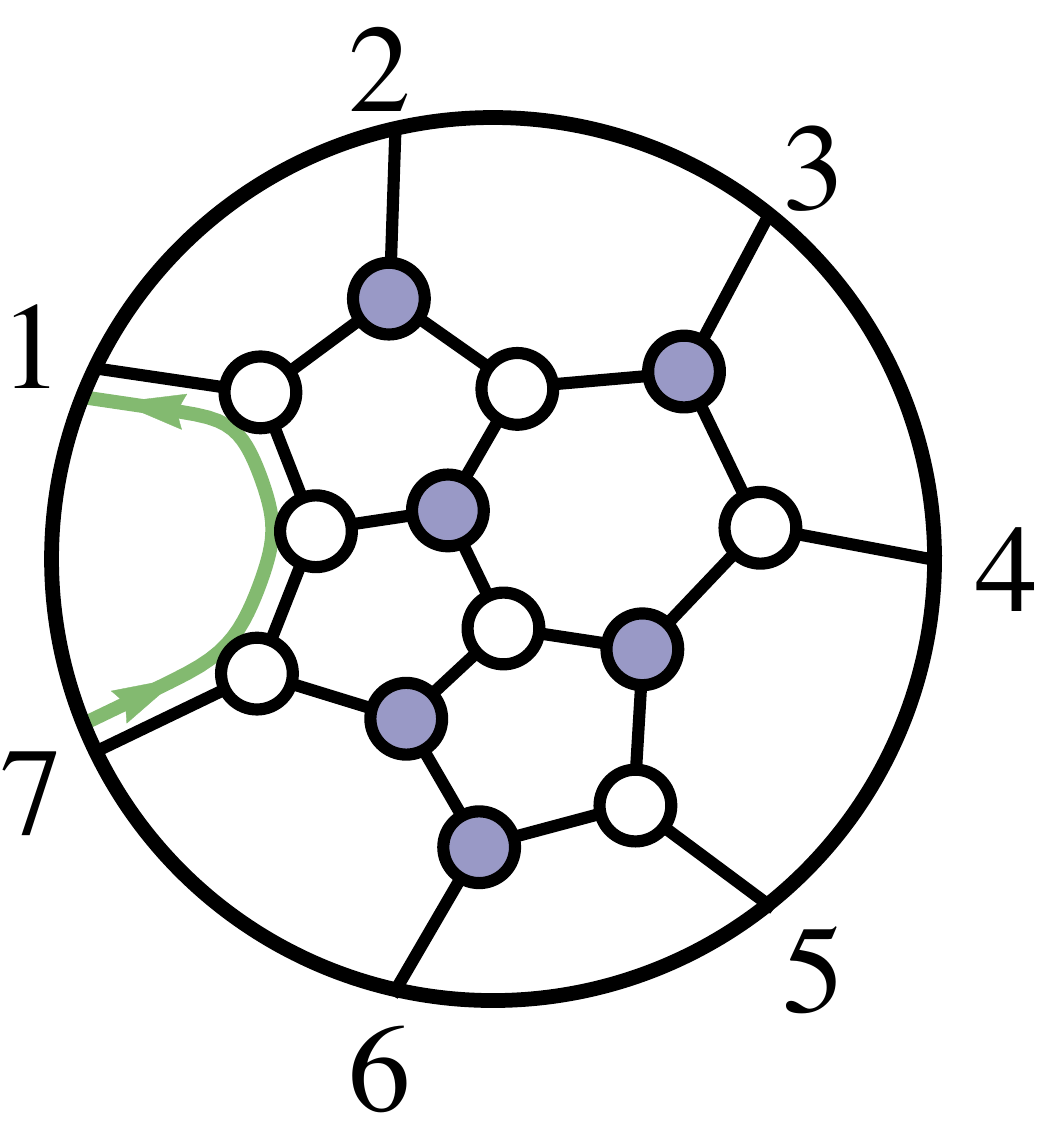}} ~~~~~\Rightarrow~~~~~ \lambda_1 \propto \lambda_7\,.
\label{fig:nosupport}
\ee
Since the momenta on legs 1 and 7 are not generically parallel, this cell is not supported by the delta functions.

This feature can also be seen from the permutations associated with such cells. Going from twistors to momentum twistors in Section \ref{s:TandMT}, we required that the $B$ plane contains the $\lambda$ 2-plane; see equation \eqref{Cfixed}. If $B$ does not contain a generic 2-plane, this imposes a constraint on  the 
external momenta; in other words, this cell does not intersect the delta function support for generic external momenta. In terms of the permutations, the requirement that $B$ contains a generic 2-plane is that $(\widetilde\sigma(i)-i)\geq 2$ for all $i$ (no column may be in the span of its nearest neighbor, and in particular no column may be zero). To see why, suppose that there exists some $i$ such that $(\widetilde\sigma(i)-i )= 0~\text{or}~1$. Then $b_i \in \text{span}\{b_{i+1}\}$ where $b_i$ is the $i^\tth$ column of $B$. Any 2-plane $\lambda\subset B$ would have to satisfy $\<i ,i+1\>=0$, which is clearly not satisfied for generic momenta. Furthermore, recall that the overall Jacobian of the transformation from twistors to momentum twistors, which is just the MHV superamplitude (see equation \eqref{momgrass5}), is singular precisely when $\<i,i+1\>=0$ for some $i$. In the example permutation above, $\widetilde\sigma(7) = 8$, so for the $\lambda$ plane to be contained in this cell it would have to satisfy $\<71\>=0$. Hence such a cell cannot have a momentum twistor dual.

Given that the number of cells that meet the support of the delta functions is identical in both spaces, it is not surprising that the permutations which label such cells are related. The permutations for supported momentum space cells $\widetilde\sigma$ can be obtained directly from momentum twistor labels $\sigma$ by the following map:
\begin{align}\label{eq:shift}
\widetilde{\sigma}(i) = \sigma(i+1) + 1 \,.
\end{align}
The inverse map from momentum space permutations to momentum twistor permutations, when it exists, was presented in eq.~(8.25) of \cite{ArkaniHamed:2012nw}. Since the momentum space permutations have physically meaningful on-shell diagram representatives, this map also provides a way to associate representative on-shell diagrams with momentum twistor cells of dimension $4$.\footnote{A diagrammatic representation directly in momentum twistor space has also been recently developed \cite{HeBai}.}

It is perhaps more surprising that non-vanishing residues in momentum space can also be labeled by $(n-5)$-index sequences similar to those which label the momentum twistor residues \eqref{5res}. From \eqref{minors}, we see that the vanishing of the $i^\tth$ minor in momentum twistor space implies that the $(i-1)^\tth$ (consecutive) minor in momentum space vanishes as well. Since there are no composite singularities in NMHV momentum twistor space, those residues are uniquely labeled by the set of vanishing minors, $\{i_1,i_2,\ldots,i_{n-5}\}_C$ (not to be confused with a permutation label). By \eqref{minors}, we can label the non-vanishing momentum space residues by a similar list of vanishing momentum space minors:
\begin{align}
\{i_1,i_2,\ldots,i_{n-5}\}_C \sim \{i_1-1,i_2-1,\ldots,i_{n-5}-1\}_B\,.
\end{align}
However, setting a collection of $(n-5)$ distinct consecutive minors to vanish in momentum space does not uniquely specify a cell in the Grassmannian.  Instead one obtains a union of cells, of which exactly one will have kinematical support.  For example, suppose $n = 7$ and we take the cell labeled by $\sigma = \{2,3,4,5,8,6,7\}$, given by the vanishing of the $6^\tth$ and $7^\tth$ minors.  Then by \eqref{eq:shift}, we have $\widetilde{\sigma} = \{4,5,6,9,7,8,10\}$, and the minors labeled by columns $(5,6,7)$ and $(6,7,1)$ vanish.  There is exactly one other cell of dimension $2n-4 = 10$ for which exactly these same minors vanish, namely $\widetilde{\sigma}' = \{4,5,6,8,9,7,10\}$.  However, $\widetilde{\sigma}'$ does not have kinematical support. This can be seen directly from the two corresponding on-shell diagrams,
\be
\centering
\begin{array}{cc}
\raisebox{-0.5\height}{\includegraphics[height=3cm]{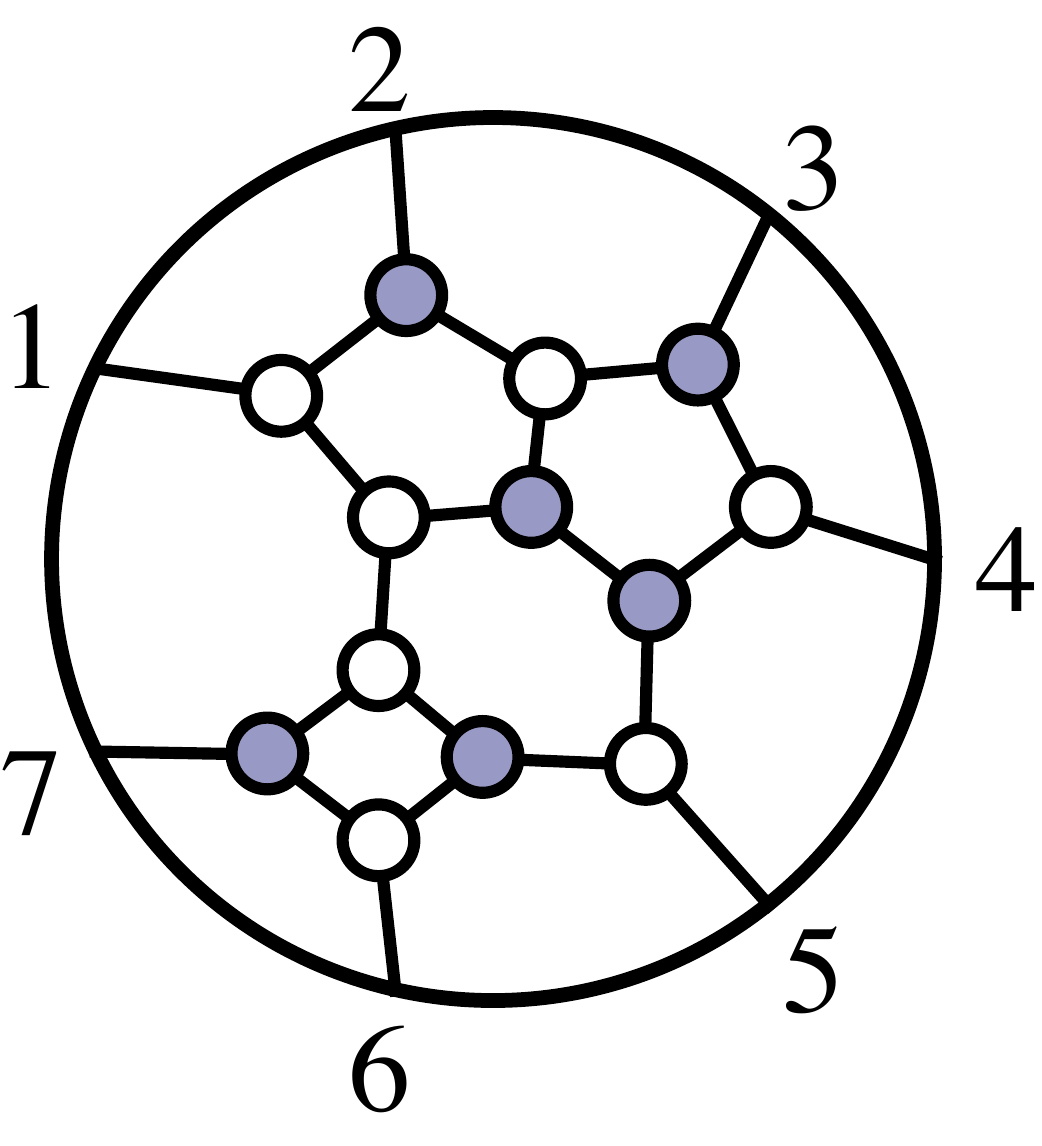}}
~~&~~
\raisebox{-0.5\height}{\includegraphics[height=3cm]{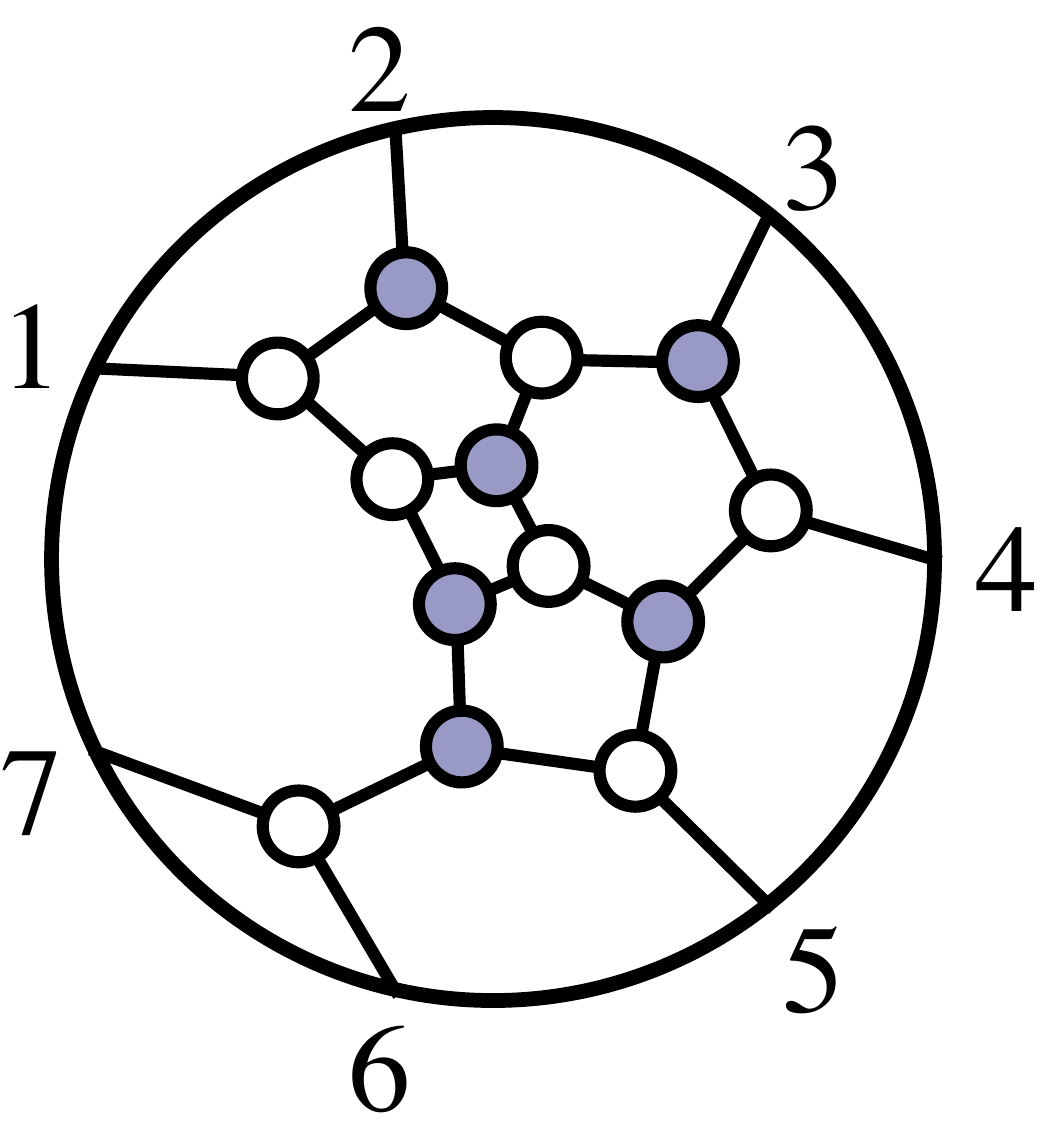}}\\
\widetilde{\sigma} = \{4,5,6,9,7,8,10\}
~~~~~&~~~~~
\widetilde{\sigma}' = \{4,5,6,8,9,7,10\}
 \,.
 \end{array}
\label{fig:nosupport2}
\ee
The second one vanishes for generic external data since it requires $\lambda_6 \propto \lambda_7$.

In general, the momentum space residue $\{i_1-1,i_2-1,\ldots,i_{n-5}-1\}_B$ is just the residue for the cell $\widetilde{\sigma}$. Recalling the Jacobian from Section \ref{s:TandMT}, we have the following relationship between residues in momentum twistor and momentum space:
\begin{align}
\A_n^{\text{MHV}}\{i_1,i_2,\ldots,i_{n-5}\}_C = \{i_1-1,i_2-1,\ldots,i_{n-5}-1\}_B\,.
\end{align}
It is also suggestive that, even though the general contours may be difficult to handle, the residues relevant for physics may be easier to study since they will not involve any composite residues. We leave this for future work.

\subsubsection{Pushing the boundaries}
\label{s:bdr}
In Section \ref{s:resapp} we used the boundary operation in two different contexts: one encoded the residue theorems that give linear relations among 5-brackets, the other selected residues of poles in the five brackets. Let us now see how these arise as boundary limits in the momentum twistor space Grassmannian.

Let us begin with a 4d cell in the NMHV Grassmannian G$(1,n)$. It is characterized by having precisely five non-vanishing entries in the representative matrix, say $c_{a,b,c,d,x}$. The boundaries of this 4d cell are the 3d cells obtained by setting one extra entry of the representative matrix to zero, i.e.~there are five 3d boundaries. Suppose we go on the boundary characterized by $c_x = 0$. Then there is not generically support on the delta functions in the Grassmannian integral, because they then enforce 
$c_a Z_a + c_b Z_b + c_c Z_c + c_d Z_d = 0$. This is a constraint on the external data that requires  the momentum twistors $Z_{a,b,c,d}$ to be linearly dependent. That is equivalent to the statement that the 4-bracket $\four{abcd}$ vanishes and, as we know from \reef{resresult}, this is precisely one of the poles in 5-bracket $\five{abcdx}$ associated with the residue of our 4d cell. Similarly, we see that the five poles of $\five{abcdx}$ are precisely in 1-1 correspondence with the 3d boundaries of the corresponding 4d cell. This justifies the terminology ``boundary operation" used in the discussion in Section \ref{s:tree}. The relative signs in \reef{bdr5bracket} 
come from the orientations of the boundaries.

Two 4d cells labeled by non-vanishing entries  $c_{a,b,c,d,x}$ and $c_{a,b,c,d,y}$, respectively, share one common 3d boundary characterized by $c_x = 0$ and  $c_y=0$. As we know from the analysis in Section \ref{s:tree}, the residue of the associated pole at $\four{abcd}$ cancels in the difference of the associated 5-brackets, $\five{abcdx}-\five{abcdy}$. Analogously, the shared boundary between the cells cancels in the sum if they are oppositely oriented.
The locality condition of having no pole $\two{abcd}$ for $(a,b,c,d)$ not of the form $(i,i+1,j,j+1)$ translates to the requirement that all shared boundaries not of the corresponding form must be oppositely oriented. These locality conditions, together with the fact that the amplitude is a sum of 4d cells, determine the formula \reef{NMHVtree}.

The residue theorems \reef{bdrOp} also have a boundary interpretation in the Grassmannian: boundaries of 5d cells give different equivalent ways of writing the same amplitude formula. We interpret the 4d cells associated with the 5-bracket residues as the boundary of a 5d cell defined by having precisely six non-vanishing entries in the representative matrices, say $c_{a,b,c,d,e,f}$. The six 4d boundaries of this 5d cell are associated with sending one of these six entries to zero. Meanwhile, the Grassmannian integral on the 5d cell is a contour integral on $\CP^1$ with six poles, corresponding to the 4d boundaries above. 
The sum of the residues at these six poles must therefore be zero by Cauchy's theorem. This is our familiar residue theorem \reef{resCon}, and we see now why it is natural to associate it with a boundary operation.

\section{3d ABJM Grassmannian}\label{sec:3DGrass}
The fascinating relation between cells of Grassmannian and scattering amplitudes of 4d $\mathcal{N}=4$ SYM has a parallel  in 3d  ABJM theory \cite{WestCSM,EastCSM}. Previously, an ABJM Grassmannian was developed for external data in momentum space \cite{LeeOG,HW,HWX,KimLee}. The purpose of this section is to apply the strategy from Section \ref{s:TandMT} to derive the ABJM Grassmannian in momentum twistor space. 

\subsection{Momentum space ABJM Grassmannian}
ABJM theory is a 3-dimensional $\mathcal{N}=6$ superconformal Chern-Simons matter theory with 4 complex fermions $\psi^A$ and 4 complex scalars $X^A$, transforming in the $\bf{4}$ and $\bf{\bar{4}}$ under the $SU(4) \sim SO(6)$ R-symmetry. The physical degrees of freedom are the matter fields, and the symmetries imply that the only  non-vanishing  amplitudes have even multiplicity, in particular one can show that $n=2k+4$ for the N$^k$MHV sector.  (For further discussion of 3d kinematics and ABJM amplitudes, see Chapter 11 of the review \cite{Elvang:2013cua}.)

A 3d momentum vector $p_i$ 
can be encoded in a symmetric $2 \times 2$ matrix $p_i^{ab}$. The on-shell condition requires it to have vanishing determinant, so we can write $p_i^{ab} = {\lambda'}_i^a {\lambda'}_i^b$, with $a,b=1,2$ being $SL(2,\mathbb{R})$ indices. One version of the 3d spinor helicity formalism uses these 2-component commuting spinors $\lambda'$ to encode the two on-shell degrees of freedom needed for a null 3d momentum vector. As in 4d, we can form the antisymmetric angle-bracket product $\<ij\> := \eps_{ab} {\lambda'}_i^a {\lambda'}_j^b$, although in 3d there are no square-spinors.  The $\mathcal{N}=6$ on-shell superspace for 3d ABJM theory involves three Grassmann variables ${\eta'}_{iI}$ with  $SU(3) \subset SU(4)$ R-symmetry indices $I=1,2,3$. We are denoting the $\lambda'_i$s (and ${\eta'}_{iI}$s) with primes to distinguish them from a different formulation of the 3d spinor helicity formalism to be introduced in Section \ref{s:MT3d}.
 
In on-shell superspace, the states of ABJM theory are organized in 
two on-shell supermultiplets
\be
\begin{split}
  \Phi
  ~=&~
  X_4+\eta'_A\,\psi^A
  -\frac{1}{2}\epsilon^{ABC}\,\eta'_A\eta'_B\,X_C
  -\eta'_1\eta'_2\eta'_3\,\psi^4\,,\\
  \bar{\Psi}
  ~=&~\bar{\psi}_4+\eta'_A\bar{X}^A
  -\frac{1}{2}\epsilon^{ABC}\,\eta'_A\eta'_B\,\bar{\psi}_C
  -\eta'_1\eta'_2\eta'_3\,\bar{X}^4\,.
\end{split}
\label{ABJMmap}
\ee
The superfield $\Phi$ is thus bosonic in nature while $\Psi$  is fermionic. 
The states of the color-ordered tree-level superamplitude in planar ABJM are arranged with alternating $\Phi$ and $\Psi$, e.g.~$\mathcal{A}_n(\Phi \Psi \Phi \Psi  \dots)$, and as a result they  do not obey the cyclic invariance $i \to i+1$ of superamplitudes in  planar $\cn=4$ SYM. Instead, the planar ABJM superamplitudes  are invariant under $i \to i+2$, up to a sign of $(-1)^{n/2+1}$.

The leading singularities of ABJM theory, and consequently the tree-level amplitudes, enjoy an $OSp(6|4)$ Yangian symmetry \cite{TillDC,ABJMBCFW}, and are given as residues of the following Grassmannian integral \cite{LeeOG,HW,HWX,KimLee}:
\eq\label{MasterInt}
\mathcal{L}_{2\tilde{k};\tilde{k}}=\int \frac{d^{2\tilde{k}^2} B'}{GL(\tilde{k})}\,\frac{\delta^{\tilde{k}(\tilde{k}+1)/2}\big(B'\cdot B^{'T}\big)
\delta^{2\tilde{k}|3\tilde{k}}\big(B'\cdot \Lambda' \big)}{m'_1 m'_2 \cdots m'_{\tilde{k}} }\,,
\eqe
where $\tilde{k}=k+2 = \frac{n}{2}$ and $\Lambda'_i=(\lambda'_i | \eta'_i)$ is the external data given in  3d momentum space. 
The denominator contains the product of the first $\tilde{k}$ consecutive minors $m'_i$ of $B'$.\footnote{We note here that for $k$ even this means that the states of the superamplitude are $\mathcal{A}_n(\Phi \Psi \Phi \Psi  \dots)$ while for $k$ odd  they are  $\mathcal{A}_n(\Psi \Phi \Psi \Phi  \dots)$. This ensures the correct little group scaling in 3d, for which the superamplitude is invariant for $\Phi$  states and changes signs for  $\Psi$  states. Alternatively, one can replace $m_1 m_2 \cdots m_{\tilde{k}}$ by $\sqrt{m_1 m_2 \cdots m_n}$. The two forms are equivalent up to signs depending on the branch of solutions to the orthogonal constraint \cite{HW}. } 
The bosonic delta-functions enforce the $\tfrac{1}{2}\tilde{k}(\tilde{k}+1)$ constraints 
\be
  0= B'\cdot {B'}^{T} = \sum_iB'_{\alpha i}B'_{\beta i}
  =  \sum_{i,j} B'_{\alpha i}B'_{\beta j} g^{ij}\,,
\ee  
with $g^{ij}=\delta^{ij}$ is the trivial metric and $\alpha, \beta = 1,2,\ldots,\tilde{k}$. Thus, in momentum space, the Grassmannian for ABJM theory is an {\bf\em 
orthogonal Grassmannian} (also known as an {\bf\em isotropic Grassmannian} in the mathematics literature) defined as the space of null $\tilde{k}$-planes in an $n$-dimensional space equipped with an internal metric $g^{ij}$. The metric is trivial in momentum space. 
We will denote an orthogonal Grassmannian as  OG$(\tilde{k},n)$.\footnote{We remark that in the theory of planar electrical networks, the orthogonal Grassmannian defined with respect to the metric $g^{ij} =\delta_{i,j+\tilde k} + \delta_{i,j-\tilde k}$ appears \cite{HS,Lam}.  Curiously, the combinatorics of ABJM amplitudes and of planar electrical networks are very closely related, but there is as yet no conceptual explanation of this.}
Because of the quadratic condition $B'\cdot {B'}^{T} = 0$, the orthogonal Grassmannian has two distinct components.  

The dimension of the integral \reef{MasterInt} is
\be
   2 \tilde{k}^2 - \tilde{k}^2 - \frac{1}{2}\tilde{k}(\tilde{k}+1) - 2\tilde{k} +3
   = \frac{(\tilde{k}-2)(\tilde{k}-3)}{2} \,,
   \label{dimInt3dA}
\ee
where the ``$+3$" is because momentum conservation is automatically encoded in the bosonic delta-functions; this will become evident in the following. 

Note that in general the metric for a Grassmannian need not be diagonal nor proportional to the identity. For example, positivity for cells of OG$(\tilde{k},n)$ is defined with the metric of alternating signs $(+,-,+,\cdots, -)$ \cite{HWX}. In the following we will see that in converting \reef{MasterInt} into momentum twistor space, we will naturally encounter more general metrics $g$ and denote the corresponding orthogonal Grassmannian as OG$_g(\tilde{k},n)$.

\subsection{3d momentum twistors}
\label{s:MT3d}
We would now like to introduce 3d momentum twistors.\footnote{See~\cite{Lipstein} for an alternative definition.} A natural way to define momentum twistor variables in 3d is to reduce it from 4d. With the 4d conformal group $SO(2,4) \sim SU(4)$, a natural way to introduce momentum twistors is to first define 4d  spacetime as a projective plane in 6d. This ``embedding space" formalism~\cite{Embedding} introduces a 6d coordinate $Y^{AB}$, which is anti-symmetric in the $SU(4)$ indices $A,B$. 4d spacetime is then defined to be the subspace $Y^2=\epsilon_{ABCD}Y^{AB}Y^{CD}=0$ with the projective identification $Y\sim rY$ (for $r$ a real or complex number, depending on the context). 
A solution to the constraint $Y^2=0$ is to write $Y^{AB}$ as a bi-twistor:
\eq
Y^{AB}_i=Z_i^{[A}Z_{i-1}^{B]}\,.
\eqe
To honor the projective constraint $Y_i\sim rY_i$, the $Z_i$s must be defined  projectively, $Z_i \sim r Z_i$.  Here $i=1,2,\dots,n$ label $n$ points $Y_i$ in the embedding space.  

Consider now the 3d analogue for which the embedding space is 5d. We can start with the 6d space and introduce an $SO(2,3)$-invariant constraint to remove the extra degree of freedom. A natural choice is to impose a $SO(2,3)\sim Sp(4)$ tracelessness condition on $Y$:
\eq
Y^{AB}\Omega_{AB}=0\,, 
\hspace{1cm} 
\Omega_{AB} = \left(\begin{array}{cc}0 & -I \\ I& 0\end{array}\right)\,.
\eqe
This also implies that the bi-twistors $Z_i$ must satisfy: 
\eq\label{ZConstraint}
Z_{i}^AZ^B_{i+1}\Omega_{AB}=0\,.
\eqe
Note that \reef{ZConstraint} is projectively well-defined, so we can construct the 3d momentum twistor as a familiar 4d momentum twistor $Z_i=(\lambda_i , \mu_i)$ subject to the constraint:
\eq\label{ZConstraint2}
\lda i, i{+}1\rda \,:=\, Z_{i}^AZ^B_{i+1}\Omega_{AB}= \langle \mu_{i} \lambda_{i+1}\rangle-\langle  \mu_{i+1}\lambda_{i}\rangle \,=\,0\,.
\eqe

We now need to identify the relation between $\lambda_i$ and $\lambda'_i$. Recall that in three dimensions, a massless momentum can be parameterized as $p_i=E_i(1, \sin\theta_i, \cos\theta_i)$, where $E_i$ is the energy. In bi-spinor notation, we can deduce:
\eq
\lambda'_{ia}\lambda'_{ib}=p_{iab}=E_i\left(\begin{array}{cc}-1+\cos\theta_i & \sin\theta_i \\ \sin\theta_i & -1-\cos\theta_i\end{array}\right)\quad \rightarrow \quad \lambda'_{ia}=i\sqrt{2E_i}\left(\begin{array}{c} -\sin\frac{\theta_i}{2} \\ \cos\frac{\theta_i}{2} \end{array}\right)\,.
\eqe 
Now since $Z_i$ is defined projectively, the components of $Z_i$ must have well-defined projective scalings. This is not possible with $Z_i=(\lambda'_i,\mu_i)$ because $p_i$ is not invariant under the scaling $\lambda'_i \rightarrow t_i \lambda'_i$. Consequently, 
$y_i=p_{i}-p_{i+1}$ cannot have any nice homogenous scaling property and neither can $\mu_i$, since the latter is defined through the incidence relation
$\mu_i^a=y_i^{ab}\lambda'_{i b}$.

The resolution is to parameterize the 3d kinematics in a fashion that is similar to 4d. We define
\eq\label{NewVariable}
\lambda_{i a}=\left(\begin{array}{c} -\sin\frac{\theta_i}{2} \\ \cos\frac{\theta_i}{2} \end{array}\right)\,,
\hspace{1cm}
 \tilde{\lambda}_{i a}=-2E_i\lambda_{i a} 
\eqe  
such that we now have $p_i=\lambda_i\tilde{\lambda}_i$. For simplicity, we set  $\tilde{E}_i=-2E_i$ in the following. Note that the number of degrees of freedom for each particle is still 2 and that $p_i$ is now invariant under the following scaling rules:
\eq
\lambda_i\rightarrow t_i\lambda_i, \quad \tilde{E}_i\rightarrow  t_i^{-2} \tilde{E}_i\,.
\eqe
Since $p_i$ (and hence $y_i$) is invariant, $\mu_i$ has the same scaling property as $\lambda_i$ through the new incidence relation  
\eq\label{Incidence2}
\mu_i^a= y_i^{ab}\lambda_{i b} =y_{i+1}^{ab}\lambda_{i b}: ~~\quad \mu_i\rightarrow t_i\mu_i\,.
\eqe
With the incidence relation (\ref{Incidence2}) and the symmetry of the $y_i$-matrices, the constraint (\ref{ZConstraint2}) is automatically satisfied
\eq
\langle \mu_{i} \lambda_{i+1}\rangle-\langle  \mu_{i+1}\lambda_{i}\rangle
=\lambda_{ia}y_i^{ab}\lambda_{i{+}1,b}
-\lambda_{i+1,a}y_{i{+}1}^{ab}\lambda_{ib}=0\,.
\eqe
 Thus we have deduced a suitable form of 3d momentum twistor which has a well defined projective property. It has many of the same properties as its 4d cousin, for example \reef{y-lambda-mu}  holds, and one can directly derive the 3d versions of the relations \reef{inc}: 
\be
  \begin{split}
  \tilde{\lambda}_{i} 
  &=~
  \frac{\two{i+1,i}\mu_{i-1} 
     + \two{i,i-1}\mu_{i+1}
     + \two{i-1,i+1}\mu_{i} 
      } 
      {\two{i-1,i}\two{i,i+1}}\,,
   \\[2mm]
  \tilde{\eta}_{iA} 
  &=~
  \frac{\two{i+1,i}\eta_{i-1,A} 
      + \two{i,i-1}\eta_{i+1,A}
      + \two{i-1,i+1}\eta_{iA} } 
      {\two{i-1,i}\two{i,i+1}}\,.
  \end{split}
  \label{inc3d}
\ee
where we define $\tilde{\eta}_i:=\eta_i'\sqrt{\tilde{E}_i}$.

In summary, we have found that the new momentum space variables
\eq
\lambda_i=\frac{1}{\sqrt{\tilde{E}_i}}\lambda'_i\,,
~~~~~~~~
 \tilde{\lambda}_i=\sqrt{\tilde{E}_i}\lambda'_i\, ,
~~~~~~~~
\tilde{\eta}_i=\eta_i'\sqrt{\tilde{E}_i}
\eqe
facilitate the introduction of 3d momentum supertwistors $\mathcal{Z}_i = (\lambda_i, \mu_i | \eta_i)$, which are just like the 4d ones but subject to the constraints \reef{ZConstraint2}. Also, in addition to the $SL(4)$-invariant $\<ijkl\>$ defined in \reef{def4bracket}, we now have a 2-bracket invariant
\be
\lda ij\rda:=  Z_{i}^AZ^B_{j}\Omega_{AB}\, .
\label{angle2}
\ee
As noted in \reef{ZConstraint2}, the projection from 4d momentum twistors to 3d ones is defined by $\lda i, i+1\rda =0$.
The 4-brackets and 2-brackets are related via a version of the Schouten identity:
\eq
\langle ijkl\rangle=\lda ij\rda\lda kl\rda+\lda ik\rda\lda lj\rda+\lda il\rda\lda jk\rda \,.
\label{4Dschouten}
\eqe
This follows from 
the identity $\epsilon_{ABCD}=-\Omega_{AB}\Omega_{CD}-\Omega_{AC}\Omega_{DB}-\Omega_{AD}\Omega_{BC}$. 
Note that the RHS of \reef{4Dschouten} has the same form as the 2d Schouten identity for angle-brackets, but the LHS has the non-vanishing contraction with the Levi-Civita symbol because the momentum twistors are 4-component objects.

As a side-remark, we note that just as the parameterization in (\ref{NewVariable})  can be viewed as a descendant from the 4d spinor helicity variables, it also has another 2d sibling:  the 2d momentum twistors that correspond to taking $\theta=0$ or $\pi$ in (\ref{NewVariable}):
\eq
\lambda_i|_{\theta_i=0}=\left(\begin{array}{c} 0 \\ 1 \end{array}\right)=: \lambda^+, 
~~~~
\tilde\lambda^+= \tilde{E}_i\lambda^+,
~~~~ 
\lambda_i|_{\theta_i=\pi}=\left(\begin{array}{c}-1 \\ 0\end{array}\right)=: \lambda^-,
~~~~
\tilde\lambda^-= \tilde{E}_i\lambda^-\,.
\eqe 
The superscript $(+,-)$ indicates which of the two distinct light-cone directions  contains the corresponding momentum: 
\eq
\lambda_i^+\tilde{\lambda}_i^+
=p_i^+
=\left(\begin{array}{cc}0 & 0 \\0 & \tilde{E}_i\end{array}\right),
\quad ~~~
\lambda_i^-\tilde{\lambda}_i^-
=p_i^-
=\left(\begin{array}{cc}\tilde{E}_i & 0 \\0 & 0\end{array}\right)\,.
\eqe
These 2d momentum variables have been used to study scattering amplitudes of planar $\mathcal{N}=4$ SYM limited to 2d kinematics~\cite{2DK1, 2DK2}.

\subsection{Derivation of ABJM momentum twistor space Grassmannian}
We are now ready to convert the integral formula in (\ref{MasterInt}) to momentum twistor space.\footnote{A similar attempt was initiated in~\cite{UnPub} with a different definition of 3d momentum twistors. As a result, projectivity was not well defined.} First we change variables 
$B'_{\alpha i} = \sqrt{\tilde{E}_i} B_{\alpha i}$, 
$\lambda'_i =\tilde{\lambda}_i / \sqrt{\tilde{E}_i}$, 
$\eta_i'=\tilde{\eta}_i/\sqrt{\tilde{E}_i}$,  so that only the variables in (\ref{NewVariable}) appear in the bosonic delta functions: 
\eq
\mathcal{L}_{2\tilde{k};\tilde{k}}=
J_E 
\int \frac{d^{2\tilde{k}^2}B}{GL(\tilde{k})\,m_1 m_2 \cdots m_{\tilde{k}} }
\,\delta^{\tilde{k}(\tilde{k}+1)/2}\big(B_{\alpha i}B_{\beta j}g^{ij}\big)
\,\delta^{2\tilde{k}}\big(B\cdot \tilde\lambda\big)
\,\delta^{(3\tilde{k})}\big(B\cdot \tilde{\eta}\big)\,,
\eqe
where the factor 
\be
J_E =
  \frac{\prod_{i=1}^{n} \tilde{E}_i^{\tilde{k}/2}}
  {\big( \tilde{E}_1 \tilde{E}_2^2 \tilde{E}_3^3 
  \cdots \tilde{E}_{\tilde{k}}^{\tilde{k}}
  \tilde{E}_{\tilde{k}+1}^{\tilde{k}-1} 
  \cdots \tilde{E}_{n-2}^2 \tilde{E}_{n-1}
  \big)^{1/2} }
\ee
comes from the scaling of the measure and the minors. 
The  $2\tilde{k}$-dimensional metric $g^{ij}$ is
\eq
g^{ij}=\left(\begin{array}{cccc}\tilde{E}_1 & 0 & 0 & 0 \\0 & \tilde{E}_2 & 0 & 0 \\0 & 0 & \ddots & 0 \\0 & 0 & 0 & \tilde{E}_{2k}\end{array}\right)\,.
\eqe
Just as in 4d, the delta function $\delta^{2\tilde{k}}(B\cdot \tilde\lambda)$ requires $B$ to be orthogonal to the $\tilde{\lambda}$-plane, and by momentum conservation ($\sum_i \lambda_i \tilde\lambda_i = 0$), 
the $B$-plane must therefore contain the $\lambda$-plane. We can use this to gauge-fix part of the $GL(\tilde{k})$ as in 4d: 
\be
B_{\alpha i}= \left(\begin{array}{c} B_{\hat\alpha i}  \\ \lambda^1_i \\ \lambda^2_i \end{array}\right)\,,
\ee
where $\hat\alpha=1,\ldots,k$ and $k=\tilde{k}-2$. With this gauge choice, the remaining delta functions become\footnote{Note that for $\alpha=\tilde{k}-1,\tilde{k}$, we have $\lambda_i B_{\hat{\beta}j}g^{ij}=\lambda_i B_{\hat{\beta}i}\tilde{E}_i=\tilde{\lambda}_i B_{\hat{\beta}i}$} 
\eqa
\delta^{\tilde{k}(\tilde{k}+1)/2}\big(B_{\alpha i}B_{\beta j}g^{ij}\big)
\,\delta^{3\tilde{k}}\big(B\cdot\tilde{\eta}\big)
\rightarrow 
\delta^{3}(\mathcal{P})\delta^{6}(\mathcal{Q})
\,\delta^{k(k+1)/2}\big(B_{\hat{\alpha}i}B_{\hat\beta j}g^{ij}\big)
\,\delta^{2k|3k}\big(B_{\hat{a}}\cdot\tilde\Lambda\big)\,,
\eqae
where $\tilde\Lambda=(\tilde\lambda | \tilde{\eta})$, $\mathcal{P}$ is the total momentum, and $\mathcal{Q}$ is the total supermomentum. 

Now we can  follow the steps from the $\mathcal{N}=4$ SYM analysis in Section  \ref{s:TandMT} to convert the Grassmannian integral to one with momentum twistor external data:
\begin{enumerate}
\item The relation between the momentum space data and the momentum twistor variables allow us to introduce a new variable $C_{\hat{\alpha} i}= B_{\hat\alpha j}Q_{ji}$, with $Q$ defined in \reef{CfromBgauge12} (see also \reef{Qmatrix}).
\item Rewrite the minors using \reef{minors}.
\item
 To invert the relation $B_{\hat\alpha j}Q_{ji}$, we use translation invariance $T_k$ to fix $B_{\hat{\alpha} 1}=B_{\hat{\alpha} 2}=0$, thus obtaining $B_{\hat\alpha \hat{j}}=C_{\hat\alpha \hat{i}}(Q^{-1})_{\hat{i}\hat{j}}$, with $\hat{i}=3,\ldots, n$. A simple expression for $Q\inv$ is given in \eqref{Qinv}, and we verify its form in Appendix \ref{app:Qinv}. This allows us to change variables from $B_{\hat\alpha \hat{j}}$ to $C_{\hat\alpha \hat{j}}$. We restore the integration variables to include $C_{\hat\alpha 1}, C_{\hat\alpha 2}$ by introducing  $\delta^{2k}(C\cdot \lambda)$. 
\end{enumerate}
This brings us to the following final form of the Grassmannian integral ($n=2k+4$):
\be\label{MasterInt2}
 \mathcal{L}_{n;k}
=
J \times 
\delta^{3}(\mathcal{P})\,\delta^{(6)}(\mathcal{Q})
\times \int \frac{d^{kn}C}{GL(k)}
\frac{\delta^{k(k+1)/2}\Big(C_{\hat\alpha \hat{i}}(Q^{-1})_{\hat{i}\hat{j}}C_{\hat\beta\hat{k}}(Q^{-1})_{\hat{k}\hat{l}}\,g^{\hat{j}\hat{l}}\Big)
\,\delta^{4k|3k}\big(C\cdot \mathcal{Z}\big)}{M_2 M_3 \cdots M_{k+3}}\,,
\ee
where $\mathcal{Z}_i=(Z_i|\eta_i)$ and
\be
J  = \big(\langle 12\rangle\langle 23\rangle\cdots \langle n1\rangle\big)^k 
  \frac{
  \prod_{i=1}^{n} \tilde{E}_i^{(k+2)/2}}
  { \tilde{E}_{k+2}^{(k+2)/2}
  \prod_{i=1}^{k+1} \Big(\tilde{E}_i^{1/2} \<i,i+1\>
   \<n-i-1,n-i\>\tilde{E}_{n-i}^{1/2} \Big)^{i}}\,.
\label{uglyJ}
\ee 
It is straightforward to verify that the $GL(1)$ weight of the integral in \reef{MasterInt2} cancels. It should be noted that here and in the remainder of this section, the angle-brackets $\<ij\>$ are now composed of the $\lambda$-spinors, not the $\lambda'$'s.

There is an unsatisfactory feature in (\ref{MasterInt2}): the sum in the constraint $C Q^{-1} C Q^{-1} g=0$ only runs over $\hat{i},\hat{j}=3,\ldots, n$, and therefore $(Q^{-1} Q^{-1} g)^{\hat{i}\hat{j}}$ cannot be interpreted as an orthogonal constraint on the Grassmannian $C$. This is because an orthogonal Grassmannian is defined with a metric $G^{ij}$ whose indices run over the full $n$-dimensional space. However, using the delta function support we can  rewrite $C Q^{-1} C Q^{-1} g$  in terms of a non-degenerate effective metric $G^{ij}$ specified by the external data. To see this note that on the support of the $\delta(C\cdot \lambda)$, we have
\eq\label{DQ}
C_{\hat{\alpha} \hat{i}}(Q^{-1})_{\hat{i}\hat{j}}=\sum_{i=2}^{\hat{j}-1}C_{\hat{\alpha} i}\langle i\hat{j}\rangle\,.
\eqe
For  example, for $k=1$, we have $\hat{i},\hat{j}=3,\ldots,6$ and
\eq
C_{\hat{\alpha} \hat{i}}(Q^{-1})_{\hat{i}\hat{j}}=
\left(\begin{array}{c} C_{\hat{\alpha} 2}\langle 23\rangle \\ 
C_{\hat{\alpha} 2}\langle 24\rangle +C_{\hat{\alpha} 3}\langle 34\rangle\\ 
C_{\hat{\alpha} 2}\langle 25\rangle +C_{\hat{\alpha} 3}\langle 35\rangle+C_{\hat{\alpha} 4}\langle 45\rangle \\ 
C_{\hat{\alpha} 2}\langle 26\rangle +C_{\hat{\alpha} 3}\langle 36\rangle+C_{\hat{\alpha} 4}\langle 46\rangle+C_{\hat{\alpha} 5}\langle 56\rangle
\end{array}\right)\,.
\label{CQ1}
\eqe
Note that $C_{\hat{\alpha} 1}$ and  $C_{\hat{\alpha} 6}$ do not appear in (\ref{CQ1}). We can use $C\cdot \lambda=0$ to get an expression in terms of a `conjugate' set of $C_{\hat{\alpha} i}$'s, e.g.~\eq
C_{\hat{\alpha} \hat{i}}(Q^{-1})_{\hat{i}\hat{j}}
=
-\left(\begin{array}{c} C_{\hat{\alpha} 1}\langle 13\rangle +C_{\hat{\alpha} 4}\langle 43\rangle+C_{\hat{\alpha} 5}\langle 53\rangle+C_{\hat{\alpha} 6}\langle 63\rangle \\ 
C_{\hat{\alpha} 1}\langle 14\rangle +C_{\hat{\alpha} 5}\langle 54\rangle+C_{\hat{\alpha} 6}\langle 64\rangle \\ 
C_{\hat{\alpha} 1}\langle 15\rangle +C_{\hat{\alpha} 6}\langle 65\rangle  \\ 
C_{\hat{\alpha} 1}\langle 16\rangle\end{array}\right)\,.
\label{CQ2}
\eqe
To reveal the symmetric form of the effective (inverse) metric $\tilde{G}^{ij}$ defined via  $C_{\hat{\alpha} i} C_{\hat{\beta} j} \tilde{G}^{ij} = ( C_{\hat{\alpha}} Q^{-1}) ( C_{\hat{\alpha}} Q^{-1})g$, we take the symmetric (in $\hat{\alpha}$ and $\hat{\beta}$) product of two copies of $C_{\hat{\alpha} \hat{i}}(Q^{-1})_{\hat{i}\hat{j}}$, one in the form \reef{CQ1} and one in the conjugate form \reef{CQ2}. For $k=1$, we find
\be
\tilde{G}^{ij}
=
\frac{1}{2}
\left(\begin{array}{cccccc}0 & 0 & -\langle 1|2|3\rangle & -\langle 1|2+3|4\rangle & -\langle 1|2+3+4|5\rangle & 0 \\ 0 & 0 & 0 & \langle 2|3|4\rangle & \langle 2|3+4|5\rangle & \langle 2|3+4+5|6\rangle \\ * & 0 & 0 & 0 & \langle 3|4|5\rangle & \langle 3|4+5|6\rangle  \\ * & * & 0 & 0 & 0 & \langle 4|5|6\rangle \\ * & * & * & 0 & 0 & 0 \\ 0 & * & * & * & 0 & 0\end{array}\right)\,,
\ee
where the terms denoted by $*$ are related to the ones explicitly written via  symmetry, $\tilde{G}^{ji} = \tilde{G}^{ij}$. The notation 
$\langle i| l |j\rangle=\langle i|p_l|j\rangle$ uses $\lambda_i g^{il} \lambda_l=p_i$ (only $l$ summed over). To rewrite the entries of the effective matrix in terms of the momentum twistors, note that
\be
\langle i| (p_{i}+p_{i+1}+\ldots+p_{j-1} )|j\rangle=\langle i| (y_i-y_j) |j\rangle=
 Z_{i}^AZ^B_{j}\Omega_{AB}\, =: \,
\lda ij\rda
\label{doublebracket}
\ee
The constraints \reef{ZConstraint2} on the external data is $\lda i, i+1\rda=0$, so in terms of the double-bracket \reef{doublebracket}, we can write the effective metric as 
\eq\label{GDef}
\tilde{G}^{ij}=\lda ij\rda\quad {\rm for}\quad 2 \le i<j \le n \,, 
\hspace{1cm}
\tilde{G}^{1j}=-\lda 1j\rda\quad {\rm for}\quad   2 \le j \le n \,.
\eqe
This defines a non-degenerate metric only for $n>4$, since for $n=4$ the only non-trivial elements are $C_1 \lda 13\rda C_3$ and $C_2\lda 24\rda C_4$ which vanishes under the support of $\delta(C\cdot Z)$. 

While the metric is non-degenerate for $n>4$, it is not manifestly cyclic symmetric. Let us first inspect the orthogonality condition:
\be
  0 = \sum_{1 \le i , j \le n} C_{\hat{\alpha} i}C_{\hat{\beta} j} \tilde{G}^{ij}
     = \sum_{3 \le i < j \le n} C_{\hat{\alpha} i}C_{\hat{\beta} j} \lda \ij\rda
         + (\hat{\alpha} \lra \hat{\beta} ) \,.
\ee
The second equality is obtained using $C\cdot Z = 0$. Now, to see that the orthogonality constraint is indeed cyclic invariant, one uses $C\cdot Z = 0$ to show that $\sum_{4 \le i < j \le n+1} C_{\hat{\alpha} i}C_{\hat{\beta} j} \lda \ij\rda= \sum_{3 \le i < j \le n} C_{\hat{\alpha} i}C_{\hat{\beta} j} \lda \ij\rda$, with the understanding that $n + 1$ equals $1$. This suffices to prove that the condition $CG C^T = 0$ is cyclic invariant. 

Next, we can simplify the sum of $n$ cyclic copies of the orthogonality condition to find  an equivalent, manifestly cyclic invariant, form of the metric:
\eq\label{GDef1}
\left\{\begin{array}{c} G^{i,i{+}2}=\frac{k}{n}\lda i,i{+}2\rda \\ G^{i,i+3}=\frac{k{-}1}{n}\lda i,i{+}3\rda \\ \vdots \\ G^{i,i{+}k{+}1}=\frac{1}{n}\lda i,i{+}k{+}1\rda\end{array}\right.
\eqe
while $G^{ij}=0$ for all other cases.

We then have the final form for the cyclically invariant momentum twistor Grassmannian integral for 3d ABJM. It is an orthogonal Grassmannian whose metric \reef{GDef1} depends on the external data:
\eq
\label{MasterInt3}
\boxed{
 \mathcal{L}_{n;k}
=
J \times 
\delta^{3}(\mathcal{P})\,\delta^{(6)}(\mathcal{Q}) \times 
\int \frac{d^{kn}C}{GL(k)}\frac{\delta^{\frac{k(k+1)}{2}}\left( C_{\hat\alpha i} G^{ij}C_{\hat\beta j} \right)\delta^{4k|3k}(C\cdot \mathcal{Z})}{M_2 M_3 \cdots M_{k+3}}\,,}
\eqe
with $n=2k+4$ and the Jacobian $J$ given in \reef{uglyJ}. 
Note that the integral is indeed projectively invariant under rescaling 
of $Z_i\rightarrow t_i Z_i$ due to the form of the effective metric $G^{ij}$ in (\ref{GDef}). 

The momentum twistor space Grassmannian integral for ABJM theory and $\mathcal{N}=4$ SYM both have $4k$ bosonic delta functions $\delta^{4k}(C \cdot Z)$, but in addition the ABJM integral \reef{MasterInt3} has the extra orthogonal constraint. In $\mathcal{N}=4$ SYM, the $\big(k(n{-}k){-}4k\big)$ remaining degrees of freedom in the momentum twistor Grassmannian are localized by the minors. For ABJM, $k(k{+}1)/2$ of the 
$\big(k(n{-}k){-}4k\big)$ degrees of freedom are localized by the orthogonal constraint, so the dimension of the integral \reef{MasterInt3} is 
\eq
2(k+2)k-k^2-4k -\frac{1}{2}k(k+1)=\frac{k(k-1)}{2}\,,
\label{dimInt3dB}
\eqe
 the same as the dimension \reef{dimInt3dA} of the momentum space Grassmannian integral \reef{MasterInt}. In particular, we note that for $n=6$ (i.e.~$k=1$) the integral localizes completely. Because the orthogonality constraint is quadratic, there are two solution branches that the integral localizes on and we must add them to obtain the $n=6$ tree-level ABJM superamplitude. This matches the observation that there is only one BCFW-diagram for the $n=6$ ABJM amplitude, but the kinematic constraint is quadratic, so the diagram yields a two-term contribution. Those are the two terms given by the two branches of the orthogonal Grassmannian. 
 In the following, we study the orthogonality condition and evaluate the integral \reef{MasterInt3} explicitly for  $n=6$.

\subsection{The 6-point ABJM amplitude in momentum twistor space}
For $n=6$, the integral \reef{MasterInt3} becomes
\be
 \mathcal{L}_{6;1} =
 J_{234} \times \delta^{3}(\mathcal{P})\,\delta^{(6)}(\mathcal{Q}) 
  \int \frac{d^6 c}{GL(1)}
  \frac{\delta\big( c_i G^{ij} c_j \big)\delta^4
 \big( c\cdot Z \big)\delta^{(3)} \big( c\cdot \eta \big)}
 {c_2 c_{3} c_{4}} \,,
 \label{L61}
\ee
where the Jacobian is given by \reef{uglyJ} and is 
\be
   J_{234} = \frac{\<12\> \<23\>\<34\>\<45\>\<56\> \<61\> \prod_{i=1}^6 \tilde{E}_i^{3/2}}
   {\tilde{E}_3^{3/2} \tilde{E}_1^{1/2} \<12\> \<45\> \tilde{E}_5^{1/2}
     \tilde{E}_2 \<23\>^2 \<34\>^2 \tilde{E}_4} \,.
     \label{J234}
\ee
If we had picked a  representation of the original momentum space Grassmannian integral with a different product of $3$ consecutive minors in the denominator, the integral $\mathcal{L}_{6;1}$ would have a denominator $c_i c_{i+1} c_{i+2}$ and the associated Jacobian $J_{i,i+1,i+2}$ would be obtained from \reef{J234} by relabeling of the lines.

\subsubsection{Orthogonality constraint and symmetry under $i \to i+2$}
\label{s:orthog}
In the momentum space Grassmannian \reef{MasterInt}, the orthogonality condition $B'\cdot B'^T=0$ implies the following relation among the minors: 
\be
   m_{i}' m_{i+1}' =  (-1)^{\tilde{k}-1} m'_{i+\tilde{k}} m'_{i+1+\tilde{k}} \,,
   \label{minorID}
\ee
with $\tilde{k}=k+2$ and indices mod $n$. The relation \reef{minorID} is key for proving that the Grassmannian integral \reef{L61} has the appropriate cyclic invariance under $i \to i+2$.

The equivalent relation for the minors of the momentum twistor space Grassmannian will depend on the external data. Let us work out what it is for $n=6$ and how it can be used to prove that our momentum twistor  Grassmannian \reef{L61} has cyclic symmetry $i \to i+2$.

Using  $C \cdot Z=0$, direct evaluation of the orthogonality condition  gives
\bea
  \nonumber
0 &=&
c_i G^{ij} c_j 
 ~=~
\frac{1}{6}\Big( c_1 \lda 13\rda c_3+c_2\lda 24\rda c_4+c_3\lda 35\rda c_5+c_4\lda 46\rda c_6+c_5\lda 51\rda c_1 +c_6\lda 62\rda c_2 \Big) \\
  &=& \frac{1}{2} c_3 \lda 35 \rda c_5 - \frac{1}{2}  c_2 \lda 26 \rda c_6\,.
  \label{n6cGc}
\eea
Since the metric $G^{ij}$ is cyclic invariant, other forms of the constraint can be obtained from cyclic symmetry: there are three distinct ones:
\be
   c_3 \lda 35 \rda c_5 = c_2 \lda 26 \rda c_6\,,~~~~
   c_4 \lda 46 \rda c_6 = c_3 \lda 31 \rda c_1\,,~~~~ 
   c_5 \lda 51 \rda c_1 = c_4 \lda 42 \rda c_2\,.
   \label{n6orthocycl}
\ee
It follows from the first two identities in \reef{n6orthocycl} that
\be
  c_1 c_2 = \frac{\lda 35 \rda \lda 46 \rda}{\lda 26 \rda \lda 31 \rda} c_4 c_5
  = \frac{\four{3456}}{\four{6123}} c_4 c_5\,.
  \label{c1c2TOc4c5}
\ee
The second equality follows from \reef{4Dschouten}. The property \reef{c1c2TOc4c5} is the equivalent of \reef{minorID} in momentum twistor space. The other relations $c_{i} c_{i+1} \propto c_{i+3} c_{i+4}$ (indices mod 6) are obtained from cyclic relabeling of \reef{c1c2TOc4c5}.

Let us now examine the cyclic symmetry $i \to i+2$. The only part that changes in the integral of \reef{L61} is the product $c_2 c_3 c_4$, which becomes 
$c_4 c_5 c_6$. It follows from a cyclic version of \reef{c1c2TOc4c5} that 
\be
  \frac{1}{c_2 c_3 c_4} = \frac{\four{1234}}{\four{4561}}  \frac{1}{c_4 c_5 c_6}
\ee
It is not hard to see that the factor $\four{1234}/\four{4561}$ is exactly compensated by the non-trivial Jacobian: by \reef{J234} and its version with $i \to i+2$
\be
  \frac{J_{456}}{J_{234}}
  = \frac{\<12\> \tilde{E}_2 \<23\>^2 \tilde{E}_3 \<34\>}{\<45\> \tilde{E}_5 \<56\>^2 \tilde{E}_6 \<61\>}
  = \frac{\dtwo{13}\dtwo{24}}{\dtwo{46}\dtwo{51}}
  =\frac{\four{1234}}{\four{4561}}\,.
\ee
Here we use the form \reef{doublebracket} of the $Sp(4)$-product to write 
$\dtwo{i,i+2} = \<i| p_{i+1} | i+2\> = \<i,i+1\> \tilde{E}_{i+1} \<i+1,i+2\>$
and the identity \reef{4Dschouten}.  
Thus, we conclude that 
\be
   \frac{J_{234}}{c_2 c_3 c_4}
  = \frac{J_{456}}{c_4 c_5 c_6}
\ee
and that the $n=6$ integral is invariant under $i \to i+2$; the non-trivial orthogonality condition and the overall Jacobian factor nicely conspire to give this result.

\subsubsection{Evaluation of the $n=6$ integral}
To evaluate the integral \reef{L61}, we first use the bosonic delta function constraints for the six-point momentum twistor space Grassmannian, which fixes four of the integrations. As in Section \ref{s:NMHVeval}, we use the 5-term Schouten identity \reef{5Schouten} to expand the constraint $C\cdot Z=0$ on a basis of four $Z$'s, although now we find it convenient to solve for $c_2,c_3,c_5,c_6$:
\be\label{cSol}
  \begin{split}
&c_2=\frac{c_1\langle 3561\rangle+c_4\langle 3564\rangle}{\langle 2356\rangle}\,,\; ~~~~
c_3=\frac{c_1\langle 5612\rangle+c_4\langle 5642\rangle}{\langle 2356\rangle}\,,
\\[1mm]
&
c_5=\frac{c_1\langle 6123\rangle+c_4\langle 6423\rangle}{\langle 2356\rangle}\,,\; ~~~~
c_6=\frac{c_1\langle 1235\rangle+c_4\langle 4235\rangle}{\langle 2356\rangle}\,.
  \end{split}
\ee
Evaluating the delta-function this way will generate a Jacobian factor of $1/\four{2356}$. The orthogonal constraint  \reef{n6cGc} becomes 
\be
   0=~ c_i G^{ij} c_j = \frac{\dtwo{42}\four{3456}}{2\four{2356}} 
   \bigg( c_4^2 - c_1^2\frac{\dtwo{31}^2\four{5612}}{\four{1234}\four{3456}} \bigg)\,.
   \label{n6ortho}
\ee 
We fix the $GL(1)$ gauge by setting $c_1=\four{2356}$; the result is going to be independent of the gauge, but this choice will simplify the other $c_i$'s.
The solution to \reef{n6ortho} is then:
\be
c_4^\pm = \pm \frac{\dtwo{31}\four{5612}\four{2356}}{\sqrt{D}}\,,
\label{c4pm}
\ee
where
$D= \<1234\>\<3456\>\<5612\>= -\prod_{i=1}^6\lda i,i+2\rda$.

The Grassmannian integral localizes on the two solutions \reef{c4pm}:
\be
 I_i = \int \frac{\delta\big( c_i G^{ij} c_j \big)\delta^4
 \big( c\cdot Z \big)\delta^{(3)} \big( c\cdot \eta \big)}
 {c_i c_{i+1} c_{i+2}} 
  =
  \sum_{s=\pm1}
  \frac{2\four{2356}}{2c_4^s\,\dtwo{42}\four{3456}} \frac{\four{2356}}{\four{2356}} 
   \frac{ \delta^{(3)}\big( c\cdot\eta \big) }{{c_i c_{i+1} c_{i+2}}}\bigg|_{c_4 = c_4^s}\,,
   \label{Ii0}
\ee
where $i$ depends on the organization of the external states (i.e.~which 3 minors we selected at the starting point in momentum space). It is implicitly understood that on the RHS, $c_1=\four{2356}$ and $c_{2,3,5,6}$ are given by \reef{cSol} with $c_4$ as in \reef{c4pm}.
The factor 
${2\<2356\>}/{2c_4^s\,\dtwo{42}\four{3456}}$
 is the Jacobian of the orthogonality condition (see footnote \ref{footie:delta}), the second $\four{2356}$ in the numerator is the gauge-fixing Jacobian, and the $1/\<2356\>$ comes from evaluating $\delta^4(c\cdot Z)$.
 The prefactors readily simplify and we get
\be
  I_i ~=~  
   \frac{\delta^{(3)}\big( c\cdot\eta \big) }{{\sqrt{D}\, c_i c_{i+1} c_{i+2}}}\bigg|_{c_4 = c_4^+} - \frac{\delta^{(3)}\big( c\cdot\eta \big) }{{\sqrt{D}\, c_i c_{i+1} c_{i+2}}}\bigg|_{c_4 = c_4^-}\,.
   \label{n6integLoc}
\ee
It is clear from \reef{cSol} and \reef{c4pm} that the individual $c_i$'s are expressions with square roots. It may be worrisome to see such denominator terms arise from the Grassmannian integral; after all, we would expect the denominators to be a product of physical poles. However, the expectation is  warranted; for each $i$, just write
\be
  \frac{1}{c_i^\pm} = \frac{c_i^\mp}{c_i^\pm c_i^\mp } \,,
  ~~~~\text{where}~~~~
  c_i^\pm :=   c_i|_{c_4 = c_4^\pm} 
  ~~\text{for}~~
  i=2,3,5,6\,.  
\ee
The combinations $c_i^+ c_i^-$ are manifestly free of any square roots. Furthermore, after applications of the Schouten identities \reef{5Schouten} and \reef{4Dschouten}, one finds that each $c_i^+ c_i^-$  has a nice factorized form involving only 4-brackets:
\be
  \begin{array}{llll}
   c_1^+ c_1^- = \<2356\>^2\,,&
   c_2^+ c_2^- = \frac{\< 5613 \> \< 6134\>\<2356\>}{\<6234\>}
   \,,&
   c_3^+ c_3^- = \frac{\< 1245\>  \< 5612\>\<2356\>}{\<2345\>}\,,\\[4mm]
   c_4^+ c_4^- = \frac{\<1235\> \< 5612\>\<2356\>^2}{ \<2456\>  \<2345\>}
      \,,~~~&
   c_5^+ c_5^- = \frac{\<6123\> \< 6134\>\<2356\>}{\<3456\>}
      \,,~~~&
   c_6^+ c_6^- = \frac{\<1235\>\< 1245\> \<2356\>}{\<2456\>}\,.
   \end{array}
   \label{cici}
\ee

Choosing $i=2$ in \reef{n6integLoc} and using \reef{cici}, we find{\footnotesize
\begin{align}
I_2 = { \sum_{s=\pm}}
\frac{\dtwo{24}^2 
 { \delta^{(3)}\big(c^{(s)}\cdot\eta \big)} 
\big[ \dtwo{35} \big(\tfour{1236}\tfour{2456}-\tfour{5612}\tfour{2346}\big) 
 + s\sqrt{D}\big(\dtwo{26}\dtwo{35}+\dtwo{25}\dtwo{36}\big) \big]
}{\tfour{1234}\tfour{5612}^2\tfour{1245}\tfour{3461}\tfour{2356}^3}.\!\!
\label{Is}
\end{align}}%
The result \reef{Is} for the 6-point amplitude has two terms because the orthogonal Grassmannian has two branches. To compare, the BCFW calculation of the 6-point amplitude  involves only one diagram, but it gives rise to two terms, just as in \reef{Is}, because the on-shell condition for 3d BCFW is not linear in the shift parameter $z$. See Chapter 11 of \cite{Elvang:2013cua} for a review of 3d BCFW and its application in ABJM theory.  

The expression \reef{Is} is probably not the ideal form of the 6-point residue, since its expected properties are not manifest. For example, the result for the $n=6$ ABJM superamplitude should have $i \to i+2$ cyclic symmetry of the integral, as discussed in section \ref{s:orthog}; dressing \reef{Is} with the Jacobian from \reef{J234} should make it invariant under $i \to i+2$, but this is not obvious. 

Another point of concern about \reef{Is} are the apparent higher-order poles 
from the denominator-factors $\four{5612}^2$ and $\four{2356}^3$.
The former is not too worrisome: using the  identity \reef{4Dschouten} gives  $\four{5612} = - \dtwo{51} \dtwo{62}=
  - \<56\> \tilde{E}_6 \<61\>^2 \tilde{E}_1 \<12\>$.
  Numerator factors can then cancel the individual angle brackets so that there are not double poles. The triple-pole at $\four{2356} =0$ would appear to be a worse   problem   because $\four{2356} = \<23\> \<56\> y_{63}^2 \propto (p_6+p_1+p_2)^2$, which is not just a simple product of angle-brackets. However, it is not hard to show that the numerator factors conspire to cancel the extra powers in $\four{2356}$ so that at most we have a simple pole at $\four{2356} =0$. A detailed argument is given in Appendix \ref{s:pole2356}.
We leave further analysis of the momentum twistor form of the $n=6$ ABJM amplitude for future work.

\subsubsection{Singularities of the residues}

As a warm up, let us briefly review how poles of the NMHV residues could be understood as boundaries of cells in the $\mathcal{N}=4$ SYM Grassmannian.   Choose a  contour $\gamma_{abcde}$ such that the integral \reef{topC} picks up the residue where $n{-}5$ $c_i$'s with $i\ne a,b,c,d,e$ vanish. The remaining five $c_i$'s are then fixed by the four bosonic delta functions $\delta^4(c\cdot Z)$ and the $GL(1)$ scaling, as discussed in Section   \ref{s:NMHVeval};  the result is the 5-bracket $[abcde]$. Its five poles  can be described in the Grassmannian integral by forcing an extra $c_i$ to be zero (as discussed in Section \ref{s:bdr}): for example, if $c_a=0$, the delta-function $\delta^4(c\cdot Z)$ says that $Z_b$, $Z_c$, $Z_d$, and $Z_e$ are linearly dependent, and hence $\four{bcde} = 0$. This is precisely one of the five poles of the residue $[abcde]$.

Now consider the ABJM momentum twistor space Grassmannian \reef{L61} for $n=6$. Since the integral is completely fixed by the bosonic delta functions and $GL(1)$, there is no contour to choose and the result is simply two terms that are conjugate to each other, one for each of the two branches determined by the orthogonal constraint. Let us gauge-fix the $GL(1)$ by setting $c_1=1$ and then analyze the  constraints of the bosonic delta-functions when $c_2=0$. 
Via \reef{n6cGc} and \reef{n6orthocycl}, the orthogonality condition with $c_2=0$ can be written  
\be
 c_5 \lda 51 \rda  = c_4 \lda 42 \rda c_2 = 0\,.
\label{orthoNow}
\ee 
So we must have $c_5=0$, or $\lda 51 \rda=0$. Examining each in turn:
\begin{itemize}
\item $c_2=c_5=0$: using $\Omega_{AB}$ to dot $Z_3$ and $Z_4$ into the constraint  $c\cdot Z = 0$, we find
\be
  \lda 13 \rda + c_6 \lda 63 \rda = 0 
  ~~~~\text{and}~~~~
  \lda 14 \rda + c_6 \lda 64 \rda = 0\,.
\ee
For this to hold true with non-vanishing $c_6$ requires $\four{6134} = 0$. The  relations in $c\cdot Z = 0$ can then be solved to find 
\be
  c_3 = \frac{\lda 51 \rda}{\lda 35 \rda}\,,
  ~~~~~
  c_4 = \frac{\four{6123}}{\four{2346}}\,,
  ~~~~~
  c_6 = \frac{\lda 13 \rda}{\lda 36 \rda}\,.
\ee
Thus $c_2=c_5=0$ leaves the four other $c_i$ non-zero but imposes the constraint $\four{6134} = 0$ on the external data. This is precisely one of the poles in \reef{Is}.

\item $c_2=0$ and $\lda 51 \rda=0$. Dotting $Z_5$ into $c\cdot Z = 0$ now gives $c_3 \lda 35 \rda=0$. 

If $c_3 = 0$, then $c\cdot Z = 0$ gives $c_4 \lda 41 \rda=0$ and $c_4 \lda 46 \rda=0$. If also $c_4 = 0$ then we get a lower-dimensional subspace ($c_2=c_3=c_4=0$). If $c_4 \ne 0$, we must have $\lda 41 \rda = \lda 46 \rda = 0$ which in addition to $\lda 51 \rda=0$ renders multiple 4-brackets to be zero.

If instead $c_3\ne 0$, we must have  $\lda 35 \rda=0$. Consistency of $c\cdot Z = 0$ then requires $\four{6134}=0$ and this combined with $\lda 51 \rda=\lda 35 \rda=0$ puts several constraints on the external data. The remaining conditions in $c\cdot Z = 0$ do not completely fix the rest of the $c_i$'s.
\end{itemize}

We conclude from the above that the three `bounderies' $c_i = c_{i+3} = 0$ in the $n=6$ orthogonal Grassmannian integral \reef{L61} correspond to poles of the form $\four{i{+}1,i{+}2,i{+}4,i{+}5} = 0$. These are exactly the three different 3-particle poles $P^2_{i{+}2,i{+}3,i{+}4}$ of the amplitudes: for example $P_{123}^2 = y_{14}^2 \propto \four{6134}$.  One the other hand, boundaries of the form $c_{i} = c_{i+1} =0$ impose constraints among two-brackets  (e.g.~ $\lda 51 \rda=\lda 41 \rda = \lda 46 \rda =0$) which are akin to soft limits. 
Thus, as in the 4d case, we find that the cell boundaries of the Grassmannian correspond to poles in the residues.

The fact that locality is partially hidden in the orthogonal constraint was already observed in the derivation of the twistor string formula for ABJM theory~\cite{Huang:2012vt},\footnote{A twistor string whose vertex operators give the corresponding formula was later presented in~\cite{Engelund:2014sqa}.} where locality is achieved as the momentum space Grassmannian is localized onto a Veronese map~\cite{ArkaniHamed:2009dg}, $B_{\alpha, i}(a_i,b_i)=a_i^{\tilde{k}-\alpha} b_i^{\alpha-1}$. This reduces a G$(k,n)$ down to a G$(2,n)$ Grassmannian, parameterized by $(a_i, b_i)$. At 6-point, this localization was achieved by the orthogonal constraint. For higher-points, only part of the orthogonal constraint is relevant to the localization to the Veronese map. Thus we anticipate that for higher-points, the effective metric will continue to play an important role for the realization of locality.

\section{Outlook}
\label{s:out}
In this paper, we presented a detailed review of the relationship between Grassmannian integral formulas of different external data and provided a new derivation to establish their equivalence. Using the new approach, we derived the momentum twistor version of the Grassmannian integral for ABJM theory. Contrary to the momentum space representation, which is an orthogonal Grassmannian with constant metric, the momentum twistor space representation corresponds to an orthogonal Grassmannian whose metric depends on the external data. 

There are a number of interesting questions that can be tackled at this point. Recently the planar amplitudes of $\mathcal{N}=4$ SYM have 
been identified as a single geometric object, the amplituhedron~\cite{Arkani-Hamed:2013jha,Arkani-Hamed:2013kca}. It is defined in the Grassmannian G($k$,$k$+4) via
\begin{equation}\label{Yspace}
Y^I_{\alpha }=C_{+,\alpha i}Z^I_{+i}\,, 
\end{equation}
where $C_{+,\alpha i}$ are cells in the positive Grassmannian\footnote{The positive Grassmannian, or non-negative Grassmannian to be precise, refers to the property that the $k \times n$ matrices are real-valued with all minors are greater or equal to zero.} G($k,n$) and $Z^{I}_{+,i}$ are $(k+4)$-component 
 vectors, $i=1,\ldots,n$,  built linearly from the momentum supertwistors (see~\cite{Arkani-Hamed:2013jha} for details).
The array of $n$ vectors $Z^{I}_{+,i}$ are viewed as elements in the positive Grassmannian  G($k{+}4,n$). The amplituhedron is then the ``volume-form" in this space, and it has logarithmic singularities at the boundaries of $Y$. 
It would be very appealing to derive this definition from the momentum twistor space Grassmannian integral. As a first step, one should be able to prove that the BCFW terms in momentum twistor space are associated with dimension $4k$ positive cells in G($k$,$n$). In principle, this is accomplished by the momentum twistor space on-shell diagrams introduced by He and Bai~\cite{HeBai}, where the individual cells are associated with diagrams that are again iteratively built from the fundamental 3-point vertices. On the other hand, from our analysis one might expect the existence of a straightforward map from cells in the momentum space Grassmannian to cells in the momentum twistor space Grassmannian. However as the minors of the two Grassmannians are related by a multiplicative string of spinor brackets, a priori it is not clear that positivity in one Grassmannian can be related to that of the other, even for the top-cell. Thus we see that positivity of the BCFW terms in momentum twistor space is non-trivial result, and should warrant further investigation.

Given that we have derived the momentum twistor space Grassmannian 
for ABJM theory, one can ask if there exists a geometric entity like the amplituhedron for ABJM? Supporting evidence for its existence includes the realization that BCFW recursion relations for the theory exists both at tree-  
and at loop-level, and when represented in terms of on-shell diagrams, stratifies the positive orthogonal Grassmannian. Unlike $\mathcal{N}=4$ SYM, where positivity ensures locality, we have already seen that the orthogonal condition plays an important role as well. What is unclear is whether or not the condition is to be viewed as a condition on the cells, or on the space $Y$ where the amplituhedron lives, perhaps both. Another interesting question is if there exist  on-shell diagrams for cells in ABJM momentum twistor space Grassmannian such as those found for $\mathcal{N}=4$ SYM~\cite{HeBai}. 
One of the remarkable results in the on-shell diagram approach for cells of the momentum space Grassmannian, is that the gluing and merging of diagrams preserves orthogonality~\cite{ArkaniHamed:2012nw,HW}. In momentum twistor space, the orthogonality is now defined with a momentum twistor dependent metric. It would be very interesting if an iterative way of constructing cells exists such that the orthogonality property of the smaller cells ensures that of the higher-dimenions ones.   

In this paper, we also studied  
residue theorems for the NMHV level in momentum twistor space and showed that an abstract homological point of view offered a clear geometric description. For amplitudes beyond NMHV, the combinatorics and geometry become much more difficult, and it is no longer the case that every momentum twistor cell of the appropriate dimension ($d=4k$) has support for generic external data. Composite singularities become the norm, even in momentum twistor space, so the residue calculation outlined in Appendix \ref{app:residues} cannot be applied directly. Moreover, whereas the entries in the $k=1$ matrix \eqref{topC} can be interpreted as homogeneous coordinates on $\CP^{n-1}$, higher $k$ Grassmannians do not have such simple geometric structure. Even the locations of poles become more complicated for $k>1$ due to the non-linear dependence of the minors on the matrix entries. Instead of cutting out hyperplanes as in Section \ref{s:resrels}, the minors vanish on generally complicated surfaces. Understanding the geometric and homological structure of such spaces in a general sense remains a subject of active research in the mathematics community.

Despite the inherent mathematical challenges, the BCFW bridge decompositions of \cite{ArkaniHamed:2012nw} suggest a possible route forward. The technique provides a robust way to generate coordinates on any cell of G$(k,n)$ with the useful property that all singularities of the integration measure are manifestly of the form $d\alpha/\alpha$ (similar to the $dc/c$ structure of \eqref{NMHVlink}). This eliminates the issue of composite residues at the cost of requiring multiple charts to cover all singularities.\footnote{Until recently it was not known how to compare the orientations of those charts, but this has since been resolved \cite{OlsonInprep}.} The existence of such a convenient representation of the Grassmannian integral offers compelling motivation to pursue higher $k$ generalizations, and it will almost certainly lead to further insights regarding residues, residue theorems, superamplitudes, and locality constraints for all $k$ and $n$.

\section*{Acknowledgements}
We would like to thank
Nima Arkani-Hamed, Jake Bourjaily, Freddy Cachazo, and Jaroslav Trnka
for useful and insightful discussions.  Y.-t.H.~would also like to think Dongmin Gang, Eunkyung Koh, Sangmin Lee, and Arthur E.~Lipstein for early collaborations on related subjects. 

H.E.~is supported in part  by NSF CAREER Grant PHY-0953232.
C.K.~is supported in part by the US Department of Energy
under grant DE-FG02-95ER40899 and she would also like to thank ASU for hospitality during the final stages of this work.
T.L.~is supported by NSF grant DMS-1160726.
T.M.O.~is supported by NSF Graduate Research Fellowship under Grant \#F031543. S.B.R is supported in part by US Department of Energy under grant DE-SC0011719.

\appendix

\section{Matrix $Q$ Details}
\label{app:Q}
\subsection{Derivation of $\det Q$}
\label{app:detQ}
Let $Q^{(0)}_{ij}$ be the full (degenerate) $n \times n$  transformation matrix from $B$ to $C$ variables, i.e.~$C_{\^\a j} = B_{\^a i}Q^{(0)}_{ij}$. It can be obtained from \eqref{DfromC} as
\begin{align}
Q^{(0)}_{ij} = \frac{\partial C_{\^\a i}}{\partial B_{\^\a j}}\,.
\end{align}
Define $Q^{(m)}_{ij}$ to be the  $(n-m)\times (n-m)$ matrix  obtained from $Q^{(0)}_{ij}$ by deleting the first $m$ rows and $m$ columns. For example, the matrix needed in \reef{detQresult} is obtained by deleting the first two rows and columns:
\begin{align}
Q = Q^{(2)} = 
\left( \begin{array}{cccccc}
\frac{\two{42}}{\two{23}\two{34}} & \frac{1}{\two{34}} & 0 & 0 & \cdots & 0 \\
\frac{1}{\two{34}} & \frac{\two{53}}{\two{34}\two{45}} & \frac{1}{\two{45}} & 0 & \cdots & 0 \\
0 & \frac{1}{\two{45}} & \ddots & \ddots & \ddots& \vdots\\
0 & 0 & \ddots &&& 0 \\
\vdots & \vdots& 0 &\frac{1}{\two{n-2,n-1}}& \frac{\two{n,n-2}}{\two{n-2,n-1}\two{n-1,n}} & \frac{1}{\two{n-1,n}}\\
0 & \cdots & & 0 &  \frac{1}{\two{n-1,n}} & \frac{\two{1,n-1}}{\two{n-1,n}\two{n1}}
\end{array}
\right)\,.
\label{Qmatrix}
\end{align}
With the definition $Q^{(n)}=1$, we will prove the following claim by induction:
\be 
  \boxed{~~
  \textbf{{Claim}:}~~~~~
  \text{For}~~i>0\,, ~~~~ 
  \det Q^{(n-i)} = (-1)^{i-1} \frac{\two{1, n-i}}{\two{n-i,n-i+1}\cdots \two{n1}}\,.~~
  \label{Qclaim}}
\ee
\textbf{{Proof}:}~ For $i=1$, we trivially have 
\begin{align}
Q^{(n-1)} = \left( \begin{array}{c}
\frac{\two{1,n-1}}{\two{n-1,n}\two{n1}}
\end{array}\right)
~~~~\implies~~~~
\det Q^{(n-1)} = \frac{\two{1,n-1}}{\two{n-1,n}\two{n1}} \,,
\end{align}
i.e.~the determinant satisfies the claim \reef{Qclaim}.  

For $i=2$, the determinant of 
\begin{align}
Q^{(n-2)} = \left( \begin{array}{cc}
\frac{\two{n,n-2}}{\two{n-2,n-1}\two{n-1,n}} & \frac{1}{\two{n-1,n}}\\
 \frac{1}{\two{n-1,n}} & \frac{\two{1,n-1}}{\two{n-1,n}\two{n1}}
\end{array}\right)\,.
\end{align}
 is calculated easily with a single application of the Schouten identity, and the answer is
\begin{align}
\det Q^{(n-2)}=-\frac{\two{1,n-2}}{\two{n-2,n-1}\two{n-1,n}\two{n1}}\,.
\end{align}
The result satisfies the claim. This establishes the base of the induction.

For the inductive argument, assume that the claim \reef{Qclaim} is satisfied for all $i<m<n$. We will prove it for $i=m$. The matrix $Q^{(n-m)}$ is of the form
\be
   Q^{(n-m)} ~=
  \raisebox{-3cm}{\includegraphics[width=7cm]{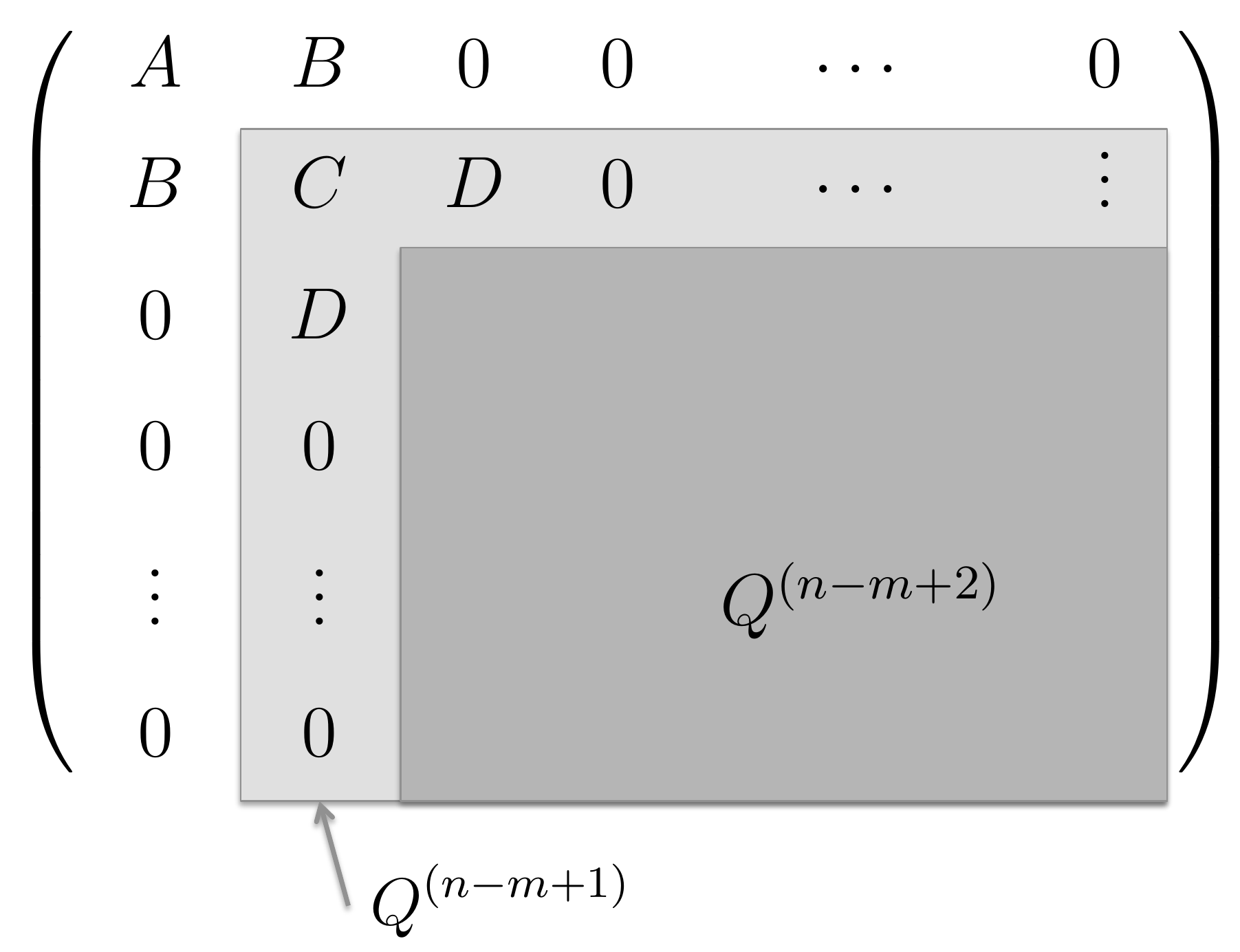}}
  \label{bigQ}
\ee
with 
\be
  A=\tfrac{\two{n-m+2,n-m}}{\two{n-m,n-m+1}\two{n-m+1,n-m+2}}\,,~~~~
  B=\tfrac{1}{\two{n-m+1,n-m+2}} \,~~~~\text{etc}.\\
\ee
To evaluate the determinant, we expand on the first row of \reef{bigQ}, and subsequently on the second row, to find
\be
  \det Q^{(n-m)} = A \det Q^{(n-m+1)} - B^2 \det Q^{(n-m+2)}\,.
\ee

By the inductive hypothesis, we may replace the two determinants on the right-hand side and obtain:
\begin{align}
\begin{split}
\det Q^{(n-m)} &=(-1)^{m}\bigg[\tfrac{\two{n-m+2,n-m}}{\two{n-m,n-m+1}\two{n-m+1,n-m+2}} \tfrac{\two{1, n-m+1}}{\two{n-m+1,n-m+2}\cdots \two{n1}} \\
&\hspace{2cm} + \tfrac{1}{\two{n-m+1,n-m+2}^2}\tfrac{\two{1, n-m+2}}{\two{n-m+2,n-m+3} \cdots \two{n1}} \bigg]
\\
&=\tfrac{(-1)^m}{\two{n-m+1,n-m+2}}\tfrac{\two{1, n-m+1} \two{n-m+2,n-m}+ \two{1, n-m+2}\two{n-m,n-m+1}}
{\two{n-m,n-m+1}\cdots\two{n1}}\\
&=(-1)^{m-1}\frac{\two{1, n-m}}
{\two{n-m,n-m+1}\cdots\two{n1}}\,.
\end{split}
\end{align}
In the last line we used the Schouten identity. Thus by induction we have proven the claim \reef{Qclaim}.
\hfill $\square$

The determinant given in \reef{detQresult} is simply the result for $i=n-2$, so
\begin{align}
|\det Q^{(2)}| = \frac{\two{12}}{\two{23}\cdots\two{n1}} =\frac{\two{12}^2}{\two{12}\cdots\two{n1}}\,.
\end{align}

\subsection{Verification of $Q^{-1}$}
\label{app:Qinv}
\noindent The square, symmetric matrix $Q$ is given by (for $\hat{i},\hat{p}\in[3,n]$)
\begin{align}
Q_{\hat{i}\hat{p}} = \left\{
\begin{array}{cl}
\frac{\ang{\hat{i}+1,\hat{i}-1}}{\ang{\hat{i}-1,\hat{i}}\ang{\hat{i},\hat{i}+1}} \,,& \hat{p}=\hat{i} \\
\frac{1}{\ang{\hat{i},\hat{i}+1}}\,, & \hat{p}=\hat{i}+1 \\
\frac{1}{\ang{\hat{i}-1,\hat{i}}}\,, &\hat{p}=\hat{i}-1\\
0\,, & \text{else.}
\end{array}\right.
\label{Q}
\end{align}
The determinant is non-vanishing as shown above, so it is invertible. We claim that its inverse is given by (for $\hat{p},\hat{j}\in[3,n]$)
\begin{align}
Q\inv_{\hat{p}\hat{j}} = \left\{
\begin{array}{cl}
-\frac{\ang{1\hat{j}}\ang{2\hat{p}}}{\ang{12}}\,, & \hat{j} \geq p \\
-\frac{\ang{1\hat{p}}\ang{2\hat{j}}}{\ang{12}}\,, & \hat{j} \leq p\,.
\end{array}\right.
\label{Qinv}
\end{align}

\noindent It is straightforward to verify the claim by a direct computation of $Q_{\hat{i}\hat{p}}Q\inv_{\hat{p}\hat{j}}$. The sum over $\hat{p}$ produces three terms due to the three non-zero entries in each row of \eqref{Q}.
\begin{align}
\begin{split}
Q_{\hat{i}\hat{p}}Q\inv_{\hat{p}\hat{j}} & = 
\frac{1}{\ang{\hat{i}-1,\hat{i}}} 
\left\{\begin{array}{cl}
-\frac{\ang{1\hat{j}}\ang{2,\hat{i}-1}}{\ang{12}}\,, & \hat{j} \geq \hat{i} \\
-\frac{\ang{1,\hat{i}-1}\ang{2\hat{j}}}{\ang{12}}\,, & \hat{j} < \hat{i}
\end{array}\right\}_{\hat{p}=\hat{i}-1}
\\ &\quad +
\frac{\ang{\hat{i}+1,\hat{i}-1}}{\ang{\hat{i}-1,\hat{i}}\ang{\hat{i},\hat{i}+1}}
\left\{\begin{array}{cl}
-\frac{\ang{1\hat{j}}\ang{2,\hat{i}}}{\ang{12}}\,, & \hat{j} \geq \hat{i} \\
-\frac{\ang{1,\hat{i}}\ang{2\hat{j}}}{\ang{12}}\,, & \hat{j} < \hat{i}
\end{array}\right\}_{\hat{p}=\hat{i}}
\\ &\quad +
\frac{1}{\ang{\hat{i},\hat{i}+1}} 
\left\{\begin{array}{cl}
-\frac{\ang{1\hat{j}}\ang{2,\hat{i}+1}}{\ang{12}}\,, & \hat{j} > \hat{i} \\
-\frac{\ang{1,\hat{i}+1}\ang{2\hat{j}}}{\ang{12}}\,, & \hat{j} \leq \hat{i}
\end{array}\right\}_{\hat{p}=\hat{i}+1}\,.
\end{split}
\label{prod}
\end{align}
Examining the limits on $\hat{j}$ in each term, we see that there are three cases that should be considered: $\hat{j}<\hat{i},\hat{j}=\hat{i},\hat{j}>\hat{i}$. We begin with $\hat{j}<\hat{i}$. Taking the relevant pieces from each of the three terms in \eqref{prod}, we find
\begin{align}
\nonumber
Q_{\hat{i}\hat{p}}Q\inv_{\hat{p}\hat{j}}|_{\hat{j}<\hat{i}} &= -\frac{\ang{2\hat{j}}\ang{1,\hat{i}-1}\ang{\hat{i},\hat{i}+1} + \ang{2\hat{j}}\ang{1\hat{i}}\ang{\hat{i}+1,\hat{i}-1}+\ang{2\hat{j}}\ang{1,\hat{i}+1}\ang{\hat{i}-1,\hat{i}}}{\ang{12}\ang{\hat{i}-1,\hat{i}}\ang{\hat{i},\hat{i}+1}} \\
&= -\frac{-\ang{2\hat{j}}\ang{1,\hat{i}+1}\ang{\hat{i}-1,\hat{i}} + \ang{2\hat{j}}\ang{1,\hat{i}+1}\ang{\hat{i}-1,\hat{i}}}{\ang{12}\ang{\hat{i}-1,\hat{i}}\ang{\hat{i},\hat{i}+1}} = 0\,,
\label{QQinv}
\end{align}
where we used the Schouten identity to combine the first two terms. Note that when $\hat{i}=n$, the third term in \eqref{QQinv} would not be present in the sum as there is no $\hat{p}=n+1$. This does not lead to any inconsistencies because $\ang{1,n+1}\equiv\ang{1,1}=0$, so the term vanishes anyway. A similar computation shows that the product also vanishes for $\hat{j}>\hat{i}$  (here the first $\hat{i}=3$ term vanishes as needed since there is no $\hat{p}=2$).

\noindent Thus we are left to consider the case when $\hat{j}=\hat{i}$. Replacing all the $\hat{j}$'s with $\hat{i}$'s, we find
\bea
\nonumber
Q_{\hat{i}\hat{p}}Q\inv_{\hat{p}\hat{j}}|_{\hat{j}=\hat{i}} &=& -\frac{\ang{1\hat{i}}\ang{2,\hat{i}-1}\ang{\hat{i},\hat{i}+1} + \ang{1\hat{i}}\ang{2\hat{i}}\ang{\hat{i}+1,\hat{i}-1}+\ang{2\hat{i}}\ang{1,\hat{i}+1}\ang{\hat{i}-1,\hat{i}}}{\ang{12}\ang{\hat{i}-1,\hat{i}}\ang{\hat{i},\hat{i}+1}} \\
&=& -\frac{-\ang{1\hat{i}}\ang{2,\hat{i}+1}\ang{\hat{i}-1,1}+\ang{2\hat{i}}\ang{1,\hat{i}+1}\ang{\hat{i}-1,\hat{i}}}{\ang{12}\ang{\hat{i}-1,\hat{i}}\ang{\hat{i},\hat{i}+1}} = 1\,,
\eea
where we have used the Schouten identity once to combine the first two terms, and again to combine the remaining terms (as above, the respective $\hat{i}=3,n$ terms cause no issues). Hence $Q_{\hat{i}\hat{p}}Q\inv_{\hat{p}\hat{j}} = \delta_{\hat{i}\hat{j}}$, so $Q\inv$ is the inverse of $Q$ as claimed.

\section{Calculations of higher-dimensional residues}
\label{app:residues}
We present here a more detailed evaluation of the NMHV residues of the momentum twistor  Grassmannian integral \reef{GrassZ}.

The starting point is the integral \reef{NMHVlink} with contour $\gamma_{abcde}$, as defined in Section \ref{s:NMHVeval}. To begin with, fix the $GL(1)$ redundancy by setting 
\begin{align}
\label{fixca}
c_a= c_a^{(0)} \ne 0\,.
\end{align}
We can use a delta-function $\delta(c_a -c_a^{(0)})$ to enforce this choice. It comes with a Jacobian factor $c_a^{(0)}$ that compensates the little-group scaling.

Next we turn our attention to the bosonic delta functions $\delta^4(c_i Z_i)$. They enforce the condition $\sum_{i=1}^n c_i\,Z_i = 0$. Since the $Z_i$'s are 4-component vectors, we can use Cramer's rule \reef{5Schouten} to write all $Z_j$ for $j\not\in \{b,c,d,e\}$ in terms of $Z_b$, $Z_c$, $Z_d$, and $Z_e$:
\begin{align}
\four{bcde}Z_j = -\Big(\four{cdej}Z_b + \four{dejb}Z_c + \four{ejbc}Z_d + \four{jbcd}Z_e\Big)\,.
\label{5SchoutenApp}
\end{align}
For generic external data, $Z_b$, $Z_c$, $Z_d$, and $Z_e$ are linearly independent, so for $\sum_{i=1}^n c_i\,Z_i = 0$ to hold, the coefficients of each must vanish. This gives four constraints
\begin{align}
\begin{split}
  c_b = \sum\limits_{j\neq b,c,d,e}c_j\frac{\four{cdej}}{\four{bcde}}
   \,=:\, 
   c_b^{(0)}\,, 
   &\qquad~~~ c_c  = \sum\limits_{j\neq b,c,d,e}c_j\frac{\four{dejb}}{\four{bcde}}
   \,=:\, c_c^{(0)}\,, \\
   c_d = \sum\limits_{j\neq b,c,d,e}c_j\frac{\four{ejbc}}{\four{bcde}}
   \,=:\, c_d^{(0)}\,, 
   &\qquad~~~ c_e  = \sum\limits_{j\neq b,c,d,e}c_j\frac{\four{jbcd}}{\four{bcde}}
   \,=:\, c_e^{(0)}\,.
\end{split}
\label{fixedcs}
\end{align}
Thus we can write
\be
  \delta^4\big(c_i Z_i\big) 
  =
  \frac{1}{\<bcde\>}\prod\limits_{i=b,c,d,e}\delta \big(c_i - c_i^{(0)}\big) \,.
  \label{deltarewrite}
\ee

Now, enforcing a delta function $\delta(z-z_0)$ can also be done by 
re-interpreting the integral as a contour integral $\oint \frac{dz}{z-z_0}$ with a contour that surrounds only $z_0$. We can therefore write the NMHV Grassmannian integral as
\begin{align}
   \mathcal{I}_{n;1}^{\gamma_{abcde}}(\mathcal{Z}) 
   =    
   \oint_{\Gamma_{abcde}} \frac{d^{1\times n}C}{\prod\limits_{i=a,b,c,d,e} \big(c_i - c_i^{(0)}\big) \prod\limits_{j\ne a,b,c,d,e}  c_j} \times
     \frac{\delta^{(4)}\big(c_i \eta_i\big)}{c_a\, c_b\, c_c\, c_d\, c_e}\,,
\label{NMHVint}
\end{align}
where $\Gamma_{abcde}$ 
is the $n$-dimensional contour that encircles each  $c_{a,b,c,d,e}^{(0)}$ as well as $c_j =0$ for each $j \ne a,b,c,d,e$. This contour extends the $(n-5)$-dimensional contour $\gamma_{abcde}$ to encircle the delta function singularities in the other five dimensions.

It is straightforward to evaluate the contour integral \reef{NMHVint} and obtain the residue. By the choice of contour $\Gamma_{abcde}$, all $c_i$'s vanish on the pole except $c_{a,b,c,d,e}$, so we find from \eqref{fixca} and \reef{fixedcs} that
\begin{align}
c_a\, c_b\, c_c\, c_d\, c_e \to \bigg(\frac{c_a^{(0)}}{\four{bcde}}\bigg)^5 \four{abcd} \four{bcde} \four{cdea} \four{deab} \four{eabc}\,.
\end{align}
Including all the Jacobian factors, the residue is therefore:
\begin{align}
  \frac{c_a^{(0)}}{\four{bcde}} \frac{\delta^{(4)}\bigg( \big( c_a^{(0)}\Big/\four{bcde}\big) \Big(\four{bcde}{\eta}_a + \four{cdea}{\eta}_b+\four{deab}{\eta}_c+\four{eabc}{\eta}_d + \four{abcd}{\eta}_e \Big)\bigg)}{\Big( c_a^{(0)}\Big/\four{bcde}\Big)^5\four{bcde}
\four{cdea}
\four{deab}
\four{eabc}
\four{abcd} }\,.
\end{align}
We can pull out the multiplicative factor from the fermionic delta function, which combines with the Jacobians to exactly cancel the extra factor in the denominator. Hence the residue is simply the result given in \reef{resresult}: 
$\mathcal{I}_{n;1}^{\gamma_{abcde}} = \five{abcde}$. Note that this is  independent of the gauge choice $c_a^{(0)}$. The expression \reef{resresult} is manifestly antisymmetric in the five labels $a,b,c,d,e$. 

We can label the residue by the $n-5$ values  $i_1,i_2,\ldots,i_{n-5} \ne a,b,c,d,e$ as $\big\{ i_1,i_2,\ldots,i_{n-5} \big\}$. This label is fully anti-symmetric and is related to the five-bracket via 
\begin{align}
\mathcal{I}_{n;1}^{\gamma_{abcde}} 
=
 \frac{1}{(n-5)!}\, \varepsilon^{ a\,b\,c\,d\,e\,i_1\, i_2\,\ldots\, i_{n-5}}\
\{i_1,i_2,\ldots,i_{n-5}\}
= \five{abcde}  
\label{5resApp} \,.
\end{align}
To see this, we recall the following results from the calculus of higher-dimensional contour integrals.\footnote{The derivation given here is similar to the example given in Section 2.3 of \cite{Mason:2009qx}; see also Section 5.1 of \cite{ArkaniHamed:2009dn}. For a mathematical reference, see Chapter 5 of \cite{Griffiths}.}  

Suppose we have a set of $m$ functions $f_i(x)$ of $m$ complex variables $x=x_1\,\ldots,x_m$, which have an isolated common zero at the origin $f_i(0)=0$ and are holomorphic in a neighborhood of a ball around the origin. Let $g(x)$ be holomorphic in the same region and non-vanishing at $x=0$. Then we have a meromorphic $m$-form
\begin{align}
\omega = \frac{g(x) dx^1\wedge \ldots\wedge dx^m}{f_1(x)\ldots f_m(x)}
\label{omega}
\end{align}
which has a simple pole at the origin. We define the residue:
\begin{align}
\text{Res}_{x=0} = (2\pi\,i)^{-m}\,\int_\Gamma \omega = \frac{g(0)}{\mathcal{J}_f(0)} \,,
\label{res}
\end{align}
where $\mathcal{J}_f(x)$ is the Jacobian determinant for the functions $f$
\begin{align}
\mathcal{J}_f(x) = \frac{\partial(f_1,\ldots,f_m)}{\partial(x_1,\ldots,x_m)} \,.
\label{Jf}
\end{align}
The contour $\Gamma$ is given by
\begin{align}
\Gamma = \{x:\, |f_i(x)| = \epsilon_i\}
\end{align}
and oriented such that
\begin{align}
d(\text{arg}f_1)\wedge \ldots\wedge d(\text{arg}f_m)
\label{contourorient}
\end{align}
is positively oriented with respect to the volume form in \eqref{omega}. In other words, any sign that would be produced by \eqref{contourorient} is compensated by reversing one of the circles in $\Gamma$. Therefore, the only signs can come from the Jacobian \eqref{Jf}.

Now we are set to compute residues of the form \eqref{res}. From \eqref{NMHVint}, we have \begin{align}
g(c)=(2\pi\,i)^{-(n-5)}~\frac{\delta^{(4)}\big(c_i {\eta}_i)}{c_a\,c_b\,c_c\,c_d\,c_e}\,,
\end{align}
where we have absorbed some numerical factors in the normalization for future simplicity. As a convention, we assign the first five functions $f_1(c),\ldots,f_5(c)$ to the delta function singularities, i.e.
\begin{align}
\begin{split}
f_1 &= c_a-c_a^{(0)}\,,\quad f_2 = c_b - c_b^{(0)}\,,\quad f_3 = c_c - c_c^{(0)}\,,\quad f_4 = c_d - c_d^{(0)}\,,\quad f_5 = c_e - c_e^{(0)}\,.
\end{split}
\end{align}
The rest of the $f_i$'s are assigned to the $n-5$ $c_j$'s that vanish at the location of the pole encircled by $\Gamma$ such that $f_i=0$ corresponds to the vanishing of the $j^\tth$ minor of $C$.
 
The contour is oriented such that we first take $f_1\to0$, then $f_2\to0$, etc, which we can use to define a labeling of the residues by the indices of the $c_i$'s; for example, we can assign them in increasing order
\begin{align}
f_6=c_1\,,f_7=c_2\,,f_8=c_3\,,\ldots \to \text{Res} = \{1,2,3,\ldots\}\,,
\end{align}
or with a different ordering $\tau$ on the labels $i\neq a,b,c,d,e$,
\begin{align}
\begin{split}
f_6=c_{\tau(1)}\,,f_7=c_{\tau(2)}\,,f_8=c_{\tau(3)}\,,\ldots \to \text{Res} &= \{\tau(1),\tau(2),\tau(3),\ldots\}
\\ &=\text{sgn}(\tau)\times \{1,2,3,\ldots\}\,,
\end{split}
\end{align}
where $\text{sgn}(\tau)$ is the signature of the permutation of the $(n-5)$ indices $i\neq a,b,c,d,e$.
The antisymmetry of the Jacobian $\mathcal{J}_f$ in \eqref{res} implies that the residue labels are antisymmetric in their indices. Similarly, the residue is antisymmetric in the labels $a,b,c,d,e$. It follows that the result is \reef{5resApp}.

\section{Pole structure of 6-point ABJM amplitude in momentum twistor space}
\label{s:pole2356}
The result \reef{Is} appears to have a triple-pole at $\four{2356}=0$. We show here that it is actually no worse than at most a simple pole.  

We first note that it follows directly from \reef{4Dschouten} that
\be
  \four{2356}=0    
  ~~~~\lra~~~~
  \dtwo{25}\dtwo{63} + \dtwo{26}\dtwo{35} = 0 \,,
  \label{star1}
\ee
and employing the Schouten identity we also have
\be
\small
  \begin{split}
  & 0 =   \dtwo{14} \four{2356}
   = -\dtwo{13} \four{5642}  - \dtwo{15}\four{6423} 
= \dtwo{13} \dtwo{52}\dtwo{64} + \dtwo{15} \dtwo{63} \dtwo{42}\,.
  \end{split}
  \label{star2}
\ee
Thus, by \reef{star1} and \reef{star2}, the $D$ that appears in the localization of the $c_i$'s become a perfect square 
\be
  D = - \dtwo{13} \dtwo{24} \dtwo{35} \dtwo{46} \dtwo{51} \dtwo{62}
  ~\xrightarrow{\< 2356\> \to 0}~
  \dtwo{13}^2 \dtwo{52}^2 \dtwo{46}^2 \,.
  \label{Dsquare}
\ee 
Now, we examine how each $c_i^\pm$ behaves in the limit  $\four{2356} = \eps \to 0$. With our gauge choice $c_1 = \four{2356}$ it is clear that $c_1^\pm = O(\eps)$ and by \reef{c4pm} we also have  $c_4^\pm = O(\eps)$. Now, the four other $c_i$'s are given by \reef{cSol} and they may appear to be finite as $\eps \to 0$; however, an extra cancellation can occur between the two terms in the numerator. To see this, consider the example of $c_3^{\pm}$:
\be
  c_3^{\pm} = \frac{\four{5612}}{\sqrt{D}} \Big( \sqrt{D} \mp   \dtwo{13} \dtwo{52}\dtwo{64} \Big)
  ~\xrightarrow{\< 2356\> = \eps \to 0}~
   \left\{ 
  \begin{array}{ll}
     O(\eps)\\
     O(1)
  \end{array}
  \right.
  \,,
\ee
where the limit follows from \reef{Dsquare} and the outcome, $O(\eps)$ and $O(1)$, depends on the relative sign $\pm$ between the two terms and the sign of $\dtwo{13} \dtwo{52}\dtwo{64}$. This will be the same for $c_2$, $c_5$, and $c_6$, and it can easily be demonstrated, using \reef{star1} and \reef{star2}, that in the limit $\< 2356\> = \eps \to 0$, we will either have  
\be
  c_2^+, c_3^+, c_5^+, c_6^+ = O(\eps) 
  ~~~~\text{and}~~~~
      c_2^-, c_3^-, c_5^-, c_6^- = O(1) 
   \label{case1}
\ee
or vice versa. Suppose $\dtwo{13} \dtwo{52}\dtwo{64}$ is such that we have the case \reef{case1}: then
\be
  \frac{\delta^{(3)}\big(c^+ .\eta) }{c_2^+c_3^+c_4^+} 
    ~\xrightarrow{\< 2356\> = \eps \to 0}~
    \frac{O(\eps^3)}{O(\eps^3)} \sim O(1)\,,
\ee
while
\be
  \frac{\delta^{(3)}\big(c^- .\eta) }{c_2^- c_3^- c_4^-} 
    ~\xrightarrow{\< 2356\> = \eps \to 0}~
    \frac{
    O(\eps^3)~~\text{or}~~
    O(\eps^2)~~\text{or}~~
    O(\eps^1)~~\text{or}~~
    O(1)    
    }{O(\eps)} \,,
\ee
i.e.~it is no worse than $O(1/\eps)$, which signifies the (expected) simple pole. 
Thus we have shown that despite the apparent $1/\eps^3$ pole in the limit $\< 2356\> = \eps \to 0$ of the $n=6$ result \reef{Is}, there is at most a simple pole. 



\begin{thebibliography}{99}


\bibitem{ArkaniHamed:2009dn} 
  N.~Arkani-Hamed, F.~Cachazo, C.~Cheung and J.~Kaplan,
  ``A Duality For The S Matrix,''
  JHEP {\bf 1003}, 020 (2010)
  [arXiv:0907.5418 [hep-th]].

\bibitem{Mason:2009qx} 
  L.~J.~Mason and D.~Skinner,
  ``Dual Superconformal Invariance, Momentum Twistors and Grassmannians,''
  JHEP {\bf 0911}, 045 (2009)
  [arXiv:0909.0250 [hep-th]].

\bibitem{ArkaniHamed:2009vw} 
  N.~Arkani-Hamed, F.~Cachazo and C.~Cheung,
  ``The Grassmannian Origin Of Dual Superconformal Invariance,''
  JHEP {\bf 1003}, 036 (2010)
  [arXiv:0909.0483 [hep-th]].


\bibitem{ArkaniHamed:2012nw} 
  N.~Arkani-Hamed, J.~L.~Bourjaily, F.~Cachazo, A.~B.~Goncharov, A.~Postnikov and J.~Trnka,
  ``Scattering Amplitudes and the Positive Grassmannian,''
  arXiv:1212.5605 [hep-th].

\bibitem{Arkani-Hamed:2013jha} 
  N.~Arkani-Hamed and J.~Trnka,
  ``The Amplituhedron,''
  arXiv:1312.2007 [hep-th].

\bibitem{Arkani-Hamed:2013kca} 
  N.~Arkani-Hamed and J.~Trnka,
  ``Into the Amplituhedron,''
  arXiv:1312.7878 [hep-th].

\bibitem{ArkaniHamed:2010gg} 
  N.~Arkani-Hamed, J.~L.~Bourjaily, F.~Cachazo, A.~Hodges and J.~Trnka,
  ``A Note on Polytopes for Scattering Amplitudes,''
  JHEP {\bf 1204}, 081 (2012)
  [arXiv:1012.6030 [hep-th]].

\bibitem{WestCSM} 
  O.~Aharony, O.~Bergman, D.~L.~Jafferis and J.~Maldacena,
  ``N=6 superconformal Chern-Simons-matter theories, M2-branes and their gravity duals,''
  JHEP {\bf 0810}, 091 (2008)
  [arXiv:0806.1218 [hep-th]].

\bibitem{EastCSM} 
  K.~Hosomichi, K.~-M.~Lee, S.~Lee, S.~Lee and J.~Park,
  ``N=5,6 Superconformal Chern-Simons Theories and M2-branes on Orbifolds,''
  JHEP {\bf 0809}, 002 (2008)
  [arXiv:0806.4977 [hep-th]].
  
\bibitem{Drummond:2008vq} 
  J.~M.~Drummond, J.~Henn, G.~P.~Korchemsky and E.~Sokatchev,
  ``Dual superconformal symmetry of scattering amplitudes in N=4 super-Yang-Mills theory,''
  Nucl.\ Phys.\ B {\bf 828}, 317 (2010)
  [arXiv:0807.1095 [hep-th]].

\bibitem{LeeOG} 
  S.~Lee,
 ``Yangian Invariant Scattering Amplitudes in Supersymmetric Chern-Simons Theory,''
  Phys.\ Rev.\ Lett.\  {\bf 105}, 151603 (2010)
  [arXiv:1007.4772 [hep-th]].

 \bibitem{HW} 
  Y.~-T.~Huang and C.~Wen,
  ``ABJM amplitudes and the positive orthogonal grassmannian,''
  JHEP {\bf 1402}, 104 (2014)
  [arXiv:1309.3252 [hep-th]].

\bibitem{HWX}
  Y.~-t.~Huang, C.~Wen and D.~Xie,
  ``The Positive orthogonal Grassmannian and loop amplitudes of ABJM,''
  arXiv:1402.1479 [hep-th].

 \bibitem{KimLee} 
  J.~Kim and S.~Lee,
  ``Positroid Stratification of Orthogonal Grassmannian and ABJM Amplitudes,''
  arXiv:1402.1119 [hep-th].

\bibitem{HS} A.~Henriques and D.~Speyer, ``The Multidimensional Cube Recurrence",
Advances in Mathematics {\bf 223} (2010), 1107--1136.

\bibitem{Lam} T.~Lam,
``Electroid varieties and a compactification of the space of electrical networks", [arXiv:1402.6261[math.CO]].

\bibitem{Elvang:2013cua} 
  H.~Elvang and Y.~-t.~Huang,
  ``Scattering Amplitudes,''
  arXiv:1308.1697 [hep-th].
  To be published as a textbook with Cambridge University Press. 


\bibitem{Parke:1986gb} 
  S.~J.~Parke and T.~R.~Taylor,
  ``An Amplitude for $n$ Gluon Scattering,''
  Phys.\ Rev.\ Lett.\  {\bf 56}, 2459 (1986).

\bibitem{Nair:1988bq} 
  V.~P.~Nair,
  ``A Current Algebra for Some Gauge Theory Amplitudes,''
  Phys.\ Lett.\ B {\bf 214}, 215 (1988).


\bibitem{MR0435831}
I.~M.~Gel'fand and G.~E.~Shilov, 
     ``Generalized functions.~Vol. 1. Properties and operations",
       (Translated from the Russian by Eugene Saletan),
   Academic Press [Harcourt Brace Jovanovich, Publishers], New
              York-London (1964 [1977]).


\bibitem{Hodges:2009hk} 
  A.~Hodges,
  ``Eliminating spurious poles from gauge-theoretic amplitudes,''
  JHEP {\bf 1305}, 135 (2013)
  [arXiv:0905.1473 [hep-th]].

 
\bibitem{OrlikTerao} Orlik, P., Terao, H.: Arrangements of Hyperplanes, Springer Verlag, 1992.

\bibitem{Brandhuber:2009xz} 
  A.~Brandhuber, P.~Heslop and G.~Travaglini,
  ``One-Loop Amplitudes in N=4 Super Yang-Mills and Anomalous Dual Conformal Symmetry,''
  JHEP {\bf 0908}, 095 (2009)
  [arXiv:0905.4377 [hep-th]].
  See also talk by P. Heslop: http://conference.ippp.dur.ac.uk/conferenceDisplay.py?confId=263

\bibitem{Elvang:2009ya} 
  H.~Elvang, D.~Z.~Freedman and M.~Kiermaier,
  ``Dual conformal symmetry of 1-loop NMHV amplitudes in N=4 SYM theory,''
  JHEP {\bf 1003}, 075 (2010)
  [arXiv:0905.4379 [hep-th]].

\bibitem{Drummond:2008cr} 
  J.~M.~Drummond and J.~M.~Henn,
  ``All tree-level amplitudes in N=4 SYM,''
  JHEP {\bf 0904}, 018 (2009)
  [arXiv:0808.2475 [hep-th]].



\bibitem{Bullimore:2009cb} 
  M.~Bullimore, L.~J.~Mason and D.~Skinner,
  ``Twistor-Strings, Grassmannians and Leading Singularities,''
  JHEP {\bf 1003}, 070 (2010)
  [arXiv:0912.0539 [hep-th]].

\bibitem{Broedel:2010rr} 
  J.~Brodel and S.~He,
  ``Dual conformal constraints and infrared equations from global residue theorems in N=4 SYM theory,''
  JHEP {\bf 1006}, 054 (2010)
  [arXiv:1004.2400 [hep-th]].

\bibitem{HeBai} 
  Y.~Bai and S.~He,
  ``The Amplituhedron from Momentum Twistor Diagrams,''
  arXiv:1408.2459 [hep-th].





\bibitem{TillDC} 
  T.~Bargheer, F.~Loebbert and C.~Meneghelli,
  ``Symmetries of Tree-level Scattering Amplitudes in N=6 Superconformal Chern-Simons Theory,''
  Phys.\ Rev.\ D {\bf 82}, 045016 (2010)
  [arXiv:1003.6120 [hep-th]].

  \bibitem{ABJMBCFW} 
  D.~Gang, Y.~-t.~Huang, E.~Koh, S.~Lee and A.~E.~Lipstein,
  ``Tree-level Recursion Relation and Dual Superconformal Symmetry of the ABJM Theory,''
  JHEP {\bf 1103}, 116 (2011)
  [arXiv:1012.5032 [hep-th]].

 \bibitem{Lipstein}
  A.~E.~Lipstein and L.~Mason,
  ``Amplitudes of 3d Yang Mills Theory,''
  arXiv:1207.6176 [hep-th].


   \bibitem{Embedding} 
  P.~A.~M.~Dirac,
  ``Wave equations in conformal space,''
  Annals Math.\  {\bf 37}, 429 (1936).
  
   G.~Mack and A.~Salam,
  ``Finite component field representations of the conformal group,''
  Annals Phys.\  {\bf 53}, 174 (1969).
  
   S.~L.~Adler,
  ``Massless, Euclidean quantum electrodynamics on the five-dimensional unit hypersphere,''
  Phys.\ Rev.\ D {\bf 6}, 3445 (1972)
  [Erratum-ibid.\ D {\bf 7}, 3821 (1973)].  

   R.~Marnelius and B.~E.~W.~Nilsson,
  ``Manifestly Conformally Covariant Field Equations and a Possible Origin of the Higgs Mechanism,''
  Phys.\ Rev.\ D {\bf 22}, 830 (1980).

\bibitem{2DK1} 
  T.~Goddard, P.~Heslop and V.~V.~Khoze,
  ``Uplifting Amplitudes in Special Kinematics,''
  JHEP {\bf 1210}, 041 (2012)
  [arXiv:1205.3448 [hep-th]].

\bibitem{2DK2}
  S.~Caron-Huot and S.~He,
  ``Three-loop octagons and $n$-gons in maximally supersymmetric Yang-Mills theory,''
  JHEP {\bf 1308}, 101 (2013)
  [arXiv:1305.2781 [hep-th]].

\bibitem{UnPub} 
D.~Gang, Y.~-t.~Huang, E.~Koh, S.~Lee and A.~E.~Lipstein, unpublished notes.  

\bibitem{Huang:2012vt} 
  Y.~-t.~Huang and S.~Lee,
  ``A new integral formula for supersymmetric scattering amplitudes in three dimensions,''
  Phys.\ Rev.\ Lett.\  {\bf 109}, 191601 (2012)
  [arXiv:1207.4851 [hep-th]].

  \bibitem{Engelund:2014sqa} 
  O.~T.~Engelund and R.~Roiban,
  ``A twistor string for the ABJ(M) theory,''
  JHEP {\bf 1406}, 088 (2014)
  [arXiv:1401.6242 [hep-th]].

 \bibitem{ArkaniHamed:2009dg} 
  N.~Arkani-Hamed, J.~Bourjaily, F.~Cachazo and J.~Trnka,
  ``Unification of Residues and Grassmannian Dualities,''
  JHEP {\bf 1101}, 049 (2011)
  [arXiv:0912.4912 [hep-th]].

\bibitem{OlsonInprep}
  T.~M.~Olson,
  ``Orientations of BCFW Charts on the Grassmannian,''
  in preparation.

\bibitem{Griffiths}
P.~A.~Griffiths and J.~Harris (1978) \textit{Principles of Algebraic Geometry}, John Wiley \& Sons.


  
\end{thebibliography}
\end{document}